\documentclass[preprint]{aastex}

\usepackage{natbib}
\bibpunct{(}{)}{;}{a}{}{,}

\usepackage{longtable}
\def  \reply{}

\shortauthors{Salo et al.}

\begin{document}




\title{Spitzer Survey of Stellar Structure in Galaxies (S$^4$G). The Pipeline 4:
Multi-component decomposition strategies and data release \\ {\reply (revised 19/11/2014)}}


\author{Heikki Salo\altaffilmark{1}, 
Eija Laurikainen\altaffilmark{1,2}, 
Jarkko Laine\altaffilmark{1}, 
Sebastien Comer\'on\altaffilmark{1,2}, 
Dimitri A. Gadotti\altaffilmark{3},
Ron Buta\altaffilmark{4},
Kartik Sheth\altaffilmark{5},
Dennis Zaritsky\altaffilmark{6},
Luis Ho\altaffilmark{7,8},  
Johan Knapen \altaffilmark{9,10},
E. Athannassoula\altaffilmark{11},
Albert Bosma\altaffilmark{11},
Seppo Laine\altaffilmark{12},
Mauricio Cisternas\altaffilmark{9},
Taehyun Kim\altaffilmark{5,13,3},
Juan Carlos Mu\~noz-Mateos\altaffilmark{5}
Michael Regan\altaffilmark{14},  
Joannah L. Hinz\altaffilmark{6},
Armando Gil de Paz\altaffilmark{15},
Karin Menendez-Delmestre\altaffilmark{16},
Trisha Mizusawa\altaffilmark{5,17},
Santiago Erroz-Ferrer\altaffilmark{9,10},
Sharon E. Meidt\altaffilmark{18},
Miguel Querejeta\altaffilmark{18}}

\altaffiltext{1}{Dept. of  Physics, University of Oulu, FI-90014, Finland}
\altaffiltext{2}{Finnish Centre for Astronomy with ESO (FINCA), University of Turku, V\"ais\"alantie 20, FI-21500 Piikki\"o, Finland}
\altaffiltext{3}{European Southern Observatory, Casilla 19001, Santiago 19, Chile}
\altaffiltext{4}{Department of Physics and Astronomy, University of Alabama, Box 870324, Tuscaloosa, AL 35487, USA}
\altaffiltext{5}{National Radio Astronomy Observatory / NAASC, 520 Edgemont Road, Charlottesville, VA 22903}
\altaffiltext{6}{University of Arizona, 933 N. Cherry Ave, Tucson, AZ  85721}
\altaffiltext{7}{Kavli Institute for Astronomy and Astrophysics, Peking
University, Beijing 100871, China}
\altaffiltext{8}{Department of Astronomy, School of Physics, Peking University, Beijing 100871, China}
\altaffiltext{9}{Instituto de Astrof\'{i}sica de Canarias, 38205 La Laguna, Spain}
\altaffiltext{10}{Departamento de Astrof\'{i}sica, Universidad de La Laguna, 38206 La Laguna, Spain}
\altaffiltext{11}{Aix Marseille Universite, CNRS, LAM (Laboratoire d'Astrophysique de Marseille) UMR 7326, 13388, Marseille, France}
\altaffiltext{12}{{\it Spitzer} Science Center - Caltech, MS 314-6, Pasadena, CA 91125, USA}
\altaffiltext{13}{Astronomy Program, Department of Physics and Astronomy, Seoul National University, Seoul 151-742, Korea} 
\altaffiltext{14}{Space Telescope Science Institute, 3700 San Martin Drive, Baltimore, MD 21218}
\altaffiltext{15}{Departamento de Astrof\'{i}sica, Universidad Complutense de Madrid, Madrid 28040, Spain}
\altaffiltext{16}{Observatorio do Valongo, Universidade Federal de Rio de Janeiro, Ladeira Pedro Antonio, 43, Saude CEP 20080-090, Rio de Janeiro - RJ - Brasil}
\altaffiltext{17}{Florida Institute of Technology, Melbourne, FL 32901}
\altaffiltext{18}{Max-Planck-Institut f\"ur Astronomie, K\"onigstuhl 17, D-69117 Heidelberg, Germany}






\begin{abstract}
  The Spitzer Survey of Stellar Structure in Galaxies (S$^4$G,
  \citealt{sheth2010}) is a deep 3.6 and 4.5 $\mu$m imaging survey of
  2352 nearby ($< 40$ Mpc) galaxies.  We describe the S$^4$G data
  analysis pipeline 4, which is dedicated to 2-dimensional structural
  surface brightness decompositions of 3.6 $\mu$m images, using
  GALFIT3.0 \citep{peng2010}.  Besides automatic 1-component
  S\'ersic fits, and 2-component S\'ersic bulge + exponential disk
  fits, we present human supervised multi-component decompositions,
  which include, when judged appropriate, a central point source,
  bulge, disk, and bar components.  Comparison of the fitted
  parameters indicates that multi-component models are needed to
  obtain reliable estimates for the bulge S\'ersic index and
  bulge-to-total light ratio ($B/T$), confirming earlier results
  \citep{laurikainen2007, gadotti2008, weinzirl2009}.  In this first
  paper, we describe the preparations of input data done for
  decompositions, give examples of our decomposition strategy, and
  describe the data products released via IRSA and via our web page
  ({\bf \tt www.oulu.fi/astronomy/S4G\_PIPELINE4/MAIN}). These
  products include all the input data and decomposition files in
  electronic form, making it easy to extend the decompositions to suit
  specific science purposes.  We also provide our IDL-based
  visualization tools (GALFIDL) developed for displaying/running
  GALFIT-decompositions, as well as our mask editing procedure
  (MASK\_EDIT) used in data preparation. In the second paper we will
  present a detailed analysis of the bulge, disk, and bar parameter
  derived from multi-component decompositions.


\end{abstract}






\section{Introduction}



How and when did the baryonic mass assemble into galactic disks?  How
does the fraction of mass confined into bulges evolve over time?  How
common are galaxies that have no classical bulges, i.e. bulges that
have their origin in the early mergers of dark matter halos and
baryonic disk systems?  These are difficult questions to answer
because galaxy evolution involves secular processes such as gas
accretion via filaments, where mass presumably ends up in bulges or
disks, or internal dynamical evolution, such as the formation of bars
which further re-distribute matter in galaxies.  Galaxies in the local
Universe are the present day manifestations of this evolution and
hence provide important clues on the evolutionary processes which took
place in the past.

The Spitzer Survey of Stellar Structure in Galaxies (S$^4$G,
\cite{sheth2010}) provides an excellent data base with which to
measure the stellar mass distribution of galaxies in the local
Universe. It is a survey of 2352 galaxies observed in the mid-IR at
3.6 and 4.5 $\mu$m, wavelengths that are largely unaffected by
internal extinction \citep{draine1984}, and trace mainly the old
stellar population (\citealt{pahre2004}; however see also
\citealt{meidt2012} and \citealt{driver2013}), so that the
mass-to-luminosity (M/L) ratio in these bands is nearly constant
inside the galaxies \citep{peletier2012}. This is particularly
important for deriving the properties of bulges and disks, because
dust and star formation are more pronounced in the disks than in the
bulges, which in the optical region affect their relative M/L-ratio
and thus the relative fraction of the bulge light
\citep{driver2013}. Dust and star formation are significant also in
the bulges of late-type galaxies \citep{fisher2006,gadotti2001}.  The
S$^4$G images are deep, reaching azimuthally averaged stellar mass
surface densities of $\sim$ 1 M$_\sun$ pc$^{-2}$, where the baryonic
mass budget at least in spiral and irregular galaxies is typically
dominated by atomic gas.  S$^4$G covers a large range of galaxy
magnitudes (over three decades in stellar mass), which makes possible
to study both late-type dwarfs and bright galaxies in a uniform
manner, and to study when the disk instabilities such as bar formation
start to play an important role.  Our sample extends to lower galaxy
luminosities than most previous samples in which bars have been
studied \citep{barazza2008, sheth2008, nair2010, melvin2014}.  {\reply
  Besides galaxy mass, another central factor affecting its structural
  evolution is its environment \citep{vanderwel2008, kormendy2012,
    weinzirl2014}. S$^4$G includes galaxies up to 40 Mpc and covers a
  wide range of different galaxy environments, including several
  galaxy groups and the Virgo and Fornax clusters (see Fig. 2 in
  \citealt{sheth2010}).  }

Plenty of information for the S$^4$G sample is already publicly
available via the IRSA archive. The data have been processed through
Pipeline 1 \citep{regan2014} (hereafter P1) which makes mosaics of the
observed individual frames, Pipeline 2 (\citealt{munoz_mateos2014};
P2) which makes masks of the foreground stars and image defects, and
Pipeline 3 (\citealt{munoz_mateos2014}; P3) which measures the basic
photometric parameters like the galaxy magnitudes and concentration
indices. In Pipeline 4 (P4), described in this study, we decompose the
two-dimensional flux distributions of the images into several
structural components using GALFIT (Peng et al. 2010).  Because even
the mid-IR wavelengths are not completely free of such contaminants as
hot dust, mass maps are also created for the images in Pipeline 5 (P5,
Querejeta et al. 2014).  The galaxies in S$^4$G have been visually
classified at 3.6 $\mu$m by \citep{buta2010, buta2014}, and we use
these classifications in the present study. Optical images are also
available for the majority of the S$^4$G sample \citep{knapen2014}.

For all of the S$^4$G galaxies for which the image quality is good
enough (e.g. no superposed bright stars, or image defects), we provide
1-component single S\'ersic, 2-component bulge-disk (S\'ersic +
exponential), and multi-component decompositions, fitting up to four
separate structure components.  Our main goal is to estimate the
parameters of the bulge and the disk in a {\reply robust} manner, which is
the motivation for our decomposition approach. In particular, it is
important to include bar-components in the decompositions because the
flux of the bar is easily mixed with the flux of the bulge
\citep{laurikainen2006}.  
{\reply Our bulge is defined as a 'photometric bulge', including the flux
in excess of that in disk and bar components; the decompositions
  themselves do not make assumptions about the physical nature of the bulge,
  whether a rotation supported classical bulge or a disk star
  formation/bar vertical buckling related pseudo bulge (see
  \citealt{kormendy2004, atha2005}).}
To measure the
scale lengths and central surface brightness of the disks in a uniform
fashion, an exponential function is used whenever possible, instead of
a generalized S\'ersic function. It is well known that galactic disks
can have more than one exponential sub-section \citep{freeman1970, erwin2005}. In this study we handle this in a fairly
conservative manner: two separate functions (added together) are used
to fit the disk in galaxies where distinct inner and outer components
of different surface brightness are present, but not in all cases in
which a disk break ('truncation' or 'anti-truncation') of some degree has been
reported in the literature. Our multi-component approach is similar to
those used previously by \cite{laurikainen2005, laurikainen2007,
  laurikainen2010}, \cite{gadotti2009}, and \cite{weinzirl2009}.  Our
motivation for offering also the single S\'ersic and bulge-disk
decompositions is that they are routinely used in large galaxy surveys
and high-redshift studies \citep{hausler2013,lackner2012,cameron2009,allen2006,driver2006,driver2013}.
Although single S\'ersic fits are not good tracers of the
properties of bulges, they are still useful in gross classification of
galaxies.

The decomposition results, released via IRSA and our web-page, are
given in such a manner that they can be easily extended having
different scientific goals in mind.  The decompositions were done via
GALFIDL, which consists of IDL-based tools for displaying and running
GALFIT (see Sect \ref{sect_galfidl}). It is important to note that due to the large amount of work
involved, P4 was started as soon as the first P1 data was
available. Because of this we did our own mask editing, and
orientation and sky background estimation\footnote{The derived sky background values and orientation parameters turned out to be in very good  agreement with P3, see Sect \ref{sect_sky}.}. These masks form
part of the final P2 masks.  Due to later changes in P1, part of the
images used in P4 contain minor shifts (or differ in size by 1-2
pixels) compared to the finalized P1 images in IRSA. Rather than repeating
the time consuming GALFIT decompositions with the updated images, we
provide together with the decomposition output files 
the sky subtracted data and mask images we used.

In this paper, we describe the decomposition method and model
components, the preparation of the data for decompositions, and
concentrate on illustrating our philosophy behind the construction of
the final multi-component decompositions. The results published in tabular
form include the outer disk orientation estimates, S\'ersic
parameters from the 1-component fits, and the final parameters from
multi-component decompositions, together with a quality flag for each
galaxy.  The data products released via IRSA include the GALFIT output
files, and all the input fits-files needed for repeating and refining the
decompositions. The P4 web pages illustrate the same models in
pictorial form, and also provide the GALFIDL code and documentation.
(The IRSA products and the P4 web page are described in the two Appendixes).
Analysis of the derived bulge, disk, and bar parameters will be presented
in paper 2 (Salo et al, in prep.).


\section{Decomposition Pipeline}

\subsection{Decomposition method and model functions}

Our decompositions use the GALFIT-software \citep{peng2002, peng2010}, which
has become the {\em de facto} standard for detailed two-dimensional
structural decompositions. It relies on parametric fitting, using the
Levenberg-Marquadt algorithm to minimize the weighted residual
${\chi^2}_{\nu}$ between observed (OBS) and model (MODEL) images,

\begin{equation}
 {\chi^2}_{\nu}              = \frac{1}{N} \sum_x \sum_y \frac{\left[OBS(x,y)-MODEL(x,y)\right]^2}{\sigma(x,y)^2} .
\end{equation}

\noindent The sum is taken over all the used (non-masked) image
pixels, and $\sigma(x,y)$ indicates the statistical uncertainty of
each pixel ({\em sigma-image}). The model image consists of a sum of
model components, i.e for bulge, disk, bar etc, convolved with the
image Point-Spread Function (PSF-image). Note that the reduced
${\chi^2}_{\nu}$ is used, with $N$ denoting the degree of freedom,
equal to the number of fitted pixels minus the number of free
parameters in the fit.

GALFIT is extremely versatile in its selection of model components.
Basically the user defines for each component its 'radial' profile
function, giving the surface brightness $\Sigma(r)$ at each isophotal
radial coordinate $r$.  The isophotal coordinates are most commonly
defined in terms of generalized ellipses \citep{atha1990},

\begin{equation}
r(x',y') = \left(|{x'-x_0}|^{C+2} + \left|\frac{y'-y_0}{q}\right|^{C+2}\right)^{\frac{1}{C+2\phantom{^2}}} .
\end{equation}

\noindent Here $x_0,y_0$ defines the center of the ellipse, $q=b/a$ is
the ratio between minor and major axis lengths. The $x',y'$ denote
coordinates in a system aligned with the ellipse, with the major axis
pointing at the position angle $PA$. For pure ellipses $C=0$, while
$C>0$ indicates boxy and $C<0$ disky isophotes\footnote{Note that in
  the original notation of \cite{atha1990} the exponent 'C+2' was
  denoted with c, a pure ellipse thus corresponding to $c=2$. Similar
  notation was used also in e.g. \cite{gadotti2011}. However, we will
  here follow the notation of GALFIT \citep{peng2002, peng2010}}. For
the pipeline decompositions, simple elliptical isophotes $C=0$ are
used for all components.  Besides generalized ellipses, GALFIT
provides several alternatives, such as definition of isophotal shape via
azimuthal or bending modes, or via coordinate rotations, which would
form a natural basis for detailed modeling of e.g. logarithmic
spirals.  To keep our models relatively simple (and uniform over the
wide range of angular sizes and surface brightnesses spanned by the
sample), we have not used these advanced GALFIT features. Keeping the
models simple makes the interpretation of the observation minus model
residuals more straightforward (see the NGC 1097 examples in \cite{sheth2010}).

The pipeline decompositions use five different choices
for the model components/radial functions:

\noindent 1) The {\bf bulge component} is  described with a S\'ersic profile ({\it ``sersic''})
\begin{equation}
\Sigma(r)= \Sigma_e \exp \left(-\kappa \left[ (r/R_e)^{1/n} -1\right]\right),
\end{equation}
\noindent where $\Sigma_e$ is the surface brightness at the effective
radius $R_e$ (isophotal radius encompassing half of the total flux of the component).
The S\'ersic-index $n$ describes the shape of the radial profile, which becomes
steeper with increasing $n$. In particular, $n=1$ corresponds to
an exponential profile and $n=4$ to a de Vaucouleurs profile. The factor
$\kappa$ is a normalization constant determined by $n$.  In GALFIT the
corresponding {\em ``sersic''}-function is used, with the integrated
magnitude $m_{bulge}$ as a free parameter (instead of $\Sigma_e$).

\noindent 2) In the case of low or moderate inclination, the {\bf disk
  component} is described with an infinitesimally thin exponential disk ({\it ``expdisk''})
,
\begin{equation}
  \Sigma(r)= \Sigma_o q^{-1} \exp (-r/h_r),
\end{equation}
\noindent where $\Sigma_o$ is the central surface brightness of the
disk observed from the perpendicular direction and $h_r$ denotes the
exponential scale length. In this case the $q= \cos i$, where $i$ is
the disk inclination. Assuming no extinction, $\Sigma_o q^{-1}$ is the
projected surface brightness at the sky plane.  The {\em
  ``expdisk''}-function in GALFIT is used, with integrated $m_{disk} =
-2.5 \log_{10} (2\pi \Sigma_0 \ {h_r}^2)$ as a free parameter (instead
of $\Sigma_0$).  {\reply Note that in cases that had more than one disk
  component, the inner disk was sometimes fit with a {\em sersic}
  or {\em ferrer2} function, to allow the profile to drop faster than
  with {\em expdisk}.}

\vskip 0.5cm

\noindent 3) For a nearly {\bf edge-on disk} (apparent axial
ratio $q \lesssim 0.2$), the function ({\em ``edgedisk''})
\begin{equation}
\Sigma(r_x,r_z)= \Sigma_o \frac{r_x}{h_r} K_1 \left(\frac{r_x}{h_r}\right) \ {\rm sech}^2 (r_z/h_z),
\end{equation}
is adopted, where $r_x$ and $r_z$ are the (positive) distances along and
perpendicular to the apparent major axis of the disk, and $K_1$ stands
for a modified Bessel function. This function corresponds to the
line-of-sight (viewing along the disk plane) integrated surface brightness
of a 3D luminosity density distribution \citep{vanderkruit1981}
\begin{equation}
L(r_x,r_z) = \frac{\Sigma_0}{2h_r} \exp(-r_x/h_r) \ {\rm sech}^2 (r_z/h_z).
\end{equation}

\vskip 0.5cm

\noindent 4) For a {\bf bar component} a modified Ferrers profile ({\it ``ferrer2''}) is assumed,
\begin{equation}
\Sigma(r)= 
\left\{ \begin{array}{ll} 
\Sigma_o \left[1- (r/r_{out})^{2-\beta}\right]^\alpha  & r < r_{out}\\
0  & r \ge r_{out}
\end{array} \right.
\end{equation}
Here $r_{out}$ defines the outer cut of the profile, while $\alpha$
defines the sharpness of this cut. The parameter $\beta$ defines
the central slope of the profile, and $\Sigma_0$ is the central
surface brightness (in the plane of the sky). 

\vskip 0.5cm

\noindent 5) When the galaxy contains an {\bf unresolved
  central component} it is fit with a PSF-convolved point source
({\em ``psf''}).  In this case the free parameter is the total
magnitude $m_{psf}$. Typically this component, if present,
is not an active or starburst nucleus, but rather a small bulge with
angular size so small that it cannot be resolved in the S$^4$G images
($R_e \lesssim FWHM= 2.1\arcsec$ of S$^4$G images).

\vskip 0.5cm

For the decomposition pipeline we chose to do three types of
decompositions: 1) one-component S\'ersic-fits, 2) two-component
bulge-disk decompositions using S\'ersic-bulges and exponential disks
(or edge-on disk if appropriate), and 3) multi-component 'final'
decompositions, optionally with additional 
bar{\reply , disk} and central components (the level of complexity of the models is discussed in more detail in
Section \ref{sect_eija_examples}).
The
first two types of models are made in an automatic manner, while the
final models always include human judgment about what components should be included.


\subsection{Preparation of data for decompositions}
\label{sect_data}

\subsubsection{What is needed?}

The S$^4$G data analysis Pipeline P1 \citep{regan2014} provides
image mosaics in both 3.6 and 4.5 $\mu$m, accompanied with {\em
weight-images}, which indicate for each pixel location the number of
original frames covering it.  Together with the header information,
these weight images provide the means for producing the {\em sigma-images}
used in GALFIT. 

Before decompositions can be started, frames masking the
foreground/background objects and various image defects are needed.
Additionally we need the galaxy centers, sky background values, and
the orientation of the galaxy relative to the sky plane, estimated from the
shape of the galaxy's outer isophotes.  In principle, this additional
input for decompositions 
are published for all S$^4$G galaxies in
\citealt{munoz_mateos2014}, available via IRSA/S$^4$G Pipeline 3.
However, at the time our Pipeline 4 decompositions were made, these
data was not yet available.  Therefore, we made our own sky background
estimates and ellipse fits.  Also, the automatically created masks
(see \citealt{munoz_mateos2014}) were visually inspected and hand-edited
when needed (these edited masks later became part of the final S$^4$G
masks). If the decompositions are rerun starting from the output files
provided in P4, it is important to use the data and mask images, as well
as the pre-defined parameters offered via P4. 

In summary, Pipeline 4 consists of scripts for editing the masks,
determining the galaxy centers, estimating the sky background,
fitting isophotal ellipses, preparing the input files, and running
GALFIT. It also includes tools for visualization of the GALFIT output
files (see Sect \ref{sect_galfidl}) and routines for storing the data on
IRSA server (Appendix A) and the P4 web pages (Appendix B).

\subsubsection{Mask images}

The raw masks for the S$^4$G 3.6 $\mu$m images were made in P2 with the
SExtractor software \citep{bertin1996}, as described in \cite{munoz_mateos2014}. Various automatic detection thresholds for point
sources were used.  However, it soon became evident that no single
criterion was sufficient to exclude all extra sources, without
sometimes affecting also the galaxy light itself, in which case the
masks needed manual editing. Also, in some cases the images contained
artifacts that needed to be removed by hand.  To speed-up this editing
process, we developed a small portable IDL-routine (MASK\_EDIT).
Basically it displays on the screen simultaneously the original and
masked images, and allows the user to remove/insert masked regions 
interactively.  As an initial step of P4, all of the raw 3.6
$\mu$m masks were visually checked and edited if needed. The resulting
masks are suitable for the purposes of our structural
decompositions. However, because the wings of the PSF are quite extended
(see Section \ref{sect_psf}) more extensive
masking might be required in some applications. 
 The
MASK\_EDIT routine, with source code and examples of use, is available
at the P4 web page.

\subsubsection{Galaxy centers, sky background, isophotal profiles}
\label{sect_sky}

After the edited masks were completed, we run the galaxies through a
semi-automatic IDL script which determines the galaxy centers, sky
background levels and galaxy orientation parameters.  The accurate
galaxy center is measured with the {\tt cntrd}-routine\footnote{{\tt
    cntrd} is part of IDL Astronomy Library \citep{landsman1993}. It locates the position where the brightness gradient is zero.},
after its approximate location is interactively defined. We also have
the option to mark the center by force, in case the automatic center
finding routine does not work satisfactorily even after repeated
trials.

\begin{figure}
\vskip -1cm \hskip -2cm \includegraphics[angle=0,width=19cm]{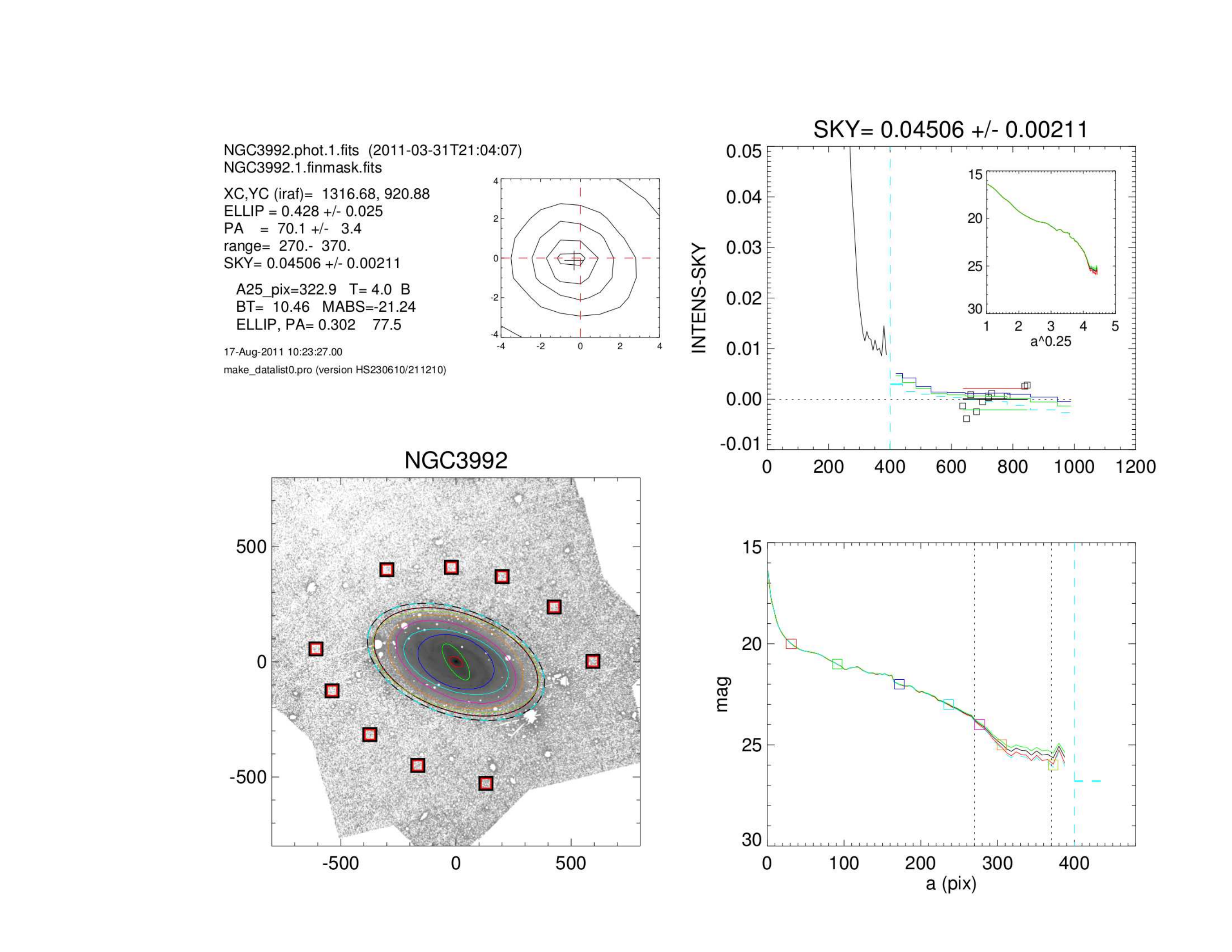}
\caption{Example of the determination of sky background for NGC 3992.
\baselineskip 0.52cm
The small {\bf upper left} frame illustrates the center location found
with {\em cntrd}-routine (black cross) relative to the image isophotes
near the center (dashed red lines indicate the nearest integer
pixels).
In the {\bf lower left} frame, red boxes indicate the local regions used for
estimating the sky background: the mean and the RMS of these local median
values were adopted for the sky background and its uncertainty (SKY and
DSKY, respectively). The dashed ellipse indicates the visually estimated 
galaxy size ($R_{gal}$). The white specks indicate masked stars.
The {\bf upper right} frame shows the intensity profile after
subtracting the SKY value (indicated in the title of the plot; note the
linear scale, intensities are from IRAF {\em ellipse} fits), marking
also the $\pm DSKY$ (vertical red/green lines) and the median sky
values in local measurement regions (boxes). The vertical dashed line
corresponds to $R_{gal}$ The insert shows the same profile, but as magnitude
versus  $a^{0.25}$, where $a$ is the isophotal major-axis
distance: a de Vaucouleurs profile would appear a straight line in this
plot.
The {\bf lower right} frame shows the intensity profile in magnitude
units (AB-magnitudes): red/green profiles correspond to
adding/subtracting DSKY to the sky background. 
%
All distances are in pixels ($0.75\arcsec$). Similar plots
for all sample galaxies are given on the P4 web site.}

{\reply The Figure + caption modfied according to referee's suggestion}
\label{sky_example1}
\end{figure}

The regions used for estimation of the sky background are identified
manually, by selecting several (typically 10-20) locations outside the
visible galaxy, while avoiding the image edges or contaminated
areas. The local sky values in these locations are obtained by taking
medians of the non-masked pixels in 40 pix $ \times$ 40 pix boxes. The
global sky background (SKY) and its uncertainty (DSKY) are then
estimated from the mean and standard deviation of these local values,
respectively (see Fig. \ref{sky_example1}).  In section
\ref{sect_errors}, using the estimated DSKY, we show that the expected
uncertainty of decomposition parameters caused by possible
uncertainties in background subtraction is negligible. We also
determine the average RMS sky variation, by taking the median of
standard deviations in different sky regions (after removing outliers
by iterative 3-sigma clipping).  We use the sky RMS estimates in
Section \ref{sect_sigma} for assessing the validity of theoretically
calculated sigma-images.

\begin{figure}
\hskip -2cm \includegraphics[angle=0,width=19cm]{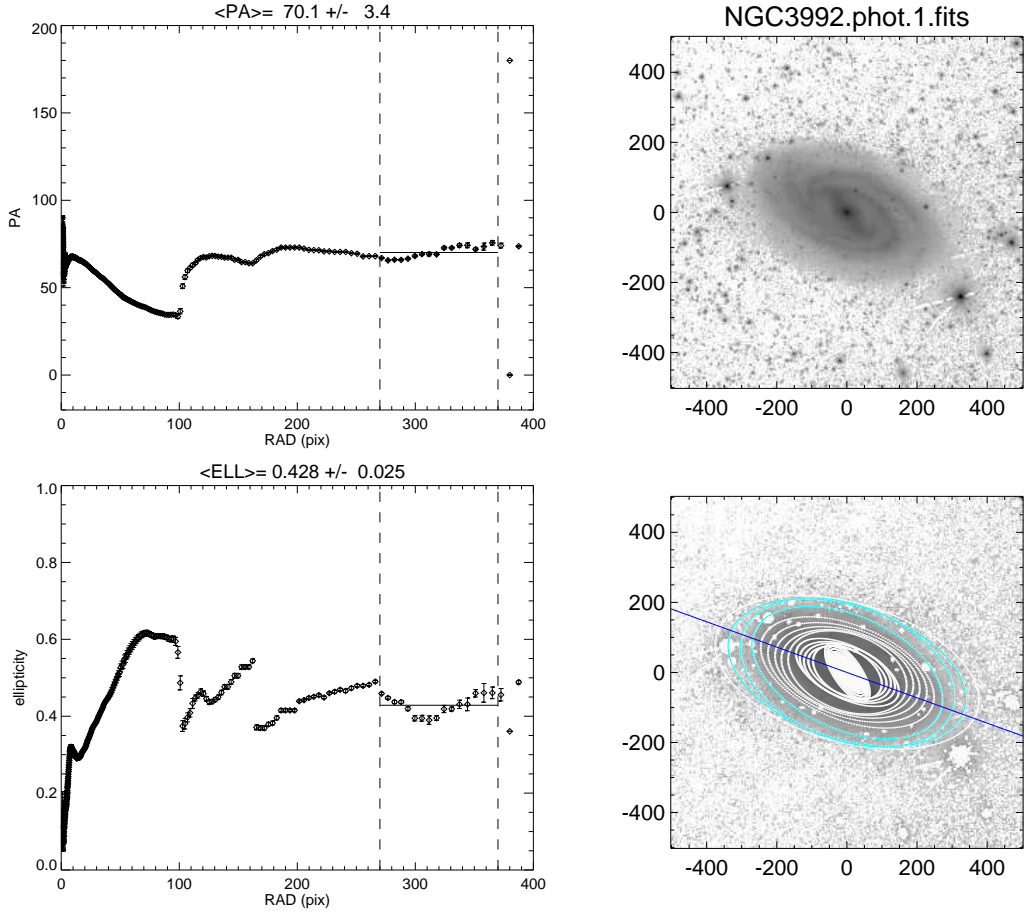}
\caption{Example of the isophotal profiles derived for NGC 3992 using
  the  IRAF {\em ellipse} routine.  The plots in the left display the PA
  and ellipticity profiles versus semimajor axis ($a$) of isophote
  ellipse: the dashed vertical lines indicate the range used in
  estimating the outer disk orientation parameters; solid horizontal
  line indicates the mean over that range. The upper right plot shows
  the observed, sky subtracted image, clipped at $ 1.3 \times R_{gal}$
  (the image region used in decompositions). In the lower right the
  isophotes are plotted on top of observed (masked) image: the
  blue isophotes correspond to the $a$ range from which the outer disk
  orientation ($\epsilon_{outer}, PA_{outer}$) was derived; the blue
  line indicates the assigned $PA_{outer}$. Similar plots for all sample
  galaxies are available on the P4 web site.}
\label{sky_example2}
\end{figure}

We calculate the isophotal profiles with a {\em pyraf} script
called from IDL,  using the standard IRAF {\tt
  ellipse} algorithm \citep{jerd1987}. As inputs for the ellipse
fitting the sky background subtracted data image and the edited mask
image are used.  We fix the ellipse center to the
previously found galaxy center and  use a logarithmic increment of 0.02
between isophote levels. As often happens with IRAF {\tt
  ellipse}, the fit does not necessarily converge over the whole galaxy
area: we have an option to re-try the fit with different
starting locations until a successful fit is obtained over the whole
galaxy region (see Fig. \ref{sky_example2}). From the isophotal
profiles, we choose a semi-major axis range from which outer
orientations ($(b/a)_{outer} = 1 -\epsilon_{outer}, PA_{outer}$) are
estimated. Also, a rough estimate of the galaxy outer radius,
$R_{gal}$, is made to define 
the image region used in the GALFIT decomposition.

\begin{figure}
\hskip -.5cm\includegraphics[angle=0,width=16cm]{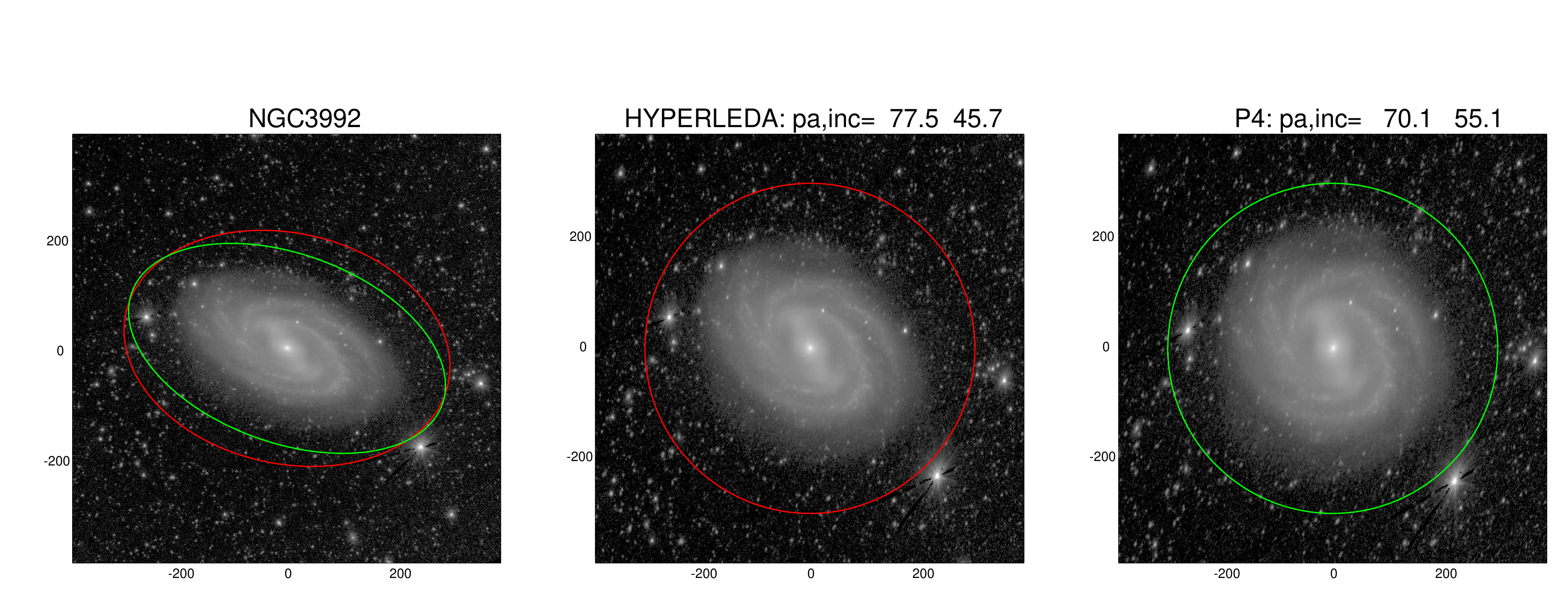}
\caption{Example of deprojections with HyperLeda and P4 orientation
  parameters ($(b/a)_{outer}$ and $PA_{outer}$) for NGC 3992. 
 The red and green ellipses
  on the original image in the left illustrate the HyperLeda and P4
  parameters, respectively (semimajor axis of the ellipse equals
  $R_{gal}$), the frame in the middle shows the deprojection with
  HyperLeda orientation, and that in the right with P4 parameters:
  here the inclination is taken simply as $i= \cos ^{-1} (b/a)$ (note
  that this differs from the inclination (``incl'') listed in
  HyperLeda, which includes a morphological type-dependent correction for
  the assumed disk thickness).  Clearly, the face-on disk is closer to
  axisymmetric when using P4 parameters: the difference would remain
  if the thickness-corrected HyperLeda inclination were used
  (47 degrees for this example). Similar plots for all sample galaxies can be found in the P4
  web site.
}
\label{painc_leda_example}
\end{figure}

Figures \ref{sky_example1} and \ref{sky_example2} give examples of
typical plots produced during these preparatory steps, illustrating
the sky background fitting and the elliptical isophote profiles.  The
estimated $\epsilon_{outer}$ and $PA_{outer}$ are marked.  In all our
decompositions, we fix the orientation of the disk component to these
outer values\footnote{The reason is to reduce the degeneracy of
  different model components in decompositions} and interpret them to
represent the galaxy viewing inclination.  Therefore, extra care is
taken to estimate the orientations reliably. For example, the
corresponding inclination $i_{disk} = \cos^{-1} (b/a)_{outer}$ is
visually checked by de-projecting the galaxy images to face-on. Figure
\ref{painc_leda_example} shows an example of such a de-projection,
also comparing the estimated inclinations with those calculated from
axial ratios given in the HyperLeda database. Typically, our
$(b/a)_{outer}$ and $PA_{outer}$ are determined at much lower surface
brightness levels than those in HyperLeda (which are mainly from
RC3 \cite{rc3}). P4 values are thus less affected by bulges, bars, or prominent
spirals, and should reflect better the orientation of the underlying
extended disk, which appears with circular outer isophotes in face-on
projection\footnote{This expectation is of course not valid for a
  vertically extended (say $T \le 0$) galaxy disk, nor in the case of
  intrinsically non-circular disks. However, the fitted GALFIT
  expdisk-function assumes an infinitesimally thin intrinsically
  axisymmetric disk, so any other treatment would be inconsistent in
  the decompositions.}.  In some cases, S$^4$G images are so deep that
the outermost isophotes are dominated by an outer stellar halo rather
than the disk. Good examples are NGC 681, NGC 1055 and NGC
4594. Possible misinterpretations of the outer isophotes were avoided
by visually inspecting all the images: when the disk (identified with
spiral arms, rings, and lenses) is clearly more inclined than
suggested by the outermost isophotes of the image, the isophotes in
the disk region were used for the estimate of galaxy orientation. For
nearly face-on galaxies, the possible stellar halos are more difficult
to distinguish, but in these cases the involved error in the
orientation is less important.  The final P4 axial ratios and position
angles, center locations, and sky background values are listed in
Table I.  For each galaxy we also include a flag indicating the
inclination uncertainty: 'ok' indicates that outer isophote axial
ratio should give a reliable estimate of $i_{disk}$, 'u' indicates
that the inclination is uncertain, while 'z' indicates that the galaxy is close
to edge-on (the axial ratio is not used for an inclination estimate).

\begin{figure}
\includegraphics[angle=0,scale=.8]{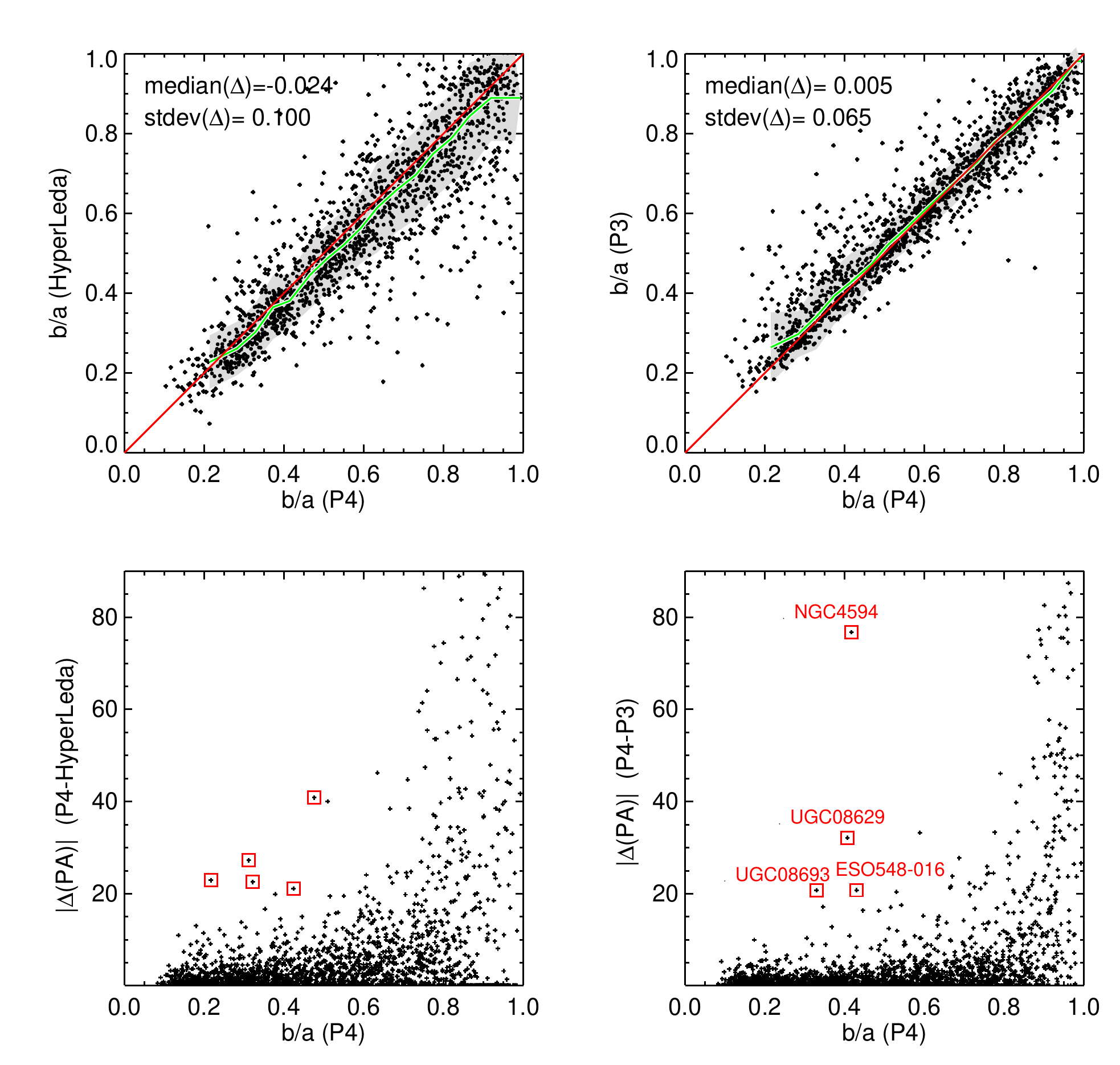}
\caption{Comparison of outer disk orientation parameters to those in
  HyperLeda, and to the P3 orientation parameters corresponding to
  fixed $\mu_{3.6}=25.5  $ mag/arcsec$^2$ isophote.  In the upper row
  axial ratios are compared: the green line indicates the running
  median of HyperLeda (or P3) axial ratio, calculated in bins of 100
  galaxies; gray indicates the RMS scatter in the bin. Red line
  indicates unit slope. The labels give the median difference and rms
  of the difference compared to P4: P4 axial ratios are generally
  larger than those in HyperLeda while the difference to P3 is
  small. The lower frames displays the absolute difference in the
  position angles: squares mark deviant points with $|\Delta PA| >
  15^\circ$ for $b/a<0.5$. In the upper frames only galaxies with
  orientation uncertainty flag 'ok' are included, while in the lower frames
  also galaxies with flag 'z' are included.}
\label{painc_leda_scatter}
\end{figure}

A scatter plot of P4 axial ratios versus HyperLeda values is presented
in Fig. \ref{painc_leda_scatter} (upper left frame; only galaxies with
flag='ok' are shown).  As anticipated, the P4 axial ratios are on the
average closer to unity than those in HyperLeda, though the difference
is not very large (median $(b/a)_{P4}-(b/a)_{HyperLeda}=0.024$). On
the other hand the standard deviation of the difference is quite large
($\sim 0.1$).  The upper right frame makes a similar comparison to P3
axial ratios \citep{munoz_mateos2014} which correspond to a fixed
surface brightness level $\mu_{3.6} =25.5$ mag/arcsec$^2$. On average,
P4 orientations are measured at about 0.9 times this distance. The
scatter is now significantly reduced and no systematic difference is
seen between P3 and P4.  The lower frames in Fig.
\ref{painc_leda_scatter} compares the position angles (now also 'z'
galaxies are included). In general the differences between P4 and
HyperLeda are fairly small: for $b/a  <0.8$ the median absolute
difference is $2^\circ$. The difference between P4 and P3 is even
smaller (median absolute difference $0.9^\circ$ for $b/a <0.8$).
Nevertheless there are some exceptions, most notably NGC4594 (The
Sombrero Galaxy): in this case the P3 fixed isophote orientation
corresponds to the extended halo, while the P4 orientation refers to the
edge-on disk.

\begin{figure}
\includegraphics[angle=0,scale=.9]{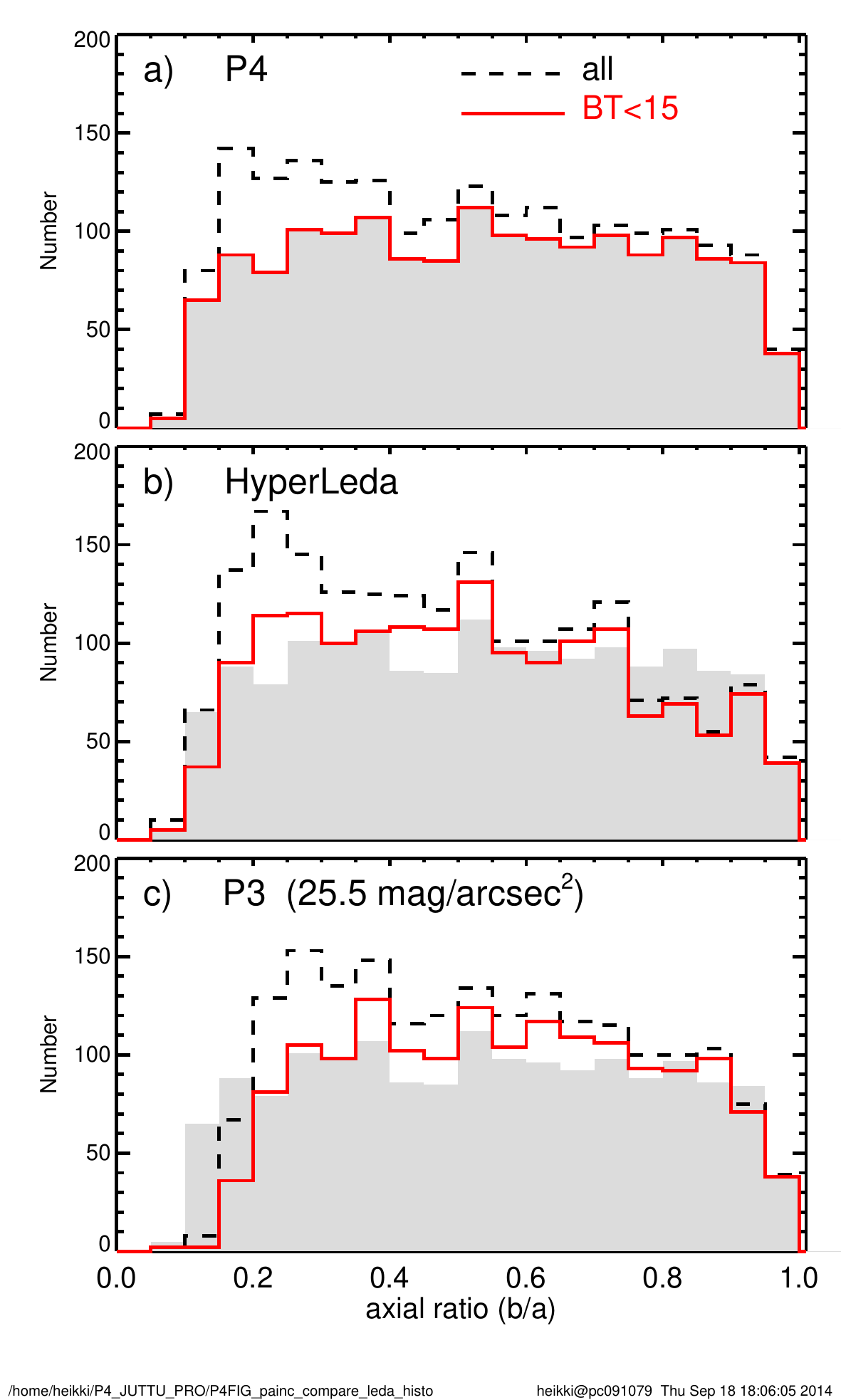}
\caption{Distribution of axial ratios, calculated with P4 (a),
  HyperLeda (b), and P3 orientation parameters. Hubble types $T \le
  -4$ and $T > 10$ are excluded (using Buta et al. 2014 Mid-IR classification).  Dashed line indicates the whole S$^4$G
  sample, with the magnitude selection $BT_{corr} < 15.5$, where
  $BT_{corr}$ is the inclination-corrected blue magnitude from
  HyperLeda. Solid line corresponds to a similar limit, but using
  non-corrected blue magnitude $BT$. To ease the comparison the P4 histogram for $BT<15$ is shown as the shaded region in each frame.} 
\label{painc_leda_histo}
\end{figure}

Since we are fixing the disk orientations in the decompositions it is
important to check the consistency of our inclinations.  Figure
\ref{painc_leda_histo}a displays the histogram of the P4 axial ratios
for Hubble types $-3 \le T \le 10$. In case of {\reply a} randomly oriented
sample of thin disks, the distribution of $b/a$ should be flat. In
case of finite vertical thickness a drop would be expected near a
lower limit $b/a = q_i$, where $q_i$ is the intrinsic aspect ratio of
the galaxies.  According to Fig. \ref{painc_leda_histo}a such a drop
is evident for $b/a \lesssim 0.15$.  However, overall the sample
contains an {\em excess} number of galaxies with small axial ratios
$b/a \lesssim 0.5$ (see the dashed line in
Fig. \ref{painc_leda_histo}a). Similar trend is seen also when using
the HyperLeda axial ratios (Fig. \ref{painc_leda_histo}b) or P3
isophotal orientations (Fig. \ref{painc_leda_histo}c).  A possible
explanation for the excess of small $b/a$ ratios is that the S$^4$G
sample has been selected \citep{sheth2010} using an
inclination-corrected blue magnitude limit ($BT_{corr}=15.5$): if this
dust correction were exaggerated, say for very late types, it would
lead to an excess of faint, highly-inclined galaxies. This explanation
is supported by the solid curves in Fig. \ref{painc_leda_histo} which
display the histograms when limiting to galaxies with non-corrected
$BT<15$: now the histogram of P4 values is quite flat. The histogram
for P3 isophotal axial ratios is rather similar, though there are
somewhat fewer small $b/a \lesssim 0.2$ values. This could be due to
the above-mentioned faint stellar halos: in case of nearly edge-on
galaxies a fixed surface brightness level could pick up the rounder
faint outer envelopes, whereas in P4 we have in such cases tried to
trace the disk isophotes.  On the other hand, compared to both P3 and
P4, the HyperLeda distribution has a clear deficit of large axial
ratios, mostly likely due to the influence of inner non-axisymmetric
structures.

\begin{figure}[h]
\includegraphics[angle=0,width=16.5cm]{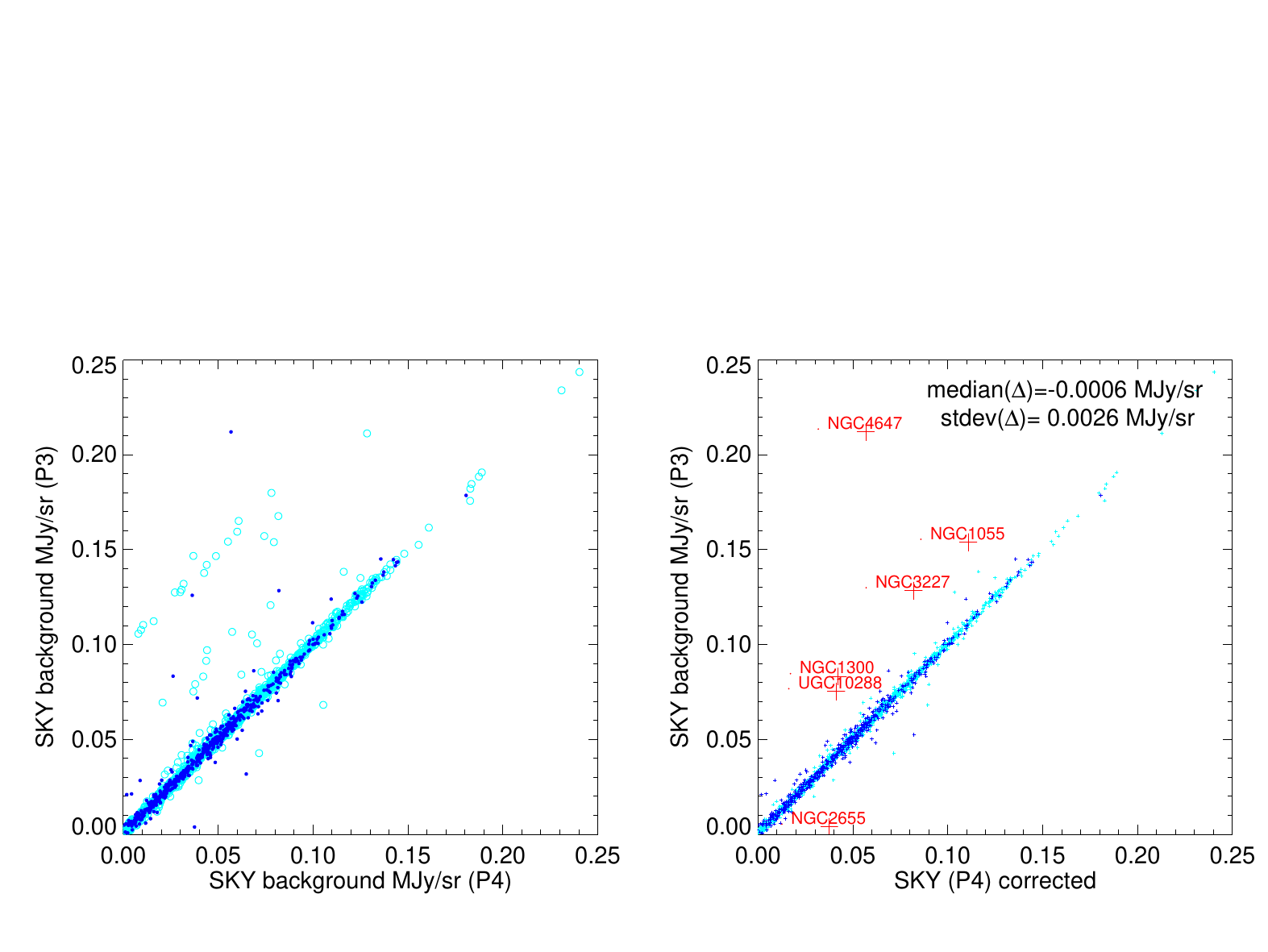}
\caption{The left frame displays the used P4 sky background estimates
  (SKY) in comparison to P3 (\citealt{munoz_mateos2014}; their variable
  'SKY1'). The large apparent differences are due to use of different
  versions of P1 mosaics. In the right we have corrected the P4 values
  to corresponds to the final P1 mosaics (=those used in
  P3). Excluding the few deviant cases (discussed in the text) the
  median difference between P3 and P4 is 0.003 MJy/sr.  Light and dark
  blue symbols indicate observations during cryogenic and warm
  Spitzer missions, respectively.}
\label{p3_comp_sky}
\end{figure}
\begin{figure}[h]
\includegraphics[angle=0,width=16cm]{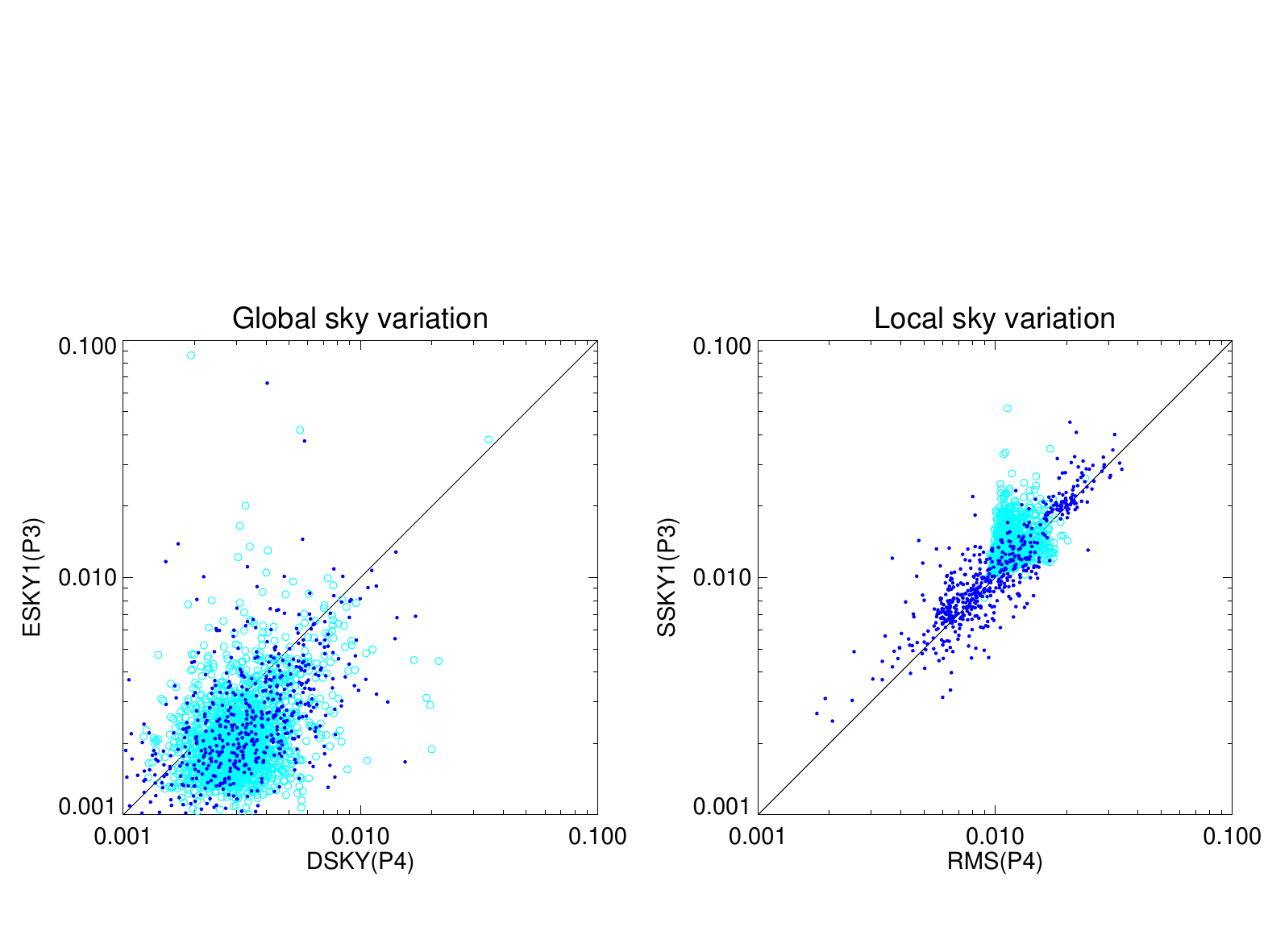}
\caption{Comparison of P4 and P3 sky variation estimates. The left
  frame displays the global variations, estimated from the standard
  deviation of sky measurements at different areas (variable 'DSKY' in
  P4 and 'ESKY1' in P3; the symbol colors are the same as in the
  previous Figure). In the right the local variation, estimated from
  the median scatter of sky values in local measurement areas
  (variable 'RMS' in P4 and 'SSKY1' in P3).}
\label{p3_comp_rms}
\end{figure}

It is interesting to compare our sky background estimates to those in
P3.  In P3 an automatic sky measurement is made using 45 sky regions
with 1000 pixels each.  The regions are chosen close to the distance
$2 R_{25}$ from the galaxy center ($R_{25}$ is the blue band 25 mag
isophotal radius from HyperLeda; if needed the distance of sky regions
is modified manually).  According to Fig. \ref {p3_comp_sky} there is
a very good agreement in the estimated sky backgrounds between P3 and
P4 (see the right frame which takes into account that different P1
mosaics are used for some of the galaxies). This good agreement is
remarkable as the measurements are made completely independently and
with different methods.  The median difference between the
sky determinations (0.0006 MJy/sr) is only about 1\% of the typical sky
background value, and its standard deviation (0.003 MJy/sr or ) is
comparable to the magnitude of global sky variations in both sets of
estimates (see Fig. \ref{p3_comp_rms}).  However, Fig. \ref {p3_comp_sky}  also
reveals some cases where the difference between P4 and P3 is
significant: inspection of the images indicates that this is due to a
bright star (NGC1055), a nearby interacting component (NGC3327,
NGC4647), or a too small FOV (NGC2655). In two cases (NGC1300,
UGC10288) the final P1 mosaic used by P3 is much improved over the
earlier version used in P4.

Fig.  \ref {p3_comp_rms} compares our sky background variation
estimates ('DSKY' denotes global variations between sky measurement
regions and 'RMS' the average of the locally determined rms-scatter)
with the corresponding estimates in P3 (\citealt{munoz_mateos2014};
their parameters ESKY1 and SSKY1, respectively).  There is a good overall
agreement in the level of estimated global variation (left frame): the
somewhat larger values for P4 are likely to follow from the larger
range of radii we used for the sky measurement regions compared to
P3. Also the local sky rms values show good agreement (right frame).


\subsubsection{Input data images}

As input for the GALFIT decompositions we use the 3.6 $\mu$m images.
Because all necessary data reduction and calibration were already done
in P1, the main preparatory steps are to subtract the estimated sky
background value and determine which image region to include in the
decomposition. In principle, GALFIT can also fit the sky background.
However, this requires that the decomposed image region contains
sufficiently large regions free of galaxy light or other
contaminants. Use of such large image regions would slow down the
decompositions considerably. Even more importantly, the S$^4$G images
often fill a substantial part of the raw frames or there are sudden
jumps in the background levels (well outside the galaxy).  To have a
control of where the sky level is estimated, we chose to do the sky
background evaluation manually, as described in Sect. \ref{sect_sky},
and to limit the decomposition to the rectangular region $\pm R_{fit}$
around the galaxy center. In practice, we choose $R_{fit}= 1.3 \times
R_{gal}$, where $R_{gal}$ is our visually estimated outer size of the
galaxy\footnote{Later comparison to P3 isophotal radii published in
  IRSA indicates that the median $<R_{\rm fit \ region}/R_{25.5}> =
  1.7$, where $R_{25.5}$ is the Pipeline 3 isophotal radius at
  $\mu_{3.6}(AB)= 25.5$. The region is thus large enough to ensure
  that also the fainter outer parts of the galaxy are included in the
  fit.}.  Finally, the image header keyword EXPTIME is set to 1 sec
(as a default GALFIT will normalize the input data values with
EXPTIME, which keyword is not relevant for P1 mosaics), and all {\em
  NaN's} (bad image values indicated with Not-a-Number value) are
replaced with a constant value, and flagged in the mask in order to
prevent them from affecting the decompositions.

\subsubsection{Sigma-images}
\label{sect_sigma}

The sigma-images quantify the statistical uncertainty of each image
pixel and thereby determine the weights applied in GALFIT
decompositions.  This uncertainty contains two contributions: the
noise contribution associated with the number of photons arriving at the
instrument ('photon noise' or 'shot noise'), and the noise originating
from the instrument itself. The photon noise is assumed to follow
a Poisson distribution, and it arises from two sources, the flux
associated with the galaxy light and the flux coming from the sky
background (zodiacal light). The main concern in the construction of
the sigma-images is that the relative contributions of the photon
noise and the instrumental noise are correctly estimated, so that
correct relative weights are used in the decomposition for the bright
central regions of the galaxies and for their faint outskirts.

The sigma-images are calculated using the pixel values and header
information in the 3.6 $\mu$m data images and the pixel values of the weight
images.  The images provided by P1 are in flux units (MJy/sr), and for
the calculation of the noise their pixel values $F$ are converted to the
number of electrons $N_e$,

\begin{equation}
N_e =  \frac{F + F_{bg}}{ F_{conv}} ~ \times T_{frame} \times N_{frames} \times g,
\end{equation}

\noindent where $F_{bg}$ is the zodiacal light background which has
been subtracted from the frame prior P1 by the automatic Spitzer
pipeline (its value is given by the header keyword {\tt SKYDRKZB}).
Note that the flux $F$ contains besides the galaxy light also the sky
background which has been subtracted in P4, $F=F_{gal}+F_{sky}$.  The
$F_{conv}$ is the conversion factor between flux units and original
digital units (header keyword {\tt FLUXCONV}, in units of MJy/sr per
DN/sec), $T_{frame}$ is the integration time/frame in seconds,
$N_{frames}$ is the number of combined frames for each pixel, and $g$
is the detector gain factor ({\tt GAIN} in units of e/DN). The number
of frames combined is coded to the pixel values $W$ of the weight
images, $N_{frames}= W/10$.  Note that $T_{frame}=30$ sec must be used
instead of the original integration time/frame given by the header
keyword {\tt FRAMTIME}: this is because during the compilation of P1
mosaics the pixel values have been normalized to this value regardless
of the original $T_{frame}$\footnote{This concerns the treatment of
  archival images observed during the cryogenic mission phase; all
  warm mission S$^4$G observations have $T_{frame}=30 secs$.}.  The
statistical uncertainty of $N_e$ in each pixel is then calculated as a
combination of Poisson noise (photon noise) and the readout noise of
the detector (RON),
\begin{equation}
\sigma^2 (N_e) = N_e +N_{frames}\times RON^2.
\end{equation}
\noindent We use RON = 15.0, 14.6, and 21 electrons, for {\tt
  FRAMTIME} = 12, 30, and 100 secs, respectively.  Note that these
values, communicated by the Spitzer Science Center Helpdesk, deviate
slightly from those given by the image header keyword {\tt RONOISE}.
The $\sigma (N_e)$ is then converted to the estimated uncertainty of the image flux (note that $\sigma (F_{gal})$ equals $\sigma(F)$ since $F_{sky}$ is constant)
\begin{equation}
\sigma_{est} (F)  = \sigma (N_e) \times F_{conv} /(T_{frame} \times N_{frames} \times g).
\label{eq_est}
\end{equation}

\begin{figure}[h]
\includegraphics[angle=0,width=16cm]{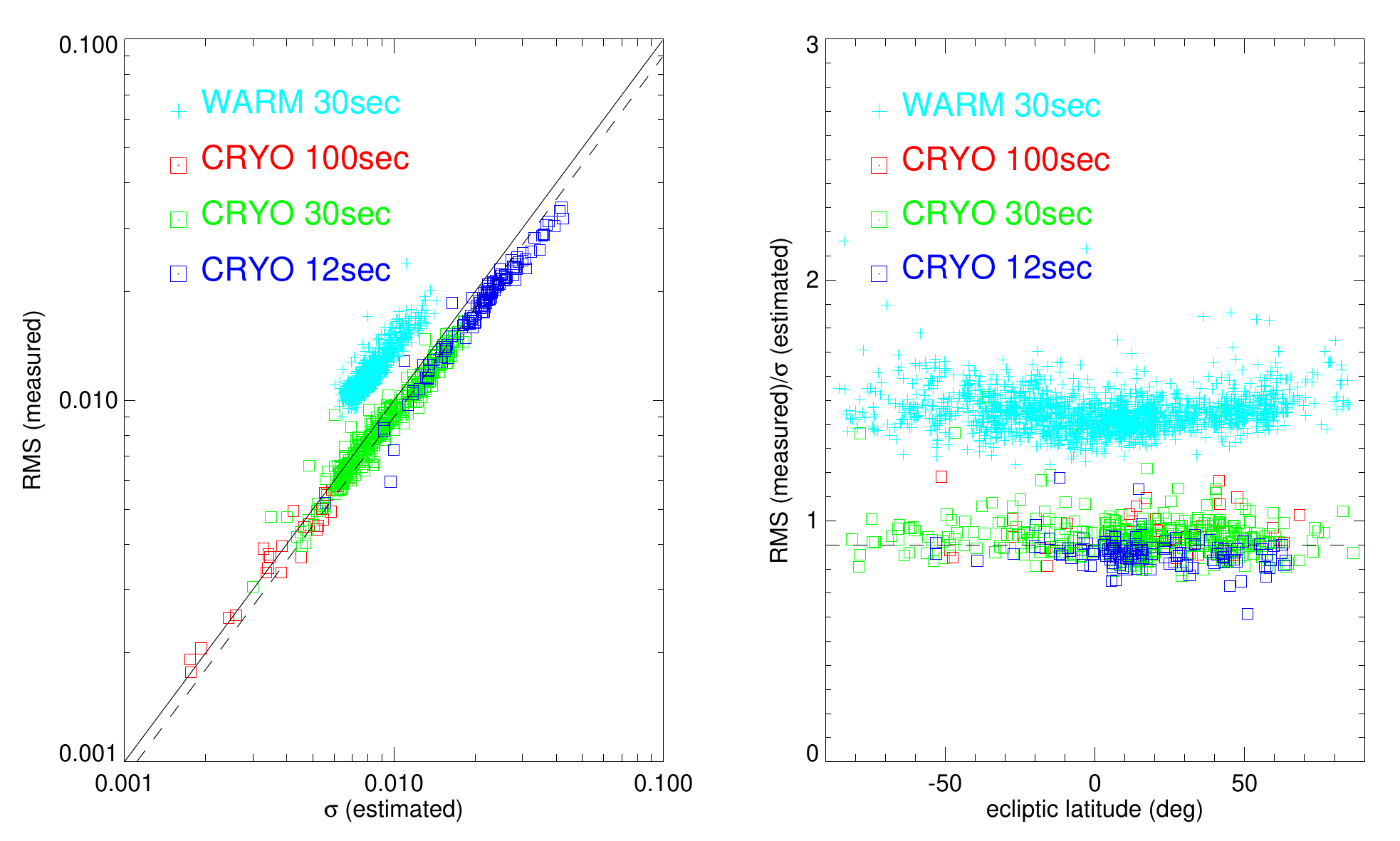}
\caption{ Comparison of measured and theoretical $\sigma$.  The left
  frame displays the measured RMS in sky regions against the estimated
  $\sigma$, which takes into account the read-out noise, and the
  Poisson noise due to sky background (including the zodiacal light
  contribution removed by the automatic Spitzer pipeline). For the
  cryogenic mission phase the agreement is fairly good, with $RMS
  \approx 0.9 \ \sigma_{est}$ (indicated by the dashed line; solid
  line indicates a one-to-one correspondence). For the warm phase the
  observed RMS is about 50\% larger than the theoretically estimated
  noise. The right frame shows the ratio of the observed and estimated
  noise as a function of ecliptic latitude.}
\label{final_sigma}
\end{figure}

In order to assess the validity of this estimate we compare it to
the actual noise measured directly from the image. 
In  Fig. \ref {final_sigma} this is done for the sky measurement regions.
%
In the left frame the measured sky RMS (an average over all sky determination
boxes) is plotted against the estimated $\sigma_{est}$ from Eq. \ref{eq_est}. Colors
distinguish between archival images from the cryogenic mission phase
(original exposure time/frame either 12, 30, or 100 secs) and the new
observations during the warm Spitzer mission (time/frame 30 secs, with
the total exposure time of 240 seconds). For the archive images the
overall agreement is quite good: there is a practically linear trend
$RMS \approx 0.9 \ \sigma_{est}$ holding for all three frame times,
with the largest noise levels corresponding to the shortest frame
times which have the largest contribution from the readout noise.  The
factor $\sim$ 0.9 is probably due to the P1 mosaicking process, during
which the images have been combined and sampled to 0.75 \arcsec pixel
size from the native pixel size of 1.2 \arcsec. Because of this
sampling the adjacent pixel values are strongly correlated, which is
not taken into account in our theoretical estimate.  Instead of trying
to account in detail for the noise propagation during the mosaicking
process we apply an empirical correction
\begin{equation}
\sigma_{decomp}(F) = 0.9 \ \sigma_{est}(F)  \ \ \ \ \ {\rm (cryogenic \ mission)}
\end{equation}
to be used in decompositions of cryogenic phase archival images.

In contrast, for the warm mission the observed RMS is nearly 50\%
larger than the theoretical estimate (Fig. \ref {final_sigma}),
indicating the presence of an additional source of noise.  Also, there is
a noticeable drop in $RMS/\sigma_{est}$ ratio near the ecliptic plane
(not present in the data from the cryogenic phase), indicating that
the photon noise contribution to the $\sigma_{est}$ (largest at
$l\approx 0^\circ$) is overestimated compared to the instrumental
contribution (constant with $l$).  Following the advice of Spitzer
Science Center Helpdesk, we include an additional instrumental noise
component ($\sigma_{conf}$), which is added quadratically to the
theoretical noise estimate. To account for the P1 mosaicking process, the
multiplicative factor of 0.9 is again included. We thus adopt
\begin{equation}
\sigma_{decomp}(F) = 0.9 \ \sqrt{{\sigma^2}_{est}(F) + \sigma_{conf}^2}   \ \ \ \ \  {\rm (warm \ mission)}
\end{equation}
for the warm mission images. The value of the empirical correction term
$\sigma_{conf}^2$ is estimated by this formula when applied to the sky
measurement regions; the same formula is then applied to all image
pixels. The adopted values of $\sigma_{conf}$ are listed in Table I above.

\begin{figure}[h]
\includegraphics[angle=0,width=16cm]{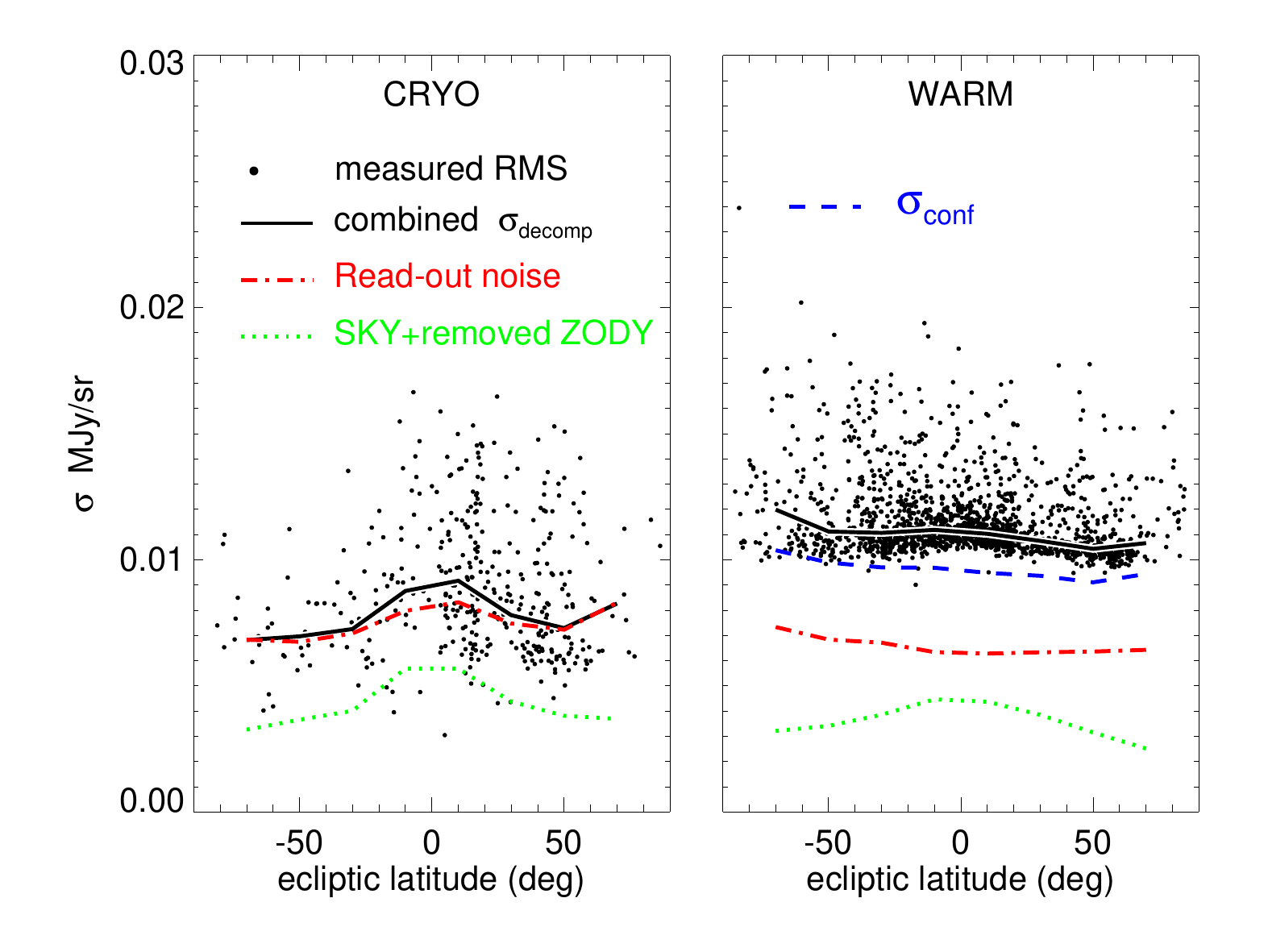}
\caption{Different contributions to the sky background noise.  The
  symbols indicate the measured average noise in sky measurement
  regions (RMS) in the cryogenic mission archival images (left frame;
  for clarity only those with 30 sec original frame time are shown)
  and in warm mission images (right frame), as a function of ecliptic
  latitude. The lines indicate various noise contributions and the
  total noise calculated with Eqs. (11) and (12): for clarity a mean
  over 20 deg bins is shown.  Note that in the left frame the peaking
  of readout noise contribution close to ecliptic plane is just a
  spurious effect.
}
\label{final_sigma_separate}
\end{figure}

Figure \ref {final_sigma_separate} illustrates the magnitudes of
different contributions to the sky background noise. For the archival
images (cryogenic phase, left frame) the noise is dominated by the
readout-noise, though the Poisson contribution due to zodiacal light
(the $F_{sky}$ we have subtracted in P4 + the SKYDRKZB subtracted
during automatic Spitzer pipeline) still has a noticeable
contribution.  For the warm Spitzer mission (right frame) the extra
noise term is even larger than the readout contribution.

\begin{figure}[h]
\includegraphics[angle=0,width=16cm]{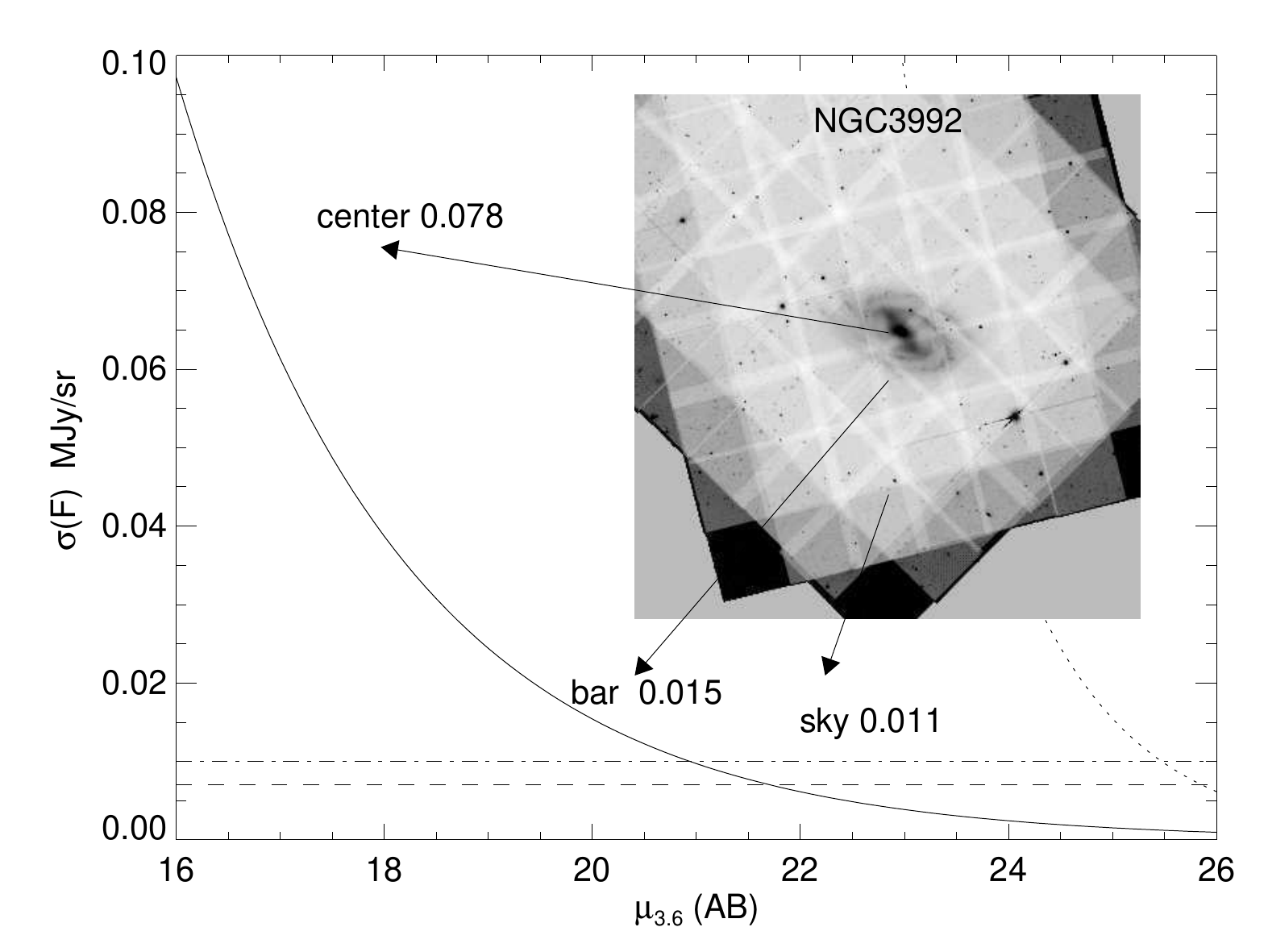}
\caption{Comparison of galaxy flux and background noise contributions.
  The solid curve is the calculated Poisson noise associated with the
  galaxy flux $F_{gal}$, as a function of surface brightness (in 3.6
  micron AB magnitudes; $F_{gal}$ is converted to surface brightness
  with Eq. \ref{eq_flux}). The horizontal lines indicate the typical
  background noise levels for the warm and cryogenic mission phases
  (dot-dashed and dotted lines, respectively; they include both the
  noise associated with sky background flux and instrumental
  contributions). The insert shows the $sigma$-map for NGC 3992.  The
  structure in the background is due to different number of frames
  covering each pixel.  Also notice how the galaxy stands out clearly on
  the $sigma$-map. The dotted line crossing the horizontal lines at
  $\mu_{3.6} \approx 25.5$ indicates the galaxy flux in MJy/sr. }
\label{final_sigma_example}
\end{figure}

Nevertheless, at the central parts of the galaxies the photon noise
associated with the galaxy light $F_{gal}$ is the largest source of
noise. This is illustrated in Fig. \ref{final_sigma_example}) which
compares the Poisson and background contributions as a function of
surface brightness.  The two horizontal lines indicate the typical
background noise levels for the cryogenic (lower) and warm (upper)
phases (includes both instrumental and noise due zodiacal light).  The
inset figure illustrates how the $\sigma$-map looks for the galaxy
NGC3992 (observed during the warm mission). Near the center ($\mu
\approx 17$), the photon noise due to $F_{gal}$ completely dominates,
though already in the bar region ($\mu =20-21)$ both photon and
instrumental contributions are important.  Altogether the sigma-images
and thus the applied relative weights between galaxy and background
regions are intermediate between those typically encountered when
decomposing ground-based optical and NIR-images. In the former case
the photon noise due galaxy light usually dominates, while in the
latter case the sigma-image is almost completely dominated by the
background noise, so that the weight is almost constant
for all pixels (this applies e.g. to \citealt{janz2014} GALFIT
decompositions of Virgo dEs based on ground-based H-band images).

In principle, the obtained $\sigma_{decomp}$ is just a statistical
estimate of the true underlying variance at each pixel. We did some
experimentation by smoothing the {\em sigma} -images (median averaging
with kernels amounting up to 20 pixels). Except in the case of a few
galaxies with very centrally peaked light profiles, this smoothing had
very little influence on the final decomposition parameters. For the
galaxies where smoothing played a role, the derived parameters were in
any case uncertain (for example, the bulge S\'ersic index obtained
unrealistically high values $>$ 10). In the end, we decided to apply
no smoothing at all. Tests related to the sigma-images are presented
in Section \ref{sect_sigma_uncertainty}.

\subsubsection{PSF-image}
\label{sect_psf}
\begin{figure}[h]
\includegraphics[angle=0,width=14cm]{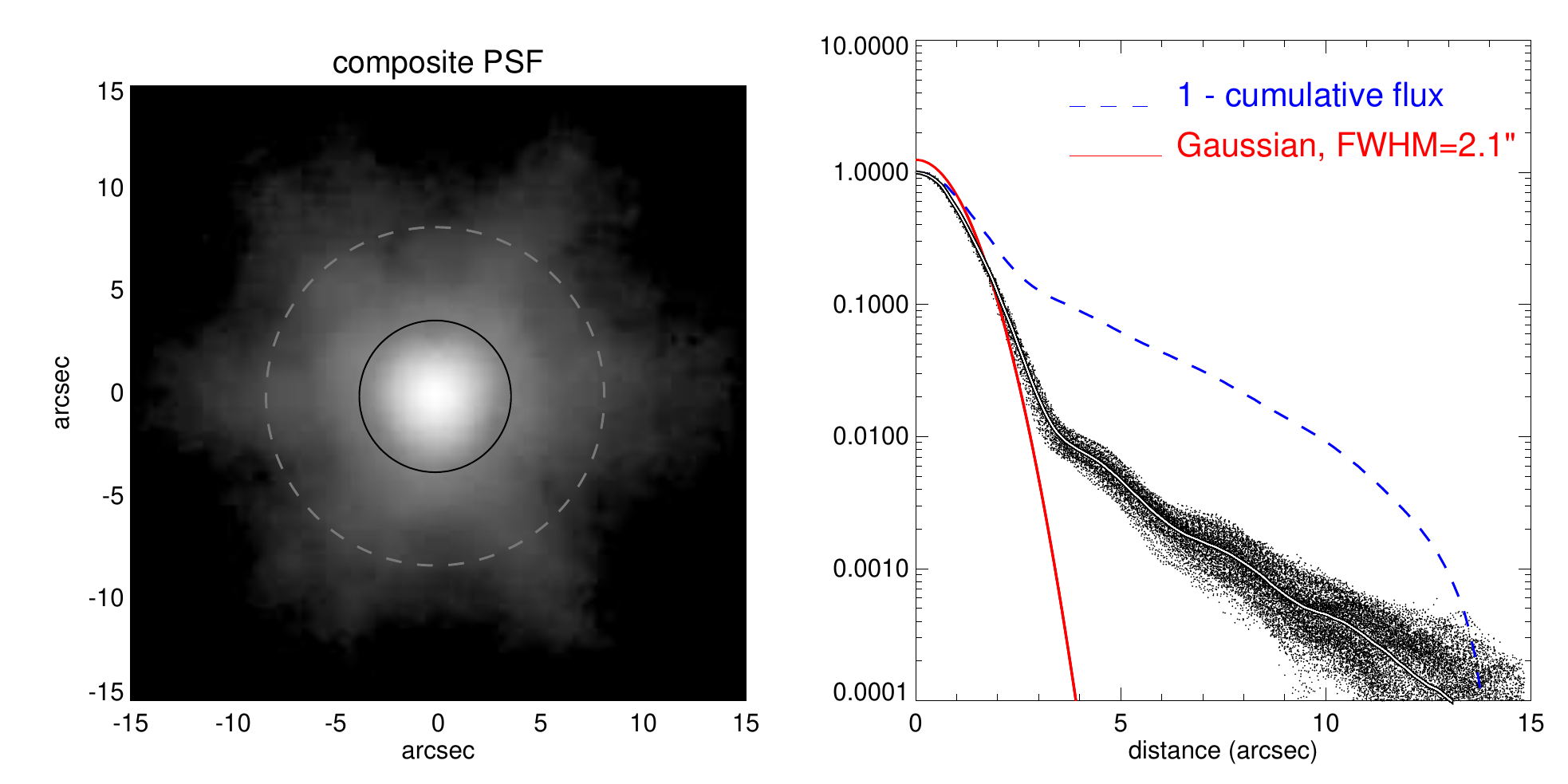}
\caption{The composite PSF used in decompositions (shown in the left
  frame in log-scale to emphasize the wings: in the right frame the
  PSF is normalized to its maximum value). The red solid curve in the
  right indicates a Gaussian with the same FWHM=2.1\arcsec, approximately
  matching the inner part of the actual PSF (black symbols). The white
  curve on top of black symbols indicates an azimuthally symmetrized
  version of the composite PSF. The dashed blue curve indicates the
  cumulative flux outside a given distance from the center: the solid
  and dashed circles in the left, with radii $3.7\arcsec$ and $8.3\arcsec$
  indicate the regions where 90\% and 98\% of the flux is concentrated
  (50\% of flux falls within 1.3\arcsec). The composite PSF,
  oversampled with a factor of 5 (pix-size 0.15 arcsec, { total size $30\arcsec \times 30\arcsec$}) was provided by Tom Jarrett.}
\label{psf}
\end{figure}

The IRAC data is not very well sampled: its native pixel resolution is
1.2\arcsec, which is close to the Gaussian spread of a point source
observed at channel 1.  As discussed in detail in \cite{peng2010},
in such a case an oversampled PSF should be used.  The IRAC PSF has
also wide wings (see Fig. \ref{psf}), so that a relatively large
convolution box size must be used in decompositions: we set this to
$40\arcsec \times 40\arcsec$ (in some cases with a very centrally
peaked light profile this region was extended to $150\arcsec \times
150\arcsec$ with considerable increase in CPU time).  Note also that
IRAC PSF depends slightly on the instrument orientation. Therefore, in
principle a separate PSF should be used with each image, determined
from point sources in the same frame, or a combination of appropriate
PSFs, in case the final image is a combination of several images
obtained at different times.  Clearly, such a procedure would be very
time consuming. Fortunately, such an accuracy is hardly needed in our
decompositions. 
The common oversampled PSF provided by T. Jarrett was used for all
images, made as a composite over several instrument rotation angles.
Fig. \ref{psf} displays the PSF, as well as a Gaussian profile (with
$FWHM=2.1\arcsec$) approximately matching the core of the composite
PSF. Also shown is an azimuthally averaged profile of the composite
PSF.  It will be shown in Section 2.5 that it is important to account
for the central core, as well as for the nearly circular wings of the PSF,
whereas the outermost spikes have less importance for the obtained
decomposition parameters.


\begin{deluxetable}{lcrrrlccrrrr}
\tablewidth{-20pt} 
\tabletypesize{\scriptsize} 
\tablecaption {Pipeline 4 parameters: galaxy center, outer orientation, and sky background}
\tablehead{
\colhead{IDE} & \colhead{}  &\colhead{x$_c$} & \colhead{y$_c$}     & \colhead{PA $\pm$ dPA}    & \colhead{ELL $\pm$ dELL} & \colhead{RANGE}  &\colhead{FLAG} &  \colhead{SKY} & \colhead{DSKY} & \colhead{RMS} & \colhead{$\sigma_{conf}$} }
\startdata
ESO011-005        &  &  780.84   &  299.02    &    42.7 $\pm$0.1   &   0.747 $\pm$0.002     &     13 - 16         & z &   0.0125  & 0.0039   &   0.0109 &   0.0092 \\ 
ESO012-010        &  &  760.68   &  294.95    &   146.2 $\pm$1.1   &   0.542 $\pm$0.012     &     75 - 90         & ok &  -0.0022  & 0.0025   &   0.0109 &   0.0096 \\ 
ESO012-014        &  &  782.57   &  459.99    &    31.0 $\pm$5.2   &   0.580 $\pm$0.040     &     67 - 75         & u &   0.0080  & 0.0033   &   0.0106 &   0.0091 \\ 
ESO013-016        &  &  478.49   &  282.78    &   -14.3 $\pm$3.1   &   0.343 $\pm$0.031     &     52 - 82         & ok &   0.0041  & 0.0027   &   0.0101 &   0.0084 \\ 
ESO015-001        &  &  291.43   &  294.80    &   125.7 $\pm$2.7   &   0.586 $\pm$0.018     &     56 - 75         & u &   0.0050  & 0.0018   &   0.0102 &   0.0086 \\ 
ESO026-001        &  &  795.50   &  495.47    &    19.3 $\pm$21.1  &   0.060 $\pm$0.028     &     52 - 60         & ok &   0.0125  & 0.0024   &   0.0105 &   0.0091 \\ 
ESO027-001        &  &  776.34   &  294.63    &    12.5 $\pm$10.1  &   0.216 $\pm$0.061     &    112 - 127        & u &   0.0109  & 0.0032   &   0.0112 &   0.0097 \\ 
...\\
UGC12791          &  &  295.88   &  289.63    &    82.6 $\pm$1.4   &   0.709 $\pm$0.037     &     45 - 60         & ok &   0.0406  & 0.0033   &   0.0110 &   0.0094 \\ 
UGC12843          &  &  290.20   &  287.90    &    17.5 $\pm$3.9   &   0.553 $\pm$0.058     &     52 - 67         & u &   0.0418  & 0.0037   &   0.0105 &   0.0086 \\ 
UGC12846          &  &  571.15   &  851.20    &    -4.7 $\pm$14.0  &   0.127 $\pm$0.031     &     60 - 67         & u &   0.0446  & 0.0007   &   0.0019 &   0.0000 \\ 
UGC12856          &  &  291.07   &  296.89    &    16.9 $\pm$1.5   &   0.615 $\pm$0.049     &     48 - 75         & u &   0.0401  & 0.0025   &   0.0111 &   0.0099 \\ 
UGC12857          &  &  284.91   &  280.12    &    33.5 $\pm$0.2   &   0.760 $\pm$0.009     &     15 - 30         & z &   0.0453  & 0.0018   &   0.0110 &   0.0096 \\ 
UGC12893          &  &  294.93   &  283.15    &    87.2 $\pm$3.8   &   0.131 $\pm$0.039     &     52 - 67         & ok &   0.0463  & 0.0022   &   0.0110 &   0.0093 \\ 
\enddata 
\label{table:ori}
\tabletypesize{\normalsize} 
\tablecomments{Galaxy center {\em $x_c,y_c$} is given in pixels, {\em ELL $\pm$ dELL} and {\em PA $\pm$dPA} are the outer
  isophote ellipicity and position angle together with their standard
  deviations in the measurement range, given by {\em RANGE} (in
  arcsecs). {\em FLAG} indicates whether the
 inclination can be reliably estimated from the ellipticity ($i_{disk} =
 \cos^{-1} (1-\epsilon_{outer}))$: {\em ok }= reliable, {\em u} =
 uncertain, {\em z=} nearly edge-on galaxy.
 {\em SKY}, {\em DSKY}, and {\em RMS} give the estimated sky level and
 its global and local variation (in MJy/sr). The last column $\sigma_{conf}$ gives the
 estimated extra instrumental noise component during the Spitzer warm mission (See Sect 2.2.5).}
\end{deluxetable}

\clearpage

\clearpage

\vskip 2cm

\clearpage



\clearpage


\subsection{Generation of input files for GALFIT decompositions}

The (ascii) input file for GALFIT specifies the galaxy data, mask,
sigma, and PSF fits-files, and the region of the data image used in
the decomposition. It also lists the components/functions used in the
decomposition model, the initial guesses for the parameters, and
specifies which of the parameters will be kept fixed, and which are
iteratively varied in order to minimize the ${\chi^2}_{\nu}$. After
convergence to a final solution, the final parameter values are
written into an output file, with similar format as the input file. If
needed, this output file can thus be used as an input for a new
iteration (see \citealt{peng2002, peng2010} for details).

The input file also specifies how to convert the image values
to magnitudes.  The data images from P1 are in flux units
(MJy/sr). A conversion from pixel values $F_i$ to (AB) surface brightness
and integrated magnitudes is done with  the formulas:
\begin{eqnarray}
\mu_{3.6} &=& -2.5 \log_{10} F_i + 5 \log_{10} {pix} + zp \label{eq_flux}\\ 
 mag_{3.6} &=& -2.5 \log_{10} \sum_i F_i +zp,
\end{eqnarray}
\noindent where $pix=0.75\arcsec$ and the zeropoint at 3.6 $\mu$m is $zp=21.097$ (P3, \citealt{munoz_mateos2014}).
Values of $pix$ and $zp$ are inserted into GALFIT input file.


All P4 input files for 1-component (S\'ersic) and 2-component
bulge+disk (S\'ersic+exponential) decompositions were generated
automatically. 
Similarly, template files were created for the
multi-component decompositions, which
contained, in addition to bulge and disk components, entries for
a Ferrers-bar, and a central unresolved PSF-component.
The user then
manually choose which components are fit and which functions used
in the final model (see Sect \ref{sect_eija_examples} for
more details).  In all our decompositions we keep the centers
of the components fixed to the galaxy center. The cases were this is
clearly not appropriate (galaxies with off-center bulges and bars) are noted in
the parameter files.

1. In {\it 1-component} input files initial guesses are needed for five free
parameters: the S\'ersic index $n$, the effective radius $R_e$, the total
magnitude $m$, the isophotal minor-to-major axial ratio $q$, and
the position angle $PA$.  The starting values of $m$ and $R_e$ were taken directly from the data
(total galaxy magnitude and half-light radius), 
for the S\'ersic index $n=2$ is inserted as an initial guess, and  $q$ and $PA$ were set to
arbitrary values (0.9 and $10^\circ$, respectively). We thus avoided using
the measured outer isophotes, to force GALFIT to search through a
wider parameter space while minimizing the ${\chi^2}_{\nu}$. Typically
1-component fits converged after 10-20 iterations. When the fit did
not converge, or if the final parameters were nonphysical (say, $n >
10$, $q<0.05$, very large or small $R_e$), a new decomposition was
started manually with new initial guesses. Usually this did not lead to any improvement, indicating that GALFIT
is indeed very efficient in avoiding spurious local minima.

2. The {\it 2-component} bulge-disk models apply a S\'ersic-function for the
bulge: they thus need guesses for the same S\'ersic parameters as
before, except that now these refer to the central component.
Accordingly, we used the initial guess $R_{e}$ (bulge) = 0.5 $\times
R_e$ (image) and $m_{bulge} = m_{image}+1$. For the disk we use
either {\em 'expdisk'} or {\em 'edgedisk'}-function, depending on the
estimated galaxy inclination. In case of low or moderate inclination
$b/a \gtrsim 0.2$ (corresponding to $i \lesssim 80^\circ$), we use the
{'\em expdisk'} function, which needs two free parameters, the scale length ($h_r$) and
the integrated magnitude of the disk, $m_{disk}$.  We chose $h_r =$ 0.25 $\times R_{gal}$ and
$m_{disk} = m_{image}+1$, thus starting with a model with fairly massive
and extended bulge. The disk orientation was  fixed to the shape
of the outer isophotes determined from the image (see Sect
\ref{sect_sky}).  In case of a nearly edge-on disk, $ b/a \lesssim 0.2$,
we use the {\em 'edgedisk'} function with four free parameters: the
central surface brightness $\mu_0$, radial scalelength $h_r$, vertical
scalelength $h_z$, and the position angle of the disk. The first
guesses are $\mu_0 (disk) = \mu_0(image)+3$, $h_r$ as for the
expdisk-model, while $h_z/h_r =0.1$.  The position angle is left free, with
$PA_{outer}$ as an initial guess.

3. In the template files for the {\it multi-component} fits the initial
bulge and disk parameters are set as for the 2-component models. For
the Ferrers-bar the free parameters are the surface brightness at the
effective radius of the bar, $\mu_e$, its outer truncation radius
$R_{bar}$ (denoted with $r_{out}$ in Eq. (7)), its axial ratio, and its position angle. As initial guesses we
choose $\mu_e$ (bar)=$\mu_e$ (image)+3, $R_{bar} = 0.25 \times
R_{gal}$, $q_{bar}=0.5$, and $PA_{bar} = PA_{disk}+90^\circ$.  For the
magnitude of the unresolved central component we used $m_{psf} =
m_{image}+5$. However, in practice we typically modified these
pre-inserted template values even before starting the search of the final
model, for example by adopting  the {\em output} parameters from
2-component decompositions for the disk and bulge.

\subsection{Visualization of GALFIT decompositions: GALFIDL Package}
\label{sect_galfidl}
In its standard use, GALFIT is executed from the operating system
command line, with an input file argument. This input file lists the
input data files and the initial guesses for the parameters, as
described above.  The final decomposition parameters are written to an
output file with a fixed name {\tt galfit.NN}, where {\tt NN} is a
running number.  Optionally, GALFIT makes a fits file containing
the clipped data image (OBS; includes the region chosen for the fit),
and total PSF-convolved model (MODEL), and the OBS-MODEL
residual. Another GALFIT option is to write a FITS file
containing model components in separate fits extensions.

In P4 we have used GALFIT via GALFIDL, which
is a set of IDL routines designed for easy visualization of the output
from GALFIT decompositions.  In addition, GALFIDL includes wrapper
routines for calling GALFIT from inside IDL, with the advantage that
the GALFIT output files and the produced plots are automatically
renamed in a systematic fashion, using the names of
the input files. We have utilized this by coding the galaxy identification
and decomposition model components to the name of each produced output file
(see Appendix A)

\begin{figure}[h]
\includegraphics[width=16cm]{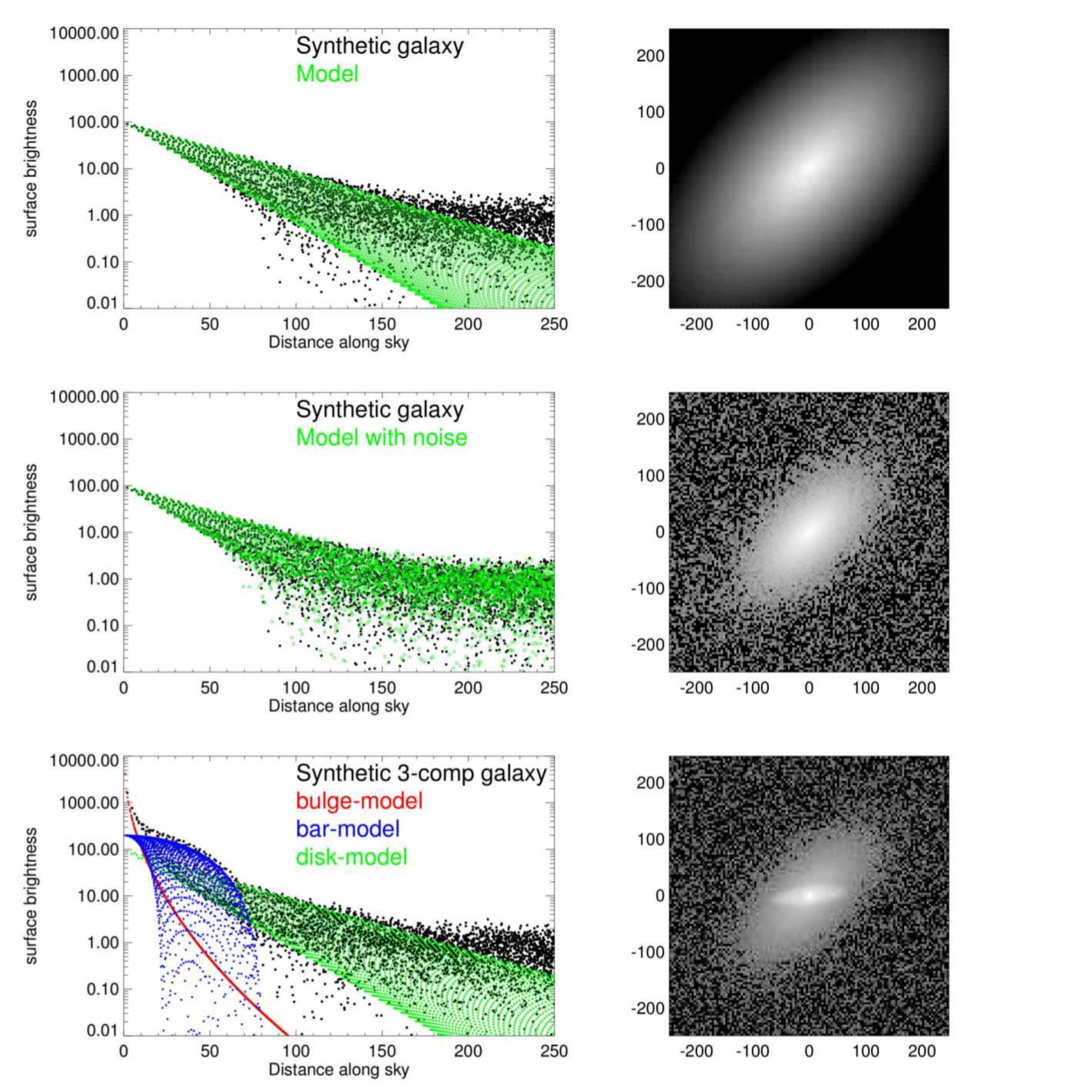}
\caption{\baselineskip 0.55cm Schematic example of the 2D-profile
  plots.  The black dots in the upper left frame show a synthetic
  observational image (exponential disk with added Poisson noise): the
  surface brightness (in arbitrary units) is plotted as a function of
  distance along the sky plane.  The green dots indicate the best
  fitting GALFIT {\em expdisk} model; the width of the wedge-shaped
  profile depends on the disk inclination (here $i=60^\circ$). Note
  that this plot might leave an impression that the outer disk is not
  properly fitted: this illusion is due to the noise distribution
  appearing skewed in magnitude plots.  The middle left frame
  illustrates the same model, after realistic noise (measured from the
  synthetic image) has been added also to the model profile, giving a
  visual confirmation that the model indeed is successful.  The images
  in the right indicate the fitted model without (upper frame) and
  with noise (middle). The lowermost frames display a fit to a
  synthetic galaxy (image in the right) composed of a Sersic-bulge
  (here $n=2$, with circular isophotes), an exponential disk, and a
  Ferrers bar.
\label{profile_2D}}
\end{figure}

The visualization options in GALFIDL follow those of the
BDbar-decomposition program we developed earlier for the NIRS0S
survey \citep{laurikainen2005}, the most central of which is
displaying a 2D plot of surface brightness vs. distance from the
galaxy center (see Fig. \ref{profile_2D}).  The advantage of this,
compared to the more commonly used azimuthally averaged profile, is that the
contributions of different model components, with different apparent
ellipticities, are easily highlighted (\citealt{laurikainen2005}; see 
also \citealt{gadotti2008}).  The other visualization options include OBS-MODEL
residual plots, profile cuts along a constant PA, comparison to
observed profiles along isophotal major axis produced by IRAF ellipse,
and plots showing schematically the different components included in
the decomposition. The next section illustrates our decomposition
strategies in more detail, concentrating on 2D-profiles. Additional plot
types are illustrated in Appendix B, which describes the output
released through a P4 web page for all S$^4$G galaxies.


\clearpage

\section{Building the final multi-component decompositions - examples}
\label{sect_eija_examples}

 The final decompositions for S$^4$G galaxies were done by fitting a
 maximum of 4 components.  
{\reply Typically the components were the bulge ({\em B}), disk
  (denoted either as {\em D} or {\em Z}, depending on whether the
  galaxy was close to edge-on), bar ({\em bar}) and the
  nucleus ({\em N}), but could be any combination of these.  } The
ingredients of the model are indicated by concatenating the
designations of the components to the final model name: this same
naming convention is used in the names of decomposition output files
stored to IRSA (Appendix A).  {\reply Note that the component
  designation is based on the intepretation of the component, not the
  function used in the fit. A disk ('D'), though most often fit with the
  {\em expdisk} function (1969 cases), may also be fitted with {\em
    ferrer2} (69 cases) or {\em sersic} functions (99 cases).
  Similarly, in six cases a bulge ('B') was fitted with an {\em expdisk}
  or {\em edgedisk}, and in one case a 'bar' with a {\em sersic}
  function.  All elliptical galaxies were fitted with a single S\'ersic
  and are designated as {\em B}).  }

In all final decompositions the orientation parameters of the outer
disk were fixed and we also fixed  $\alpha$ and $\beta$  in the Ferrers function ($\alpha$=2, $ \beta$=0). All other
parameters were left free for fitting.  However, to find the
structure components properly it was convenient to temporarily fix
many of the model parameters at the beginning, and then release them
one by one. For some galaxies, the length of the bar was kept fixed
even in the final model. This was the case if GALFIT persistently gave
a clearly incorrect  bar length when compared to visual evaluation (in such
a case the ${\chi^2}_{\nu}$ minimization was attempting to fit some other
feature than a bar). 




%

\noindent 
Altogether over 20 different combinations of components were used
in the final decomposition models; Table \ref{table:final_taconomy}
collects an inventory of the main categories.  This diversity of
models is motivated by our desire to measure the bulge (if
present) and the underlying disk parameters in a reliable manner. 
{Note that our definition of 'bulge' is quite broad, based
  on the excess flux in the central parts of the galaxy over that
  associated with the disk and bar components ('photometric bulge').
  The decompositions themselves do thus not attempt to judge the
  physical character of this component, whether a merger-related,
  velocity-dispersion supported classical bulge, or a
  rotationally-supported 'pseudo-bulge' \citep{kormendy1982},
  representing either a secularly formed central stellar disk
  component \citep{kormendy1993} or a bar-related inner boxy/peanut
  component formed via bar vertical buckling \citep{combes1981,
    atha2005}. However, in Paper 2 we address the deduced bulge
  parameters ($n$, $B/T$) in the context of often-used
  classical/pseudo bulge indicators \citep{kormendy2004} and also make
  comparisons to compilations of pseudo-bulges identified based on
  their {\em HST} morphology and star formation properties
  \citep{fisher2010}.}

In (non edge-on) galaxies with two distinct disk components (desgnated
with {\em dd}, the inner disk was fit either with an exponential or a
S\'ersic function, depending on the flattening of the profile. 
{\reply Such
inner disk components differ from our photometric 'bulges' by their
much shallower profiles; they are also usually associated with a
distinct inner spiral structure.  Small central components were fit
with the PSF, indicated as {\em 'N'} in the model names.  However,
because of the limited resolution of S$^4$G images, many of those
structures, particularly in late-type spirals, might actually be small
bulges rather than nuclear point sources. Typically these components
contribute less than a few percent of the total flux.}

\begin{deluxetable}{lllr}
\tablewidth{0pt} \tabletypesize{\scriptsize} 
\tablecaption {Main categories of final decomposition models.}  
\tablehead{}
\startdata 
Disk: moderate inclination\hskip .2cm ~  & & & 1889   \cr
\hline 
&BD \hskip 2cm ~ & 311 \cr
&BDbar & 213 \cr
&ND& 214  \cr
&NDbar& 184  \cr 
&Dbar & 458  \cr
&DD & 125  \cr 
&D & 367  \cr 
\hline 
Disk: nearly edge-on & & & 362  \\ 
\hline 
&BZ & 55  \cr
&NZ & 62  \cr
&Zbar & 8  \cr 
&ZZ & 113  \cr
&Z & 126  \cr
\hline
{\reply elliptical}: & B & & 26  \cr 
\hline 
\\
ALL \hskip 2cm ~ & & & 2277 
\enddata 
\tablecomments{ Final decompositions were made for
  2277 galaxies: in case of low or moderate inclination (apparent
  $\epsilon \lesssim 0.8$), the disk component was
  fitted with {\em expdisk}-function, while for nearly edge-on
  galaxies ( $\epsilon \gtrsim 0.8$), the {\em
    edgedisk}-function was used. In models {\bf BD} and {\bf BZ}, a
  bulge component was identified besides a disk, and it was modeled
  with a {\em S\'ersic-function}. These models {\em may} also contain
  additional disk components or unresolved central components (modeled
  with {\em psf}).  The models {\bf BDbar} include those bulge+disk
  systems which contained also a bar (modeled with {\em
    ferrer2}). In {\bf ND} or {\bf NZ} models the central component is
modeled with {\em PSF} instead of {\em S\'ersic}-function. This may
represent either a true central point source or (more commonly) an
unresolved bulge.  The models {\bf NDbar} include also a bar. 
The models {\bf Dbar} and {\bf Zbar} have no inner
{\em S\'ersic} or {\em psf}-components, but include a bar-component.
They may also contain an outer disk component. The {\bf DD} models
contain both an inner and outer disk (and no bulge nor bar), while {\bf D} models refer to
pure disks. Similarly {\bf Z} models apply a single {\em
  edgedisk}-function, while {\bf ZZ} models contain both thin and
thick disk-components.  
}
\label{table:final_taconomy}
\end{deluxetable}

\subsection{Non-barred galaxies:}

The decompositions were made starting from simple 1 and 2-component
models, and then adding as many components as necessary. For
non-barred galaxies the process leading to the final model was the
following:

(1) Accepting the automatic 1-component model (single S\'ersic) as the
final model.  This was the case for elliptical galaxies (see NGC 3962
in Fig. \ref{ngc3962_eija}).

(2) Accepting the automatic 2-component bulge/disk decomposition as a
final model.  A typical example is NGC 3938 (Fig. \ref{ngc3938_eija}).

(3) Adopting a bulge/disk model, after interactively finding modified
initial parameters that converged to an acceptable final fit. 


(4) Adding a nucleus component or an  inner
disk (e.g. NGC 1357, Fig. \ref{ngc1357_eija}) to the bulge/disk model.

(5) When the galaxy had no obvious bulge we started from a single
exponential disk, and if necessary, a second disk and/or nucleus was
added (see  NGC 723, Fig. \ref{ngc0723_eija}). 
 
When the outer profile was affected by a possible stellar
halo, the outermost part of the profile was not fitted.  The best model
was vetted by looking at the original image, the residual image
after subtracting the model, the 2D surface brightness profile, and
the ellipticities of the structures. The value of final ${\chi^2}_{\nu}$ was
not used as a criterion in assessing the relative merits of the models
(often a simpler final model was preferred even if a more complicated
model would have yielded slightly smaller reduced ${\chi^2}_{\nu}$).  

{\reply Also, it is well known that many elliptical galaxies have
  small inner disks \citep{rest2001}, and it has been shown by \cite{huang2013}
  that many elliptical galaxies are better fitted with
  multiple S\'ersic profiles. Nevertheless,  such a detailed approach was not
  taken in this study, in which the emphasis is in the analysis of disk
  galaxies (paper 2).}  It is worth noticing that while using deep
  images like those in $S^4G$, in an automatic fit the bulge profile
  even in late-type spirals is easily degenerate with the outer part
  of the disk.  In automatic fits this may lead to an unrealistically
  large S\'ersic $n$ and $R_e$ for the bulge, of which NGC 1357 is a
  good example (Fig. \ref{ngc1357_eija}).

\subsection{Barred galaxies:}

For barred galaxies a similar step-wise approach was followed. NGC 936
(Fig. \ref{ngc0936_eija}) and NGC 5101 (Fig. \ref{ngc5101_eija}) are
good examples demonstrating the importance of preventing the bar from
mixing with the bulge flux. Adding a bar component to a simple
bulge/disk model drastically changes the obtained properties of the
bulge (for NGC 936 $B/T$ drops from 0.46 to 0.19; for NGC 5101 from
0.99 to 0.22). NGC 5101 has also a type II profile in the disk
break/truncation classification associated to a broad outer ring
(Laine et al. 2014).  Using the edge of the ring as a manifestation of
a different flux distribution in the outer disk might be a bit
misleading. Because of such ambiguities in the interpretation, we
typically fit the type II disk profiles with a single exponential
component.  However, there are other barred galaxies in our sample,
such as IC 4901 (Fig. \ref{ic4901_eija}), in which two exponential
components (+ Ferrers function for the bar) were used for fitting the
disk. In this particular galaxy using two exponentials is necessary,
and those clearly correspond to distinct surface brightness
components.  Note that the outer disk of NGC 5101 is clearly lopsided (see
the residual plot in Fig.  \ref{ngc5101_eija}). Such asymmetries are
not taken into account in the pipeline decompositions (in case of
strongly distorted galaxy no final model was made).  See
\cite{zaritsky2013} for a detailed study of galaxy lopsidedness using S$^4$G
images.

\subsection{Pure disk galaxies:}

A third main group of galaxies in our sample are those having no
obvious bulge. They may have a single exponential disk (NGC 3377A in
Fig. \ref{ngc3377A_eija}), or more than one disk components (NGC 723
in Fig. \ref{ngc0723_eija}). The structure fit as an inner disk in NGC
723 consists of broad, prominent, and tightly wound spiral
arms. Bulgeless galaxies may have bars, of which NGC 3517
({Fig. \ref{ngc3517_eija}) is an example.  To get an homogeneous
  estimate for the scale length of the disk for these galaxies, the
  outer disks were always fit with an exponential function, even in
  galaxies where the disk would have been somewhat better fit by a
  S\'ersic function with $n$ slightly less than unity.  Generally, the
  assumption of an exponential disk is good, but there are also cases,
  like ESO026-001 (Fig. \ref{eso026-001_eija}) in which a S\'ersic
  function would actually be a better choice.

\subsection{Edge-on galaxies:}

The GALFIT models for the nearly edge-on galaxies assume that the disk
is viewed completely edge-on.  A bulge, and in some cases also a bar
or an additional thick disk component were included
(Fig. \ref{eso533-004_eija}). In these models also the vertical
thickness was an output parameter.  However, these models are
tentative, and are meant solely as starting points for better,
scientifically oriented decompositions.  There already exists detailed
modeling of edge-on galaxies in S$^4$G, based on fitting their
vertical profiles to hydrodynamical thin-thick disk models
\citep{comeron2011,comeron2012}. Their radial luminosity profiles have been analyzed in\citep{comeron2012, comeron2014, martin_navarro2012}.

\subsection{Scope of pipeline decompositions}
The P4 models for the spiral galaxies $T > 0$ are generally good, giving
reliable estimates for parameters such as the bulge-to-total flux
ratio ($B/T$), the scale length of the disk ($h_r$) and its  central
surface brightness ($\mu_0$). However, despite the fact that up to
four components were fit, the pipeline decompositions for the
early-type disk systems (T$\lesssim $1), because of their complex structures,
are often insufficient.  These systems may have nuclear bars, ovals
and lenses, which are not included in our models in any systematic
fashion. Because of this, the pipeline $B/T$ flux-ratios, particularly for S0
galaxies, can be over-estimated. Including all these structures will
require even more complex decompositions, such as those
done in the near-IR by \citep{laurikainen2005, laurikainen2006, laurikainen2009, laurikainen2010}.
Such time consuming modeling goes beyond the scope of our current P4
decompositions; nevertheless, the P4 decomposition output files provide
good starting point for further fine-tuning.

\clearpage



\begin{figure}
\includegraphics[angle=0,scale=.9]{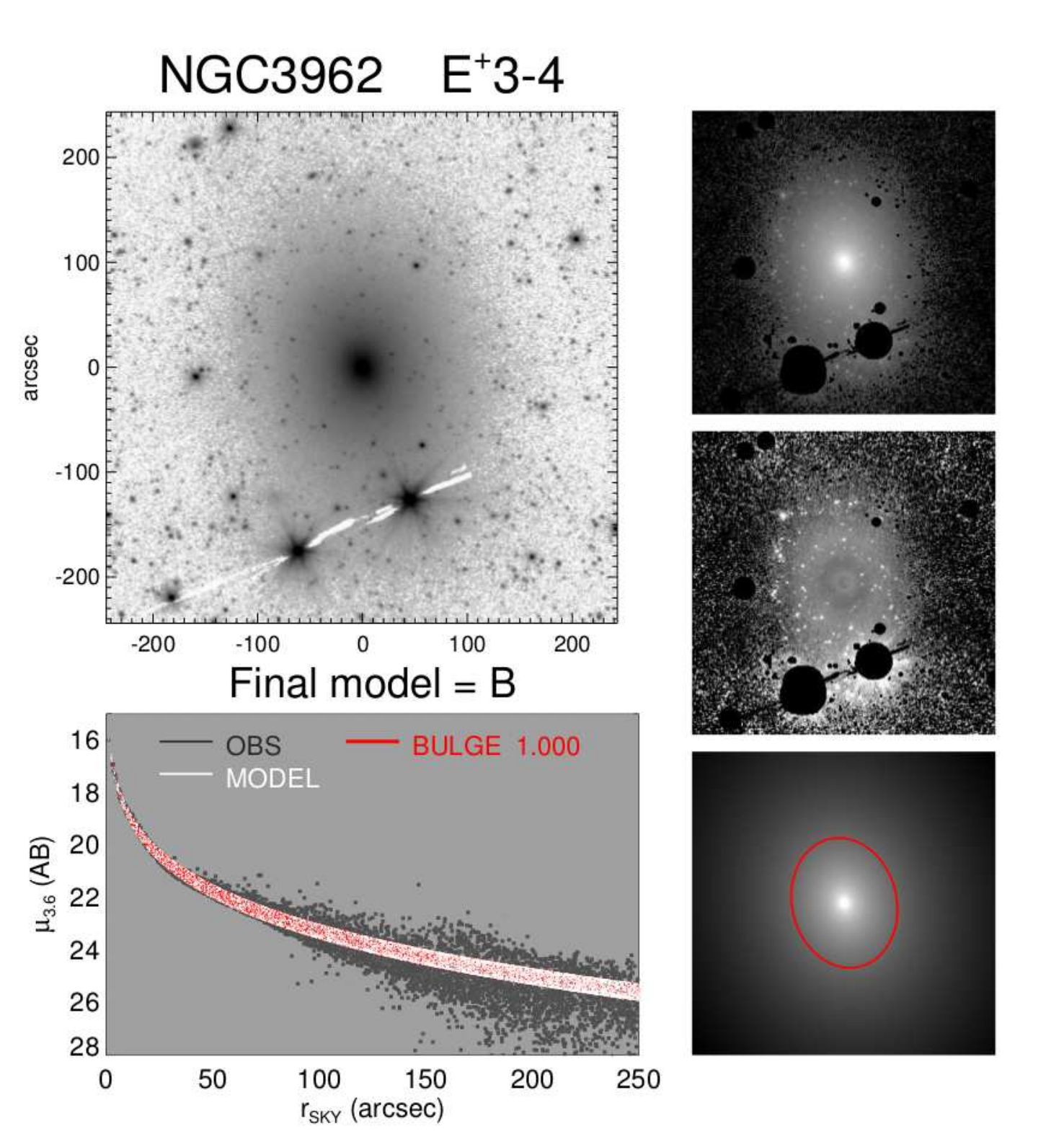}

\caption{ 
\baselineskip 0.55cm
Figs. \ref{ngc3962_eija} -\ref{eso026-001_eija} show the
  examples of final decompositions described in the text.  In the big
  panel the galaxy image is shown in an inverted logarithmic scale
  (magnitude range $27 > \mu_{3.6}(AB) > 18$), clipped to display the
  main morphological characteristics. The three small panels show the
  masked original image (upper right panel), the model image (lower
  right), and the residual OBS-MODEL in the middle (range $\pm 1 mag$;
  white indicates excess light over the model). The lower left frame
  shows the 2D profiles of the observed and model images (black and
  white dots), together with the model components (colors; labels
  indicate the relative fraction of flux in this component; in this
  particular case there is only one component). The same components
  are also marked, with the same colors, on the lower right model image:
  the semimajor axis of the ellipse corresponds to $2R_e$ of the component in
  question.  The mid-IR classification from Buta et al. (2014) is also
  indicated.  In this particular example for NGC 3962, the single
  S\'ersic fit provides an acceptable final model. The overall profile is
  close to a de Vaucouleurs profile (S\'ersic $n =5.6$) in accordance
  with the morphological classification (E). Nevertheless, the slight
  bends in the profile and the structure in the residual image
  suggests that if desired, it would have been possible to get an even
  slightly better fit by including multiple components.}
\label{ngc3962_eija}
\end{figure}

\begin{figure}
\includegraphics[angle=0,scale=.99]{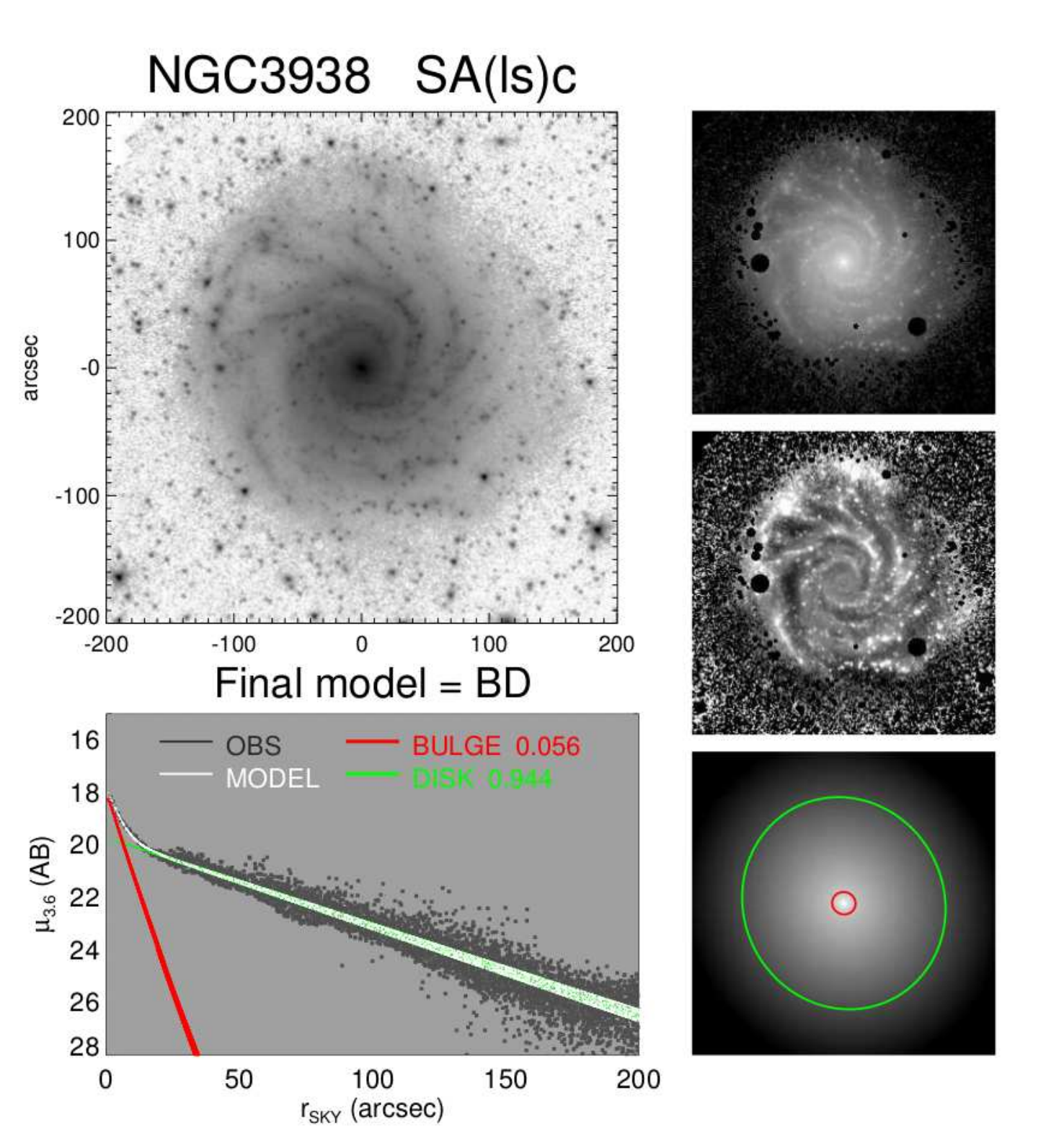}
\caption{NGC 3938: Example of a galaxy in which the automatic S\'ersic
  bulge + exponential disk fitting gave an acceptable final
  solution. In the profile plot the spirals appear as
  small undulations on the generally well-fitted disk.
  The labels in the profile plot
  indicate the relative fraction of flux in various components.}
\label{ngc3938_eija}
\end{figure} 

\clearpage



\begin{figure}
\includegraphics[angle=0,scale=.8]{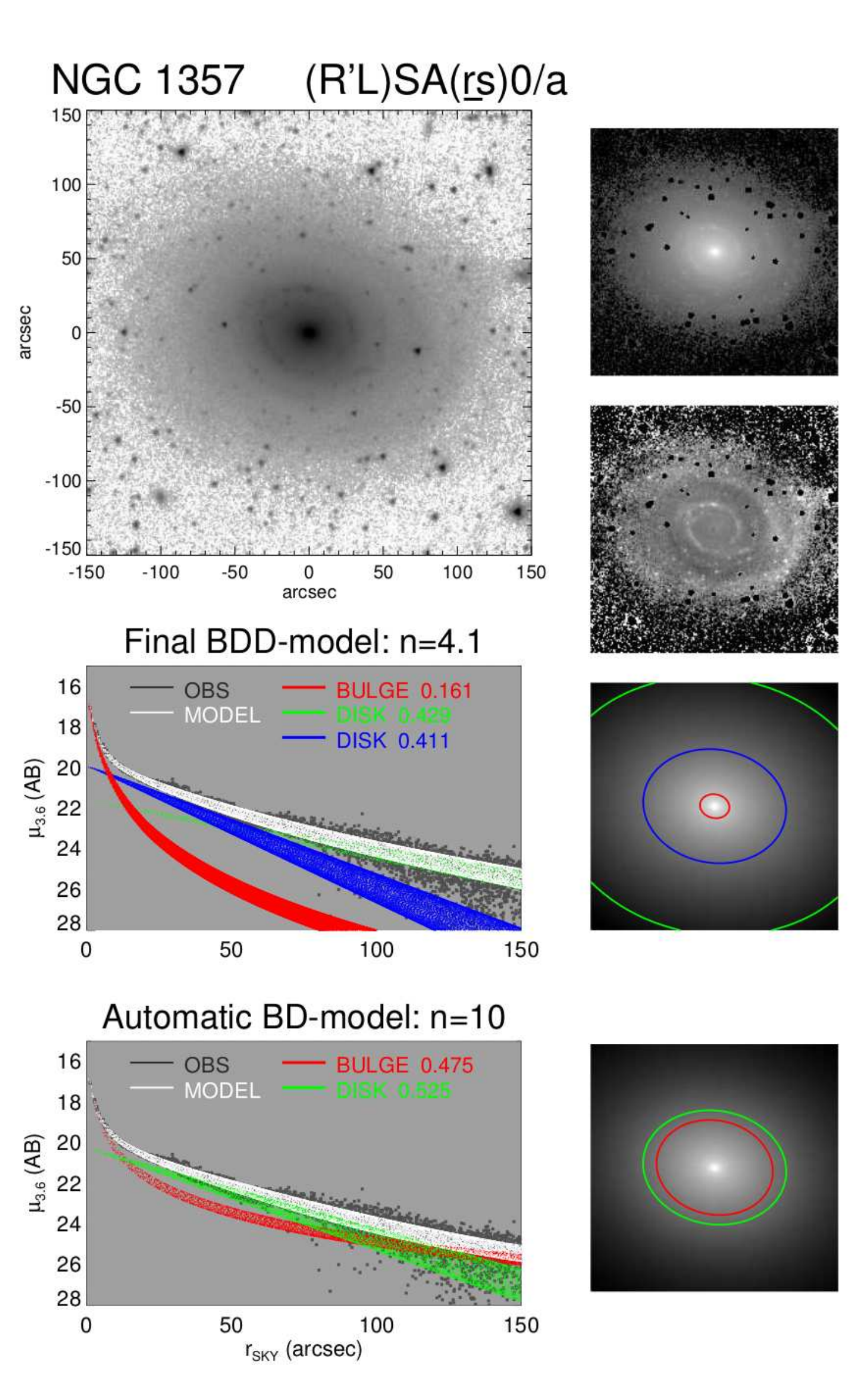}
\caption{\baselineskip 0.5cm
NGC 1357: The surface brightness profile of this galaxy shows a small bulge and
a large, fairly exponential disk. However, the automatic bulge-disk fit would give
an unreliably large bulge extending through the whole galaxy (lowermost row). A more
reasonable fit is obtained by adding another exponential disk component to
the inner part of the galaxy (upper profile). This inner component corresponds 
to the region of tightly wound spiral arms with higher surface brightness.
}
\label{ngc1357_eija}
\end{figure} 

\clearpage

\begin{figure}
\includegraphics[angle=0,scale=.99]{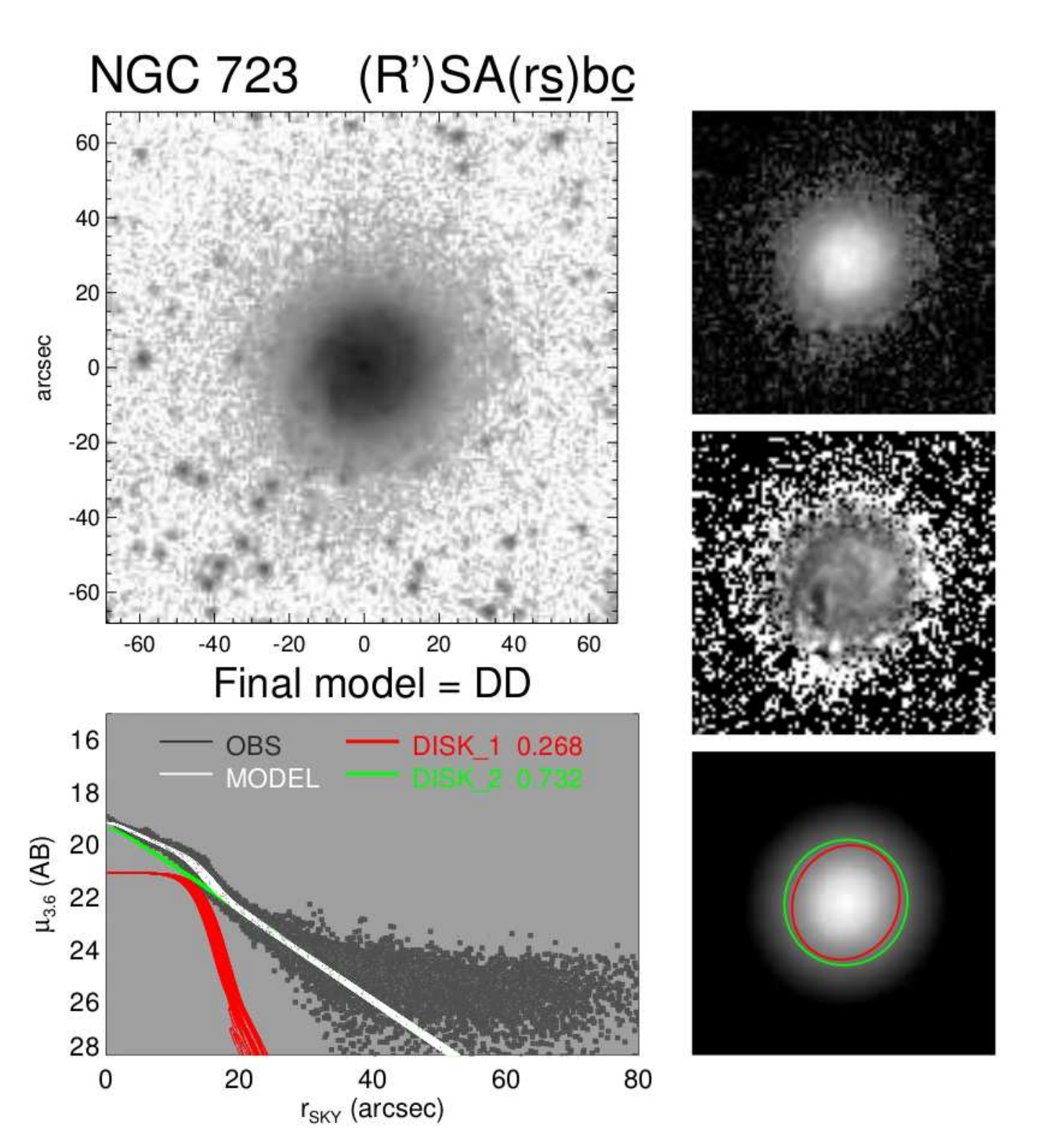}
\caption{NGC 723: This is a disk galaxy with no {\reply centrally
    concentrated} bulge component, but it is clearly not a single
  exponential galaxy.  The bump in the surface brightness profile
  corresponds to the strong high surface brightness spiral structure
  in the inner parts of the galaxy, here fit with a nearly flat part
  (S\'ersic function with $n=0.12$).}
\label{ngc0723_eija}
\end{figure}

\begin{figure}[ht]
\includegraphics[angle=0,scale=.8]{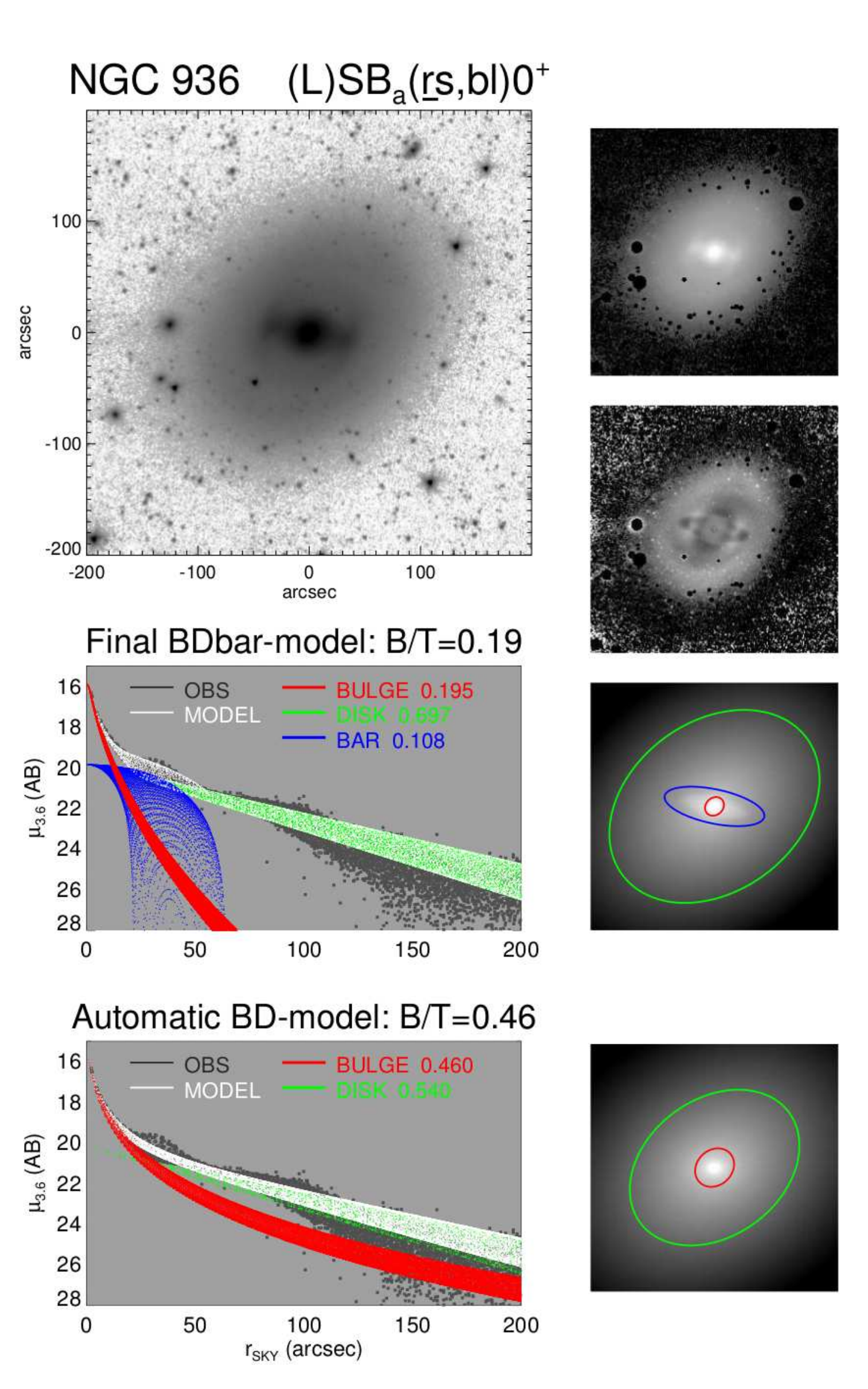}
\caption{\baselineskip 0.5cm
NGC 936: An example of a barred galaxy where inclusion of a bar
  component to the model makes a large difference in the bulge
  parameters. The lowermost row shows the automatic bulge/disk model,
  whereas the upper panels include a bar component. In the simple
  model, the bar flux is degenerate with the bulge flux. Multi-component decomposition is thus essential for getting a
  realistic bulge-to-total flux ratio: the B/T = 0.46 in the {\em BD}-model,
  but drops to 0.19 in the {\em BDbar} model .}
\label{ngc0936_eija}
\end{figure} 

\clearpage

\begin{figure}
\includegraphics[angle=0,scale=.8]{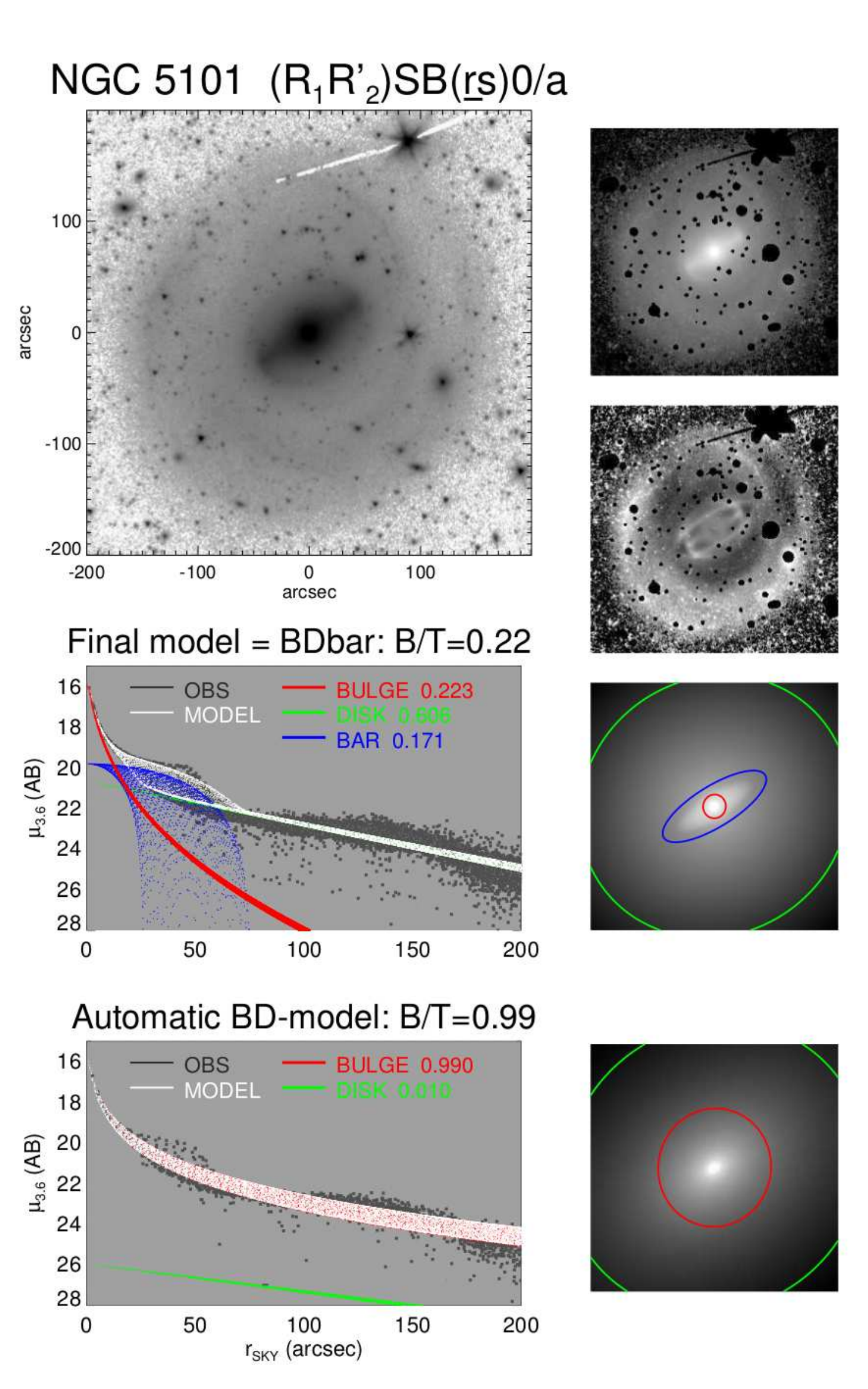}

\caption{\baselineskip 0.5cm
NGC 5101: Another example of {\em BDbar} decompositions where
  inclusion of the bar components is essential in getting realistic
  bulge parameters. Note also that in the final model the underlying
  disk is fit with an exponential function, although the outermost
  profile is downbending (Type II break/truncation). However, the steeper
  outer slope seems to be associated with a broad double outer ring, rather
  than a fundamentally distinct outer disk component.
\label{ngc5101_eija}}
\end{figure}

\begin{figure}
\includegraphics[angle=0,scale=.99]{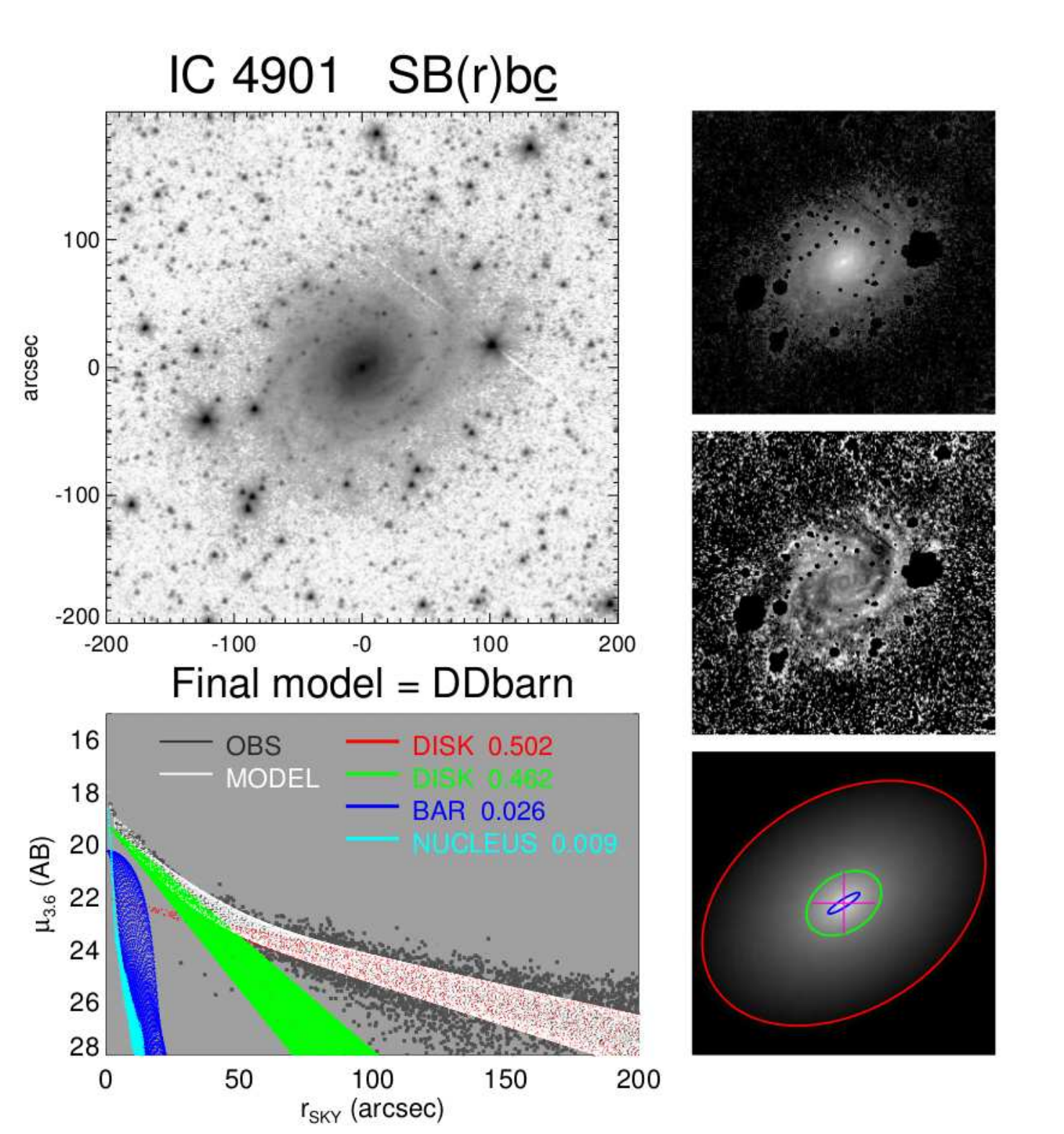}
\caption{ IC 4901: An example of a barred galaxy, in which the disk is
  fitted with two exponential functions. Additionally, a small central psf-component is included, marked as a cross on the model image. The inner disk corresponds to
  the higher surface brightness part of the disk outside the bar,
  where the spiral arms are also prominent.}
\label{ic4901_eija}
\end{figure} 

\clearpage

\vskip 1cm

\begin{figure} [ht]

\includegraphics[angle=0,scale=.99]{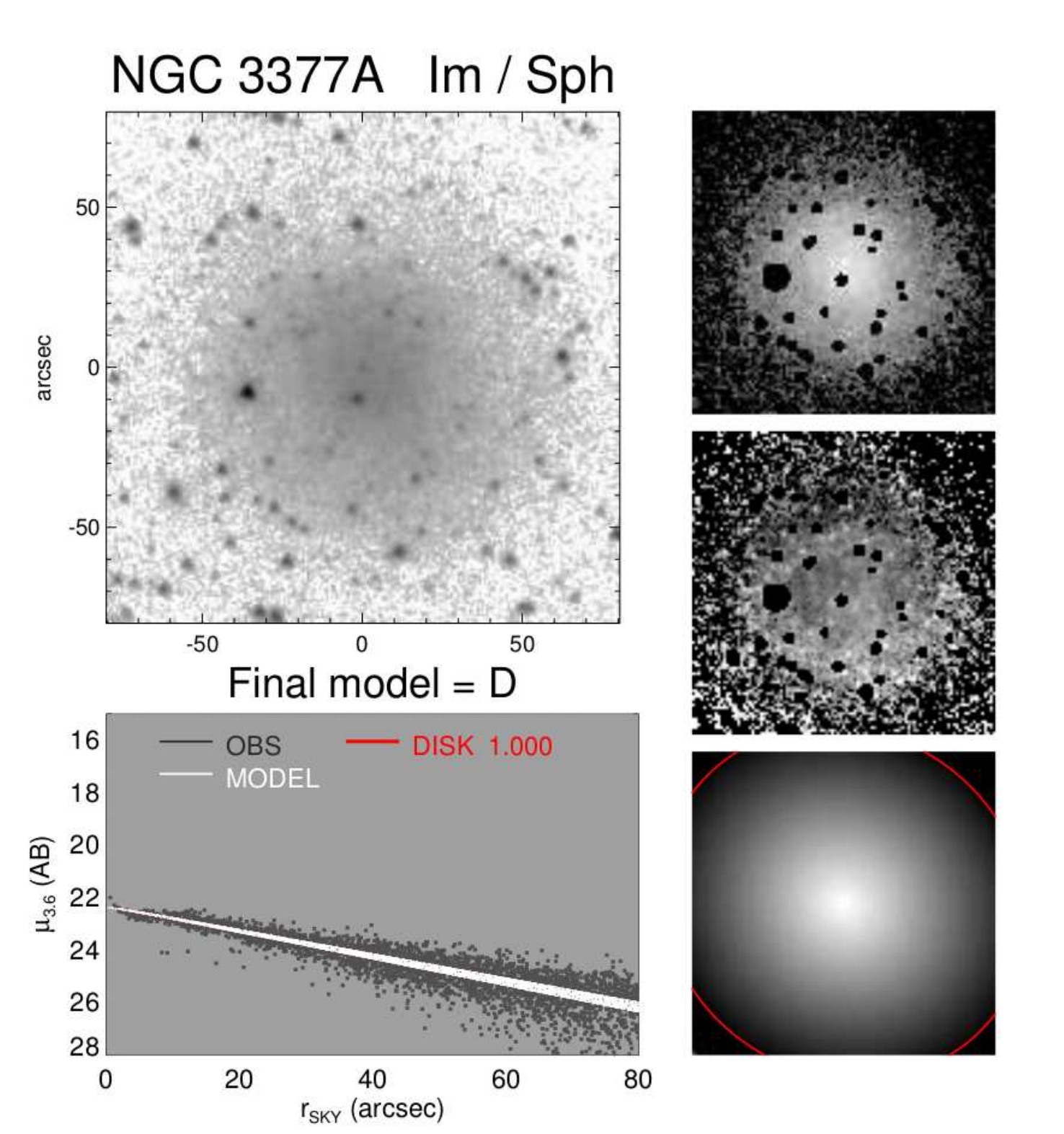}

\caption{NGC 3377A: A bulgeless disk galaxy, well fit with a
  single exponential function.}
\label{ngc3377A_eija}
\end{figure}

\begin{figure}
\includegraphics[angle=0,scale=.99]{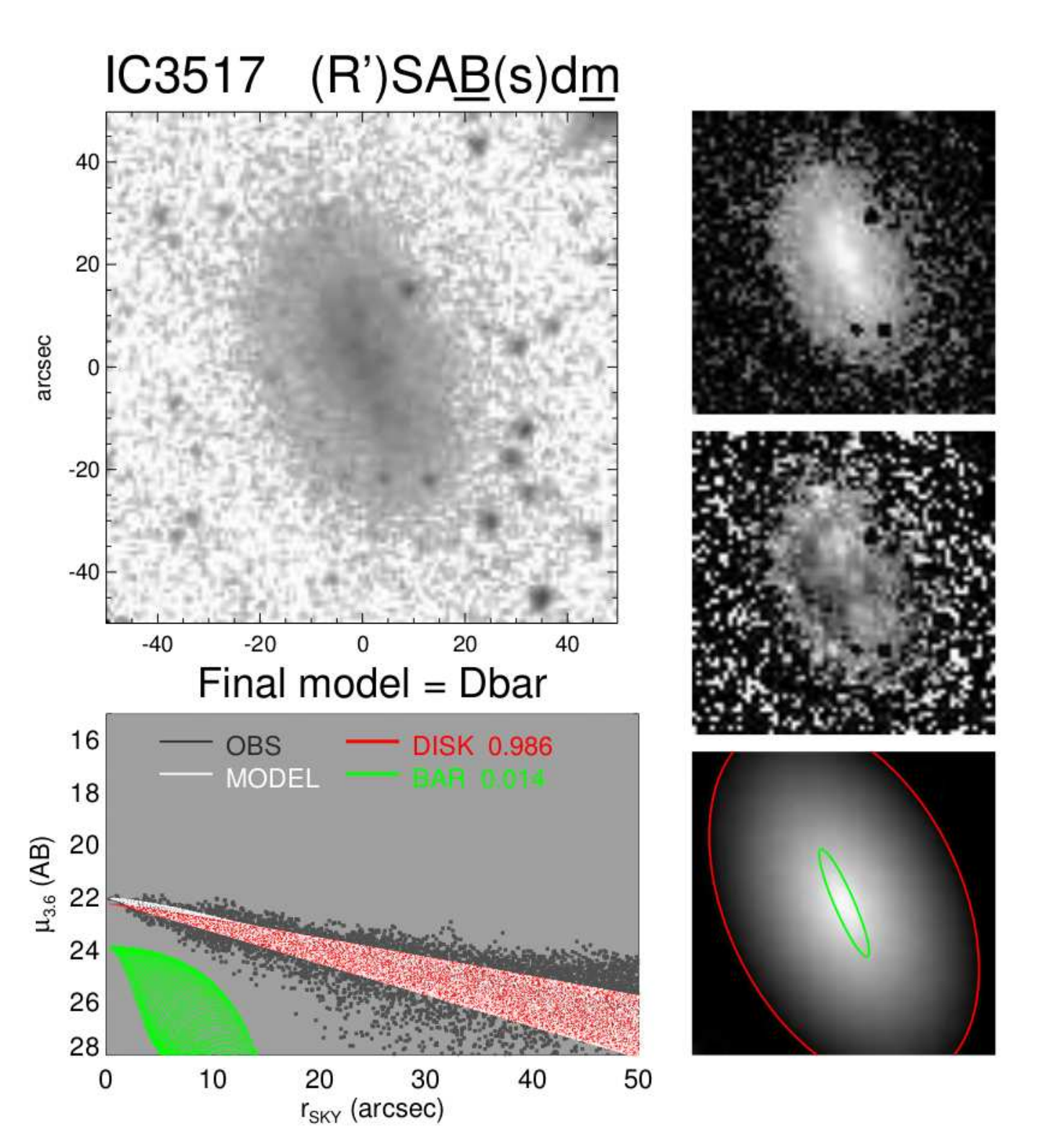}
\caption{IC3517: Another example of a bulgeless disk galaxy. The
  surface brightness profile is well fit with a single exponential
  function. However, the image shows also an elongated inner
  structure, which can be fitted with a Ferrers function.}
\label{ngc3517_eija}
\end{figure} 


\begin{figure}
\includegraphics[angle=0,scale=.99]{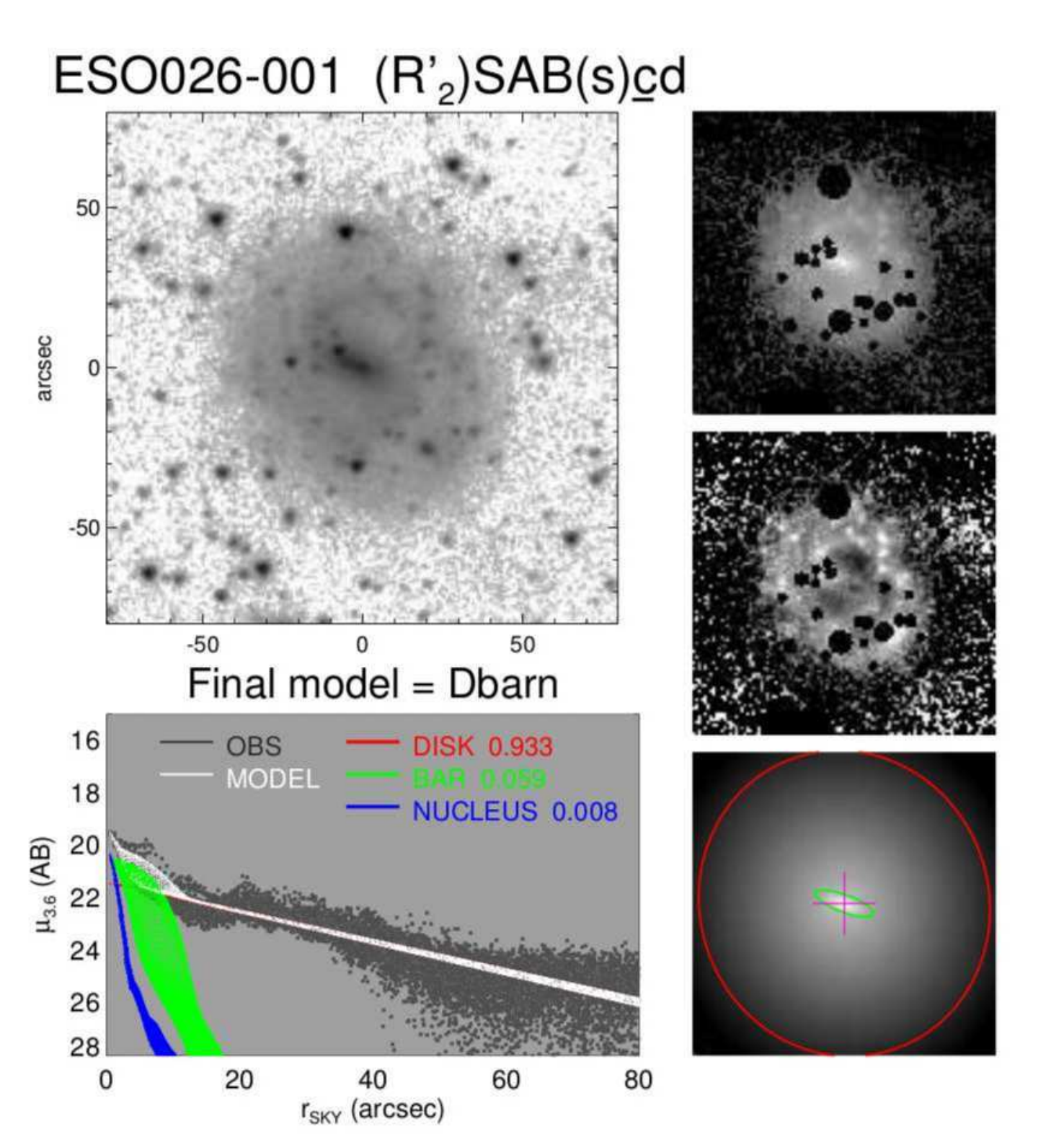}
\caption{ESO026-001: An example of a galaxy in which the disk is fitted
  with an exponential function, although a S\'ersic function with $n<$1
  would have given a more precise fit to the disk. }
\label{eso026-001_eija}
\end{figure}

\begin{figure}
\includegraphics[angle=0,scale=.99]{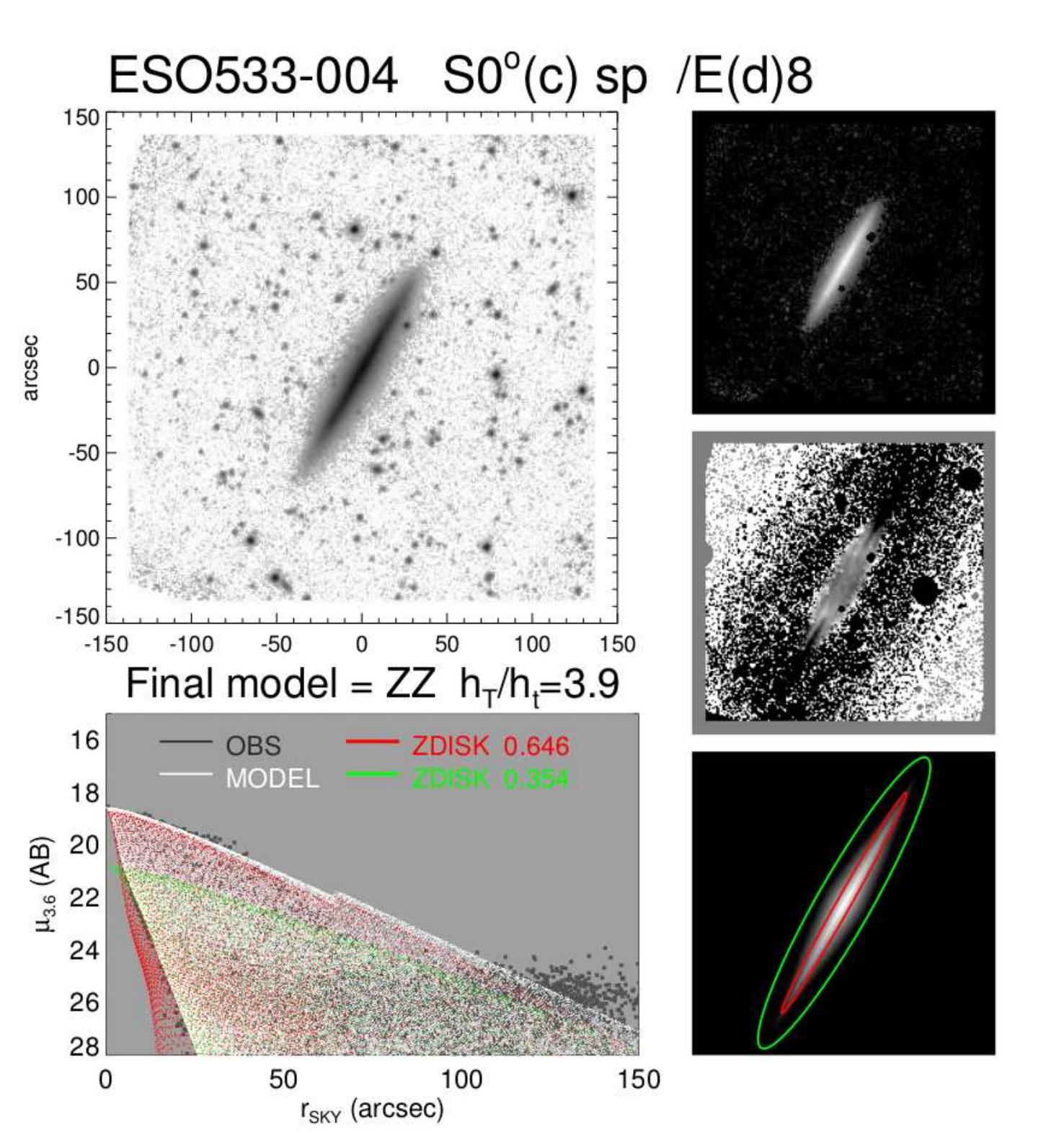}
\caption{ESO533-004: An example of an edge-on galaxy fitted with two
  disk components: the ratio of thick disk to thin disk scale heights
  is $h_T/h_t=3.9$ and the thick disk contains 35\% of the total model
  luminosity. For comparison, the detailed vertical profile fits in
  Comeron et al. 2012 indicated $h_T/h_t=4.3-4.6$ depending on the
  radial location, with about 45\% of light in the thick dist component.}
\label{eso533-004_eija}
\end{figure}

\clearpage


\section{Uncertainties of the decomposition parameters}
\label{sect_errors}

The formal uncertainties of the decomposition parameters have little
significance, as they refer to purely statistical uncertainty due to
image noise based on the assumption that the model
is accurately describing the true underlying light
distribution. Taking into account the complex morphology of most
galaxies, this assumption is clearly not valid (see Peng et al. 2010
for detailed discussion of errors)\footnote{The formal uncertainties
  calculated by GALFIT are listed in the headers of pipeline output
  files in IRSA}.  Related to this, the final value of the reduced
${\chi^2}_{\nu}$ is a poor indicator of the goodness of the fit (even for a
good model it is typically much larger than unity) and is thus not used as a
decisive factor in choosing the preferred final model.  In practice,
the choice of the final model components plays a crucial role: for
example as seen in Section \ref{sect_eija_examples}, omission of the
bar component when a bar is present may lead to seriously biased bulge
parameters.  In this Section we perform a systematic comparison of
bulge and disk parameters between 2-component and final
multi-component models. We also first examine the potential
uncertainties related to the preparation of data before the
decompositions, namely the used PSF-function, the effect of sky
subtraction uncertainty and the sigma-image.

\subsection{PSF}

As illustrated in Fig. \ref{psf}, the IRAC PSF has extended wings.
Moreover, the PSF and the orientation of its asymmetric extensions
vary from image-to-image, which has not been taken into account in our
decompositions. To check the importance of the PSF wings, we compared 
differences in decomposition parameters obtained
when the adopted composite PSF was replaced with a Gaussian PSF having
the same $FWHM=2.1\arcsec$.  Fig. \ref{fig_psf_sersic} compares the
resulting effect on the S\'ersic parameters in 1-component models.
Clearly, decompositions with the Gaussian PSF yield
$n$ values that are systematically too small, differences reaching even
tens of percents for some of the galaxies (though the median deviation
is less than  5$\%$). 
However, these rather large deviations are not representative
of the true uncertainties, but rather give an idea of the 
magnitude of the error if the tails of the PSF were altogether ignored.  A
better measure of the actual uncertainty in P4 decompositions is
obtained by comparing with an azimuthally symmetrized version of the
composite PSF.  Clearly, now the differences in $n$ are much smaller
(see the red symbols in Fig. \ref{fig_psf_sersic}).

We also checked the influence that the PSF has on the multi-component
models. For that purpose we rerun all final decompositions that
included both bulge and disk components (+ possible bar and center
components; total of 524 models after excluding nearly edge-on
galaxies), using both the Gaussian PSF and the symmetrized PSF.  Table
\ref{table:psf_error} lists the median of relative differences in
bulge $B/T, n$, $R_e$, disk scale lengths $h_r$, and bar-to-total
ratio Bar/T, when compared to the results obtained using the standard
composite PSF.  The largest differences are seen for the S\'ersic
parameters while using the Gaussian PSF, whereas $h_r$ is barely
affected. On the hand, the differences in parameters between those
obtained using the composite PSF and using the symmetrized version are
negligible. Based on these results we conclude that the spikes of the
PSF have no significant effect as long as the nearly circular wings of
the PSF are included.  The use of single composite PSF for all S$^4$G
images should thus be acceptable.

\subsection{Sky subtraction}

In principle, poor sky subtraction can severely affect the
decomposition results, in particular the parameters of the disk.  To
constrain the possible magnitude of such uncertainties, we rerun the
multi-component decompositions that included both bulge and disk
components (+ possible bar and center components; same 524 models as
above). Two additional sets of sky values, $SKY' = SKY \pm DSKY$, were used,
where  DSKY was the standard deviation of the different sky regions.
 Fig. \ref{fig_psf_disk} shows the effect on the
scalelength of the disk. Although individual changes can in few cases
be large ($ h_r ($mod)$/ h_r($ori$) > 1.2$ in 9 cases when too small a
sky is subtracted), the median differences are less than 2\% (and even
smaller in the other parameters of interest, see Table
\ref{table:sky_error}).
The sky subtraction is not a concern in the current decompositions.

\subsection{Sigma-image}
\label{sect_sigma_uncertainty}
The weights applied to various pixels have an important role in
decompositions, in particular when the galaxy structure is
complicated, so that the differences between the applied model and the
true structure are large.  As mentioned in Section \ref{sect_sigma}
the $\sigma$-image itself is a statistical estimate of the 
underlying $\sigma$ in each pixel, so it might be reasonable to
smooth it before applying it in the decompositions.  In Fig.
\ref{fig_final_sigma} (left column) we examine the effect of
sigma-image smoothing on the derived bulge parameters. A median filter
is applied with a width of 5 pixels.
  Clearly the effect is
quite small except for a few deviant cases marked on the plot. In
these cases the bulge parameters are sensitive also to changes in the
PSF or the sky background level.  

For comparison, Fig \ref{fig_final_sigma} (right column) also
illustrates the changes in bulge parameters if a constant sigma is
assumed at all image pixels. A constant sigma exaggerates the relative
weight of the central regions compared to the outskirts.  Besides a
large scatter, also a systematic increase of the estimated $n$ is
obvious: the median $n_{mod}/n_{ori} =1.2$ (the mean ratio is 1.4).
What typically happens is that the fit tries to reproduce the central
peak with an increased $n$, even if the outer disk then becomes too much
bulge dominated.  Indeed, the bias (and the scatter) is particularly
large for earlier type disks (open circles in the plot indicate $T\le
5$; median $n_{mod}/n_{ori}=1.25$).  This comparison reminds us that when
decomposition parameters from different studies are compared, it is
also important to pay attention that similar weights have been
applied.

\subsection{Two-component versus multi-component decompositions?}
\label{sect:2comp_multi}

Automatic 2-component S\'ersic-exponential (or S\'ersic-S\'ersic)
models are often applied to large data surveys. This is a natural
approach as the data quality (depth/angular resolution) might be
insufficient for more detailed modeling so that the large effort in
multi-component decompositions does not seem justified.  Moreover, it
has been recently claimed \citep{tasca2011} that 2-component
decompositions (S\'ersic + exponential) are sufficient also for
fitting  barred galaxies.  Their argument was based on obtaining
similar average $B/T$ ratios for barred and non-barred galaxies in
their 2-component bulge/disk models. They reasoned that if the
omission of the bar were a problem it should manifest as a higher B/T
for barred galaxies.  However, to accurately address this matter one
has to compare the different types of decompositions (2-component and
multi-component) for well-defined samples of barred/non-barred
galaxies.

Such a comparison between different decomposition models is shown in
Fig. \ref{fig_bd_final_bar_nonbar}.  Again, those galaxies for which
the final model contains both a bulge and a disk are studied.  For the
non-barred galaxies (those with no {\em bar}-component; leftmost
column in the Figure) the bulge parameters (S\'ersic $n$, $B/T$,
$R_e/h_r$) in automatic 2-component runs are almost identical to those
in the final models.  This agreement is expected because over 80\% of
the final non-bar models are just S\'ersic-expdisk models (15\% have
two disk components, and 2\% have an extra central component), and typically the automatically found {\em BD}
models did not need any refinement. For barred galaxies (those with a
{\em bar}-component in the final model; middle column), the obtained
median values depend drastically on whether the bar is included.  This
result emphasizes that the examples of decompositions given in Section
\ref{sect_eija_examples}, highlighting the importance of modeling the
bar (e.g. Figs. \ref{ngc0936_eija} and \ref{ngc5101_eija}) were not
exceptional cases.  Overall, ignoring the bar increases the estimated
$B/T$ ratios by a factor of 2-3 because of gross (even by as much as a
factor of 5) overestimate of $R_e$ and $n$. For example, for spirals
in the range $1 \le T \le 5$ the 2-component decompositions suggest $n
\gtrsim 4$ whereas the multi-component runs indicate $n \approx
1-2$. Altogether, in the final models the difference in bulge
parameters obtained in the multi-component decompositions for barred
and non-barred galaxies is fairly small (right column in Fig. \ref{fig_bd_final_bar_nonbar}).

 The conclusion that multi-component decomposition models are
 essential to measure realistic bulge parameters for barred galaxies
 is not new \citep{laurikainen2006, laurikainen2007, gadotti2008,
   weinzirl2009}).  A similar conclusion, based on synthetic images,
 was reached also by \cite{laurikainen2005}.

In Fig. \ref{fig_final_nirsos} we compare the combined bar/non-barred
sample of the previous figure with the decompositions in \cite{laurikainen2007}.  Because of the large fraction of barred galaxies, the
difference in the obtained bulge properties between the 2-component
and multi-component decompositions remains significant, even when
barred and non-barred galaxies are considered together.  
{\reply We find an excellent agreement between the current
  multi-component results and those in \cite{laurikainen2007},
  obtained with a different decomposition code ({\em BDBAR}; however
  BDBAR uses IDL {\em Curvefit} and is thus based on the same
  Levenberg-Marquadrdt minimization as GALFIT), and based on different
  near-IR image data.}  It is worth noticing that in these
decompositions the S\'ersic $n$ for Hubble types Sa-Sc is nearly
$n\sim$1, whereas in the decompositions by \cite{tasca2011},
for the same Hubble types, the S\'ersic index is peaked at
$n\sim$4. Small values of the S\'ersic index, similar to ours for
these Hubble types, are reported also by \cite{graham2008}.

\begin{deluxetable}{ccccccccc}
\tablewidth{-20pt} 
\tabletypesize{\scriptsize} 
\tablecaption {The effect of a modified PSF on final decomposition model
parameters}
\tablehead{ 
{} &{} &{ GAUSSIAN \ PSF \hfill} {} & &  {}& {SYMMETRIZED \ PSF} \\ \tableline \\
\tableline \\
\colhead{}  & & \colhead{median($D$)}   &\colhead{median($|D|$)} & \colhead{}     & \colhead{median($D$)}    & \colhead{median($|D|$)}
}
\startdata
$B/T$    &  &       -1.5 \%  &        4.5 \%   &    &        -0.1 \%      &      0.2 \%  \\
$n$      &  &       -3.0 \%  &        7.8 \%   &    &         0.0 \%      &      0.5 \%  \\
$R_e$    &  &        8.9 \%  &       9.8 \%   &    &         0.1 \%      &       0.3 \%  \\
$h_r$    &  &        0.0 \%  &        0.3 \%   &    &         0.0 \%      &      0.0 \%  \\
$Bar/T$  &   &       -1.5 \% &        4.4 \%   &    &        -0.1 \%      &     0.2 \%  \\
%
%
%
\enddata 
\label{table:psf_error}
\tablecomments{$D$ stands for the relative difference
(e.g. $D=[n($mod$)-n($ori$)]/n($ori$)$), where 'ori' refers 
to the standard
composite PSF. Medians are used to characterize the typical deviations
and the scatter, to eliminate spurious cases where the
decompositions with Gaussian PSF converged to a different type of
solution.}
\end{deluxetable}

\clearpage

\begin{figure}
\includegraphics[angle=0,scale=.95]{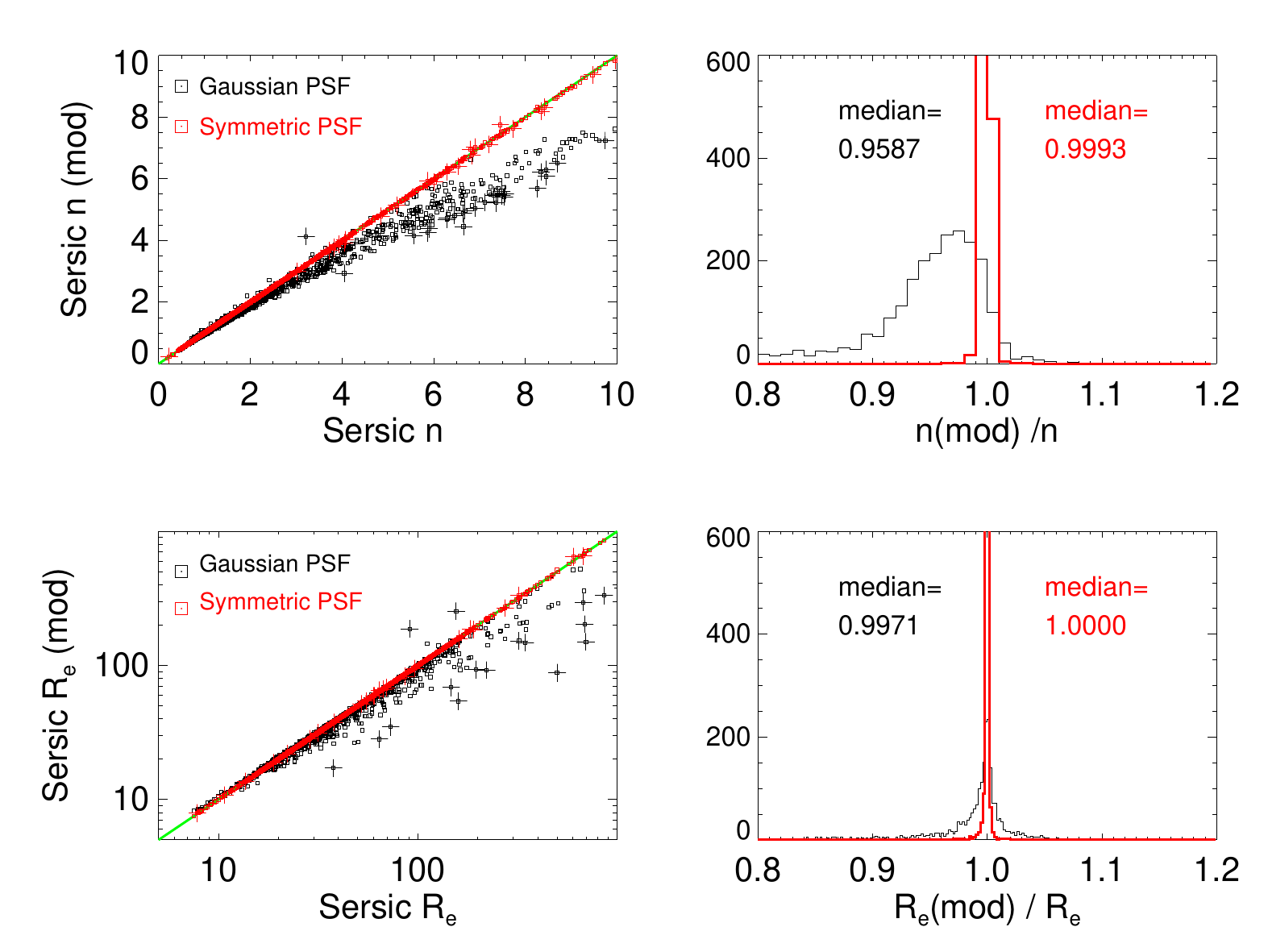}
\caption{ The effect of the PSF on decomposition parameters. The plots on
  the left show the S\'ersic index and effective radius in 1-component
  fits using modified PSFs instead of the standard composite PSF:
  black symbols indicate results using a Gaussian PSF (wings
  truncated), and red points when using a symmetrized composite
  PSF. Larger black (red) symbols indicate points deviating by more
  than 25$\%$ (10$\%$) from the unit line.  In the right panels the
  histograms of the relative changes in the parameters are shown:
  black and red colors have the same meaning as in the left frames.}
\label{fig_psf_sersic}
\end{figure} 
\clearpage

\begin{figure}
\includegraphics[angle=0,scale=.85]{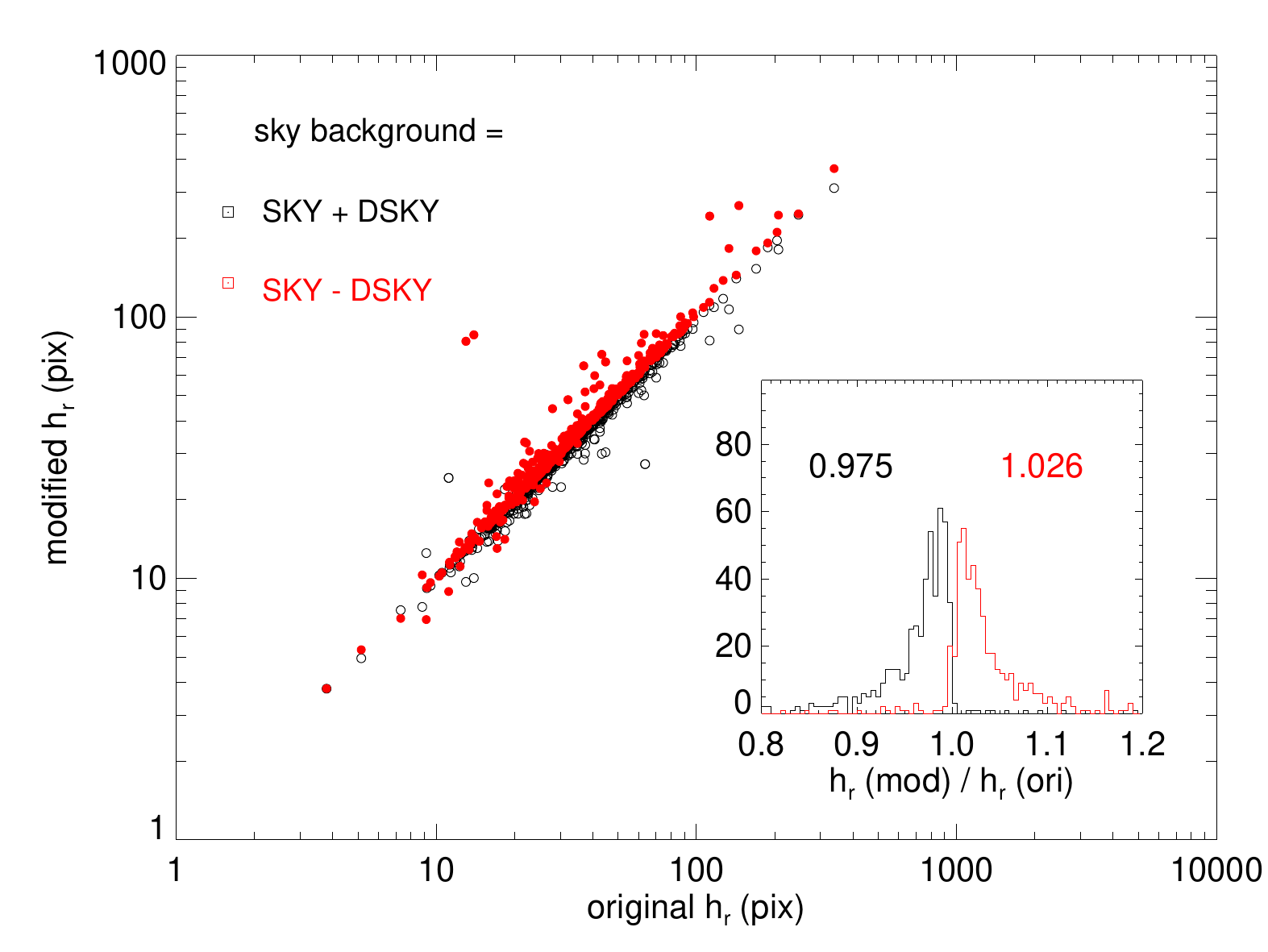}
\caption{ The effect of sky background subtraction on the disk
  scalelength. All final decomposition models including both a bulge
  and disk (and possibly additional bar and central components) were
  rerun using images where the assumed sky background was modified by $\pm
  DSKY$, where $DSKY$ is the conservative estimate of
  global sky variations in the image (see Section \ref{sect_sky}).}
\label{fig_psf_disk}
\end{figure}

\begin{deluxetable}{ccccccccc}
\tablewidth{-20pt} 
\tabletypesize{\scriptsize} 
\tablecaption {The effect of sky subtraction on final decomposition model parameters
parameters}
\tablehead{ 
{} &{} &{ SKY + DSKY \hfill} {} & &  {}& {SKY - DSKY} \\ 
\colhead{}  & & \colhead{median($D$)}   &\colhead{median($|D|$)} & \colhead{}     & \colhead{median($D$)}    & \colhead{median($|D|$)}
}
\startdata
$B/T$    &  &        0.2 \%  &        1.5 \%    &    &        -0.1 \%       &      1.6 \%   \\
$n$      &  &       -1.4 \%  &        1.8 \%    &    &         1.8 \%       &      2.0 \%   \\
$R_e$    &  &       -1.2 \%  &        1.4 \%    &    &         1.7 \%       &      1.8 \%   \\
$h_r$    &  &       -2.5 \%  &        2.6 \%    &    &         2.6 \%       &      2.8 \%  \\
$Bar/T$  &   &       0.2 \%  &        1.5 \%   &    &        -0.1 \%       &      1.6 \%  \\
\enddata 
\label{table:sky_error}
\tablecomments{$D$ stands for the relative difference
(e.g. $D=(n(mod)-n(ori))/n(ori)$), where 'ori' refers to 
the standard
sky subtraction.}
\end{deluxetable}
\clearpage

\begin{figure}
\includegraphics[angle=0,scale=.85]{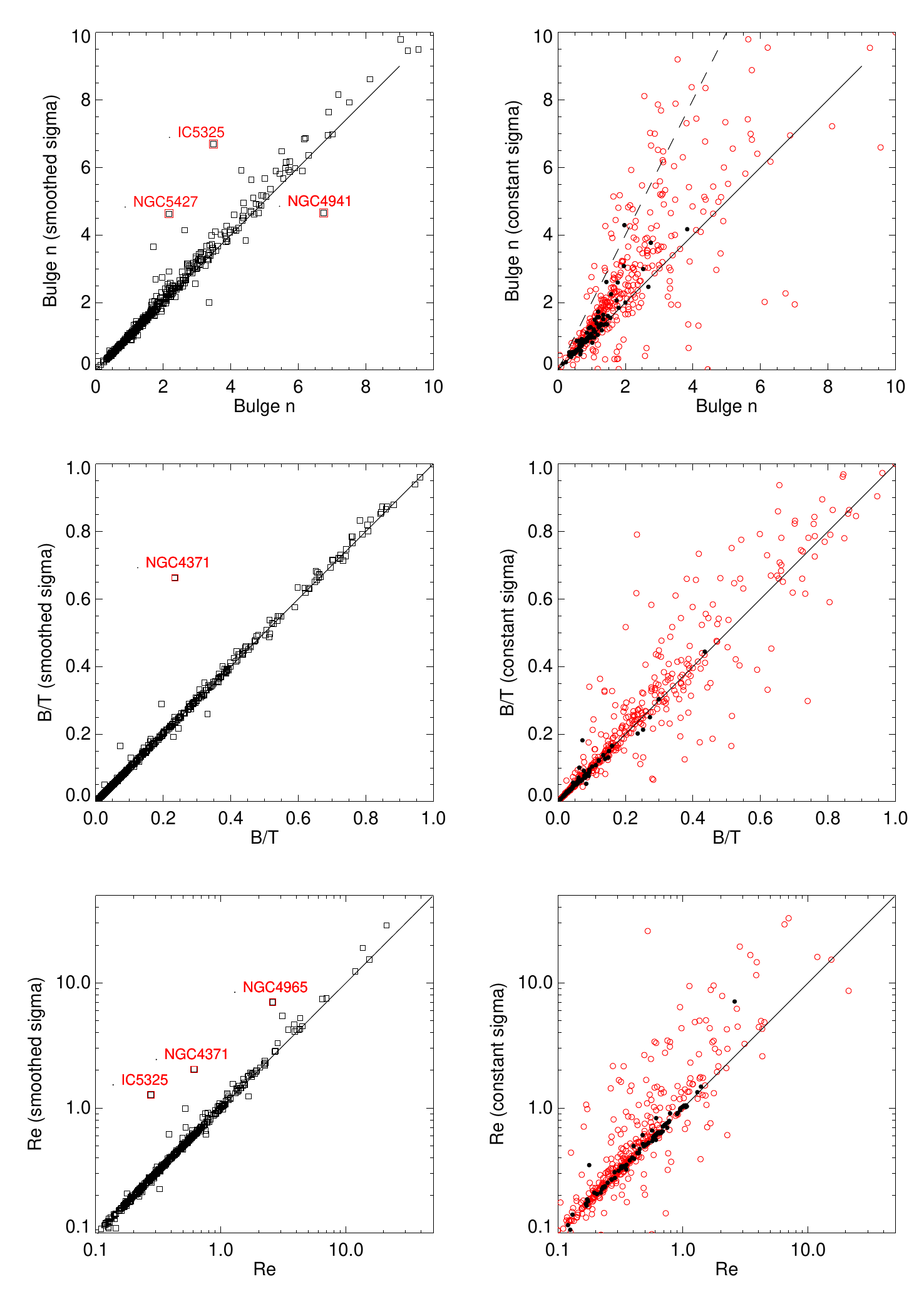}
\caption{
\baselineskip 0.5cm
Sensitivity of estimated bulge parameters (Sersic $n$,
  bulge-to-total flux ratio $B/T$, effective radius $R_e$ in kpc's) on
  the used $\sigma$-image.  In the left column, we have smoothed the P4
  $\sigma$-images with $5 pix \times 5 pix$ median filter, 
  while in the right it has been replaced with a constant
  $\sigma$. The scatter plots show the modified parameter values
  versus the original ones.  In the right, the red open and black filled circles
  refer to galaxies with mid-IR type $T \le 4$ and $T \ge 5$,
  respectively. Lines corresponding to one-to-one correspondence are drawn in each frame: in the right uppermost frame the dashed line indicates $n_{mod}/n_{ori}=2$.}
\label{fig_final_sigma}
\end{figure} 
\clearpage







\begin{figure}
\includegraphics[angle=0,scale=.8]{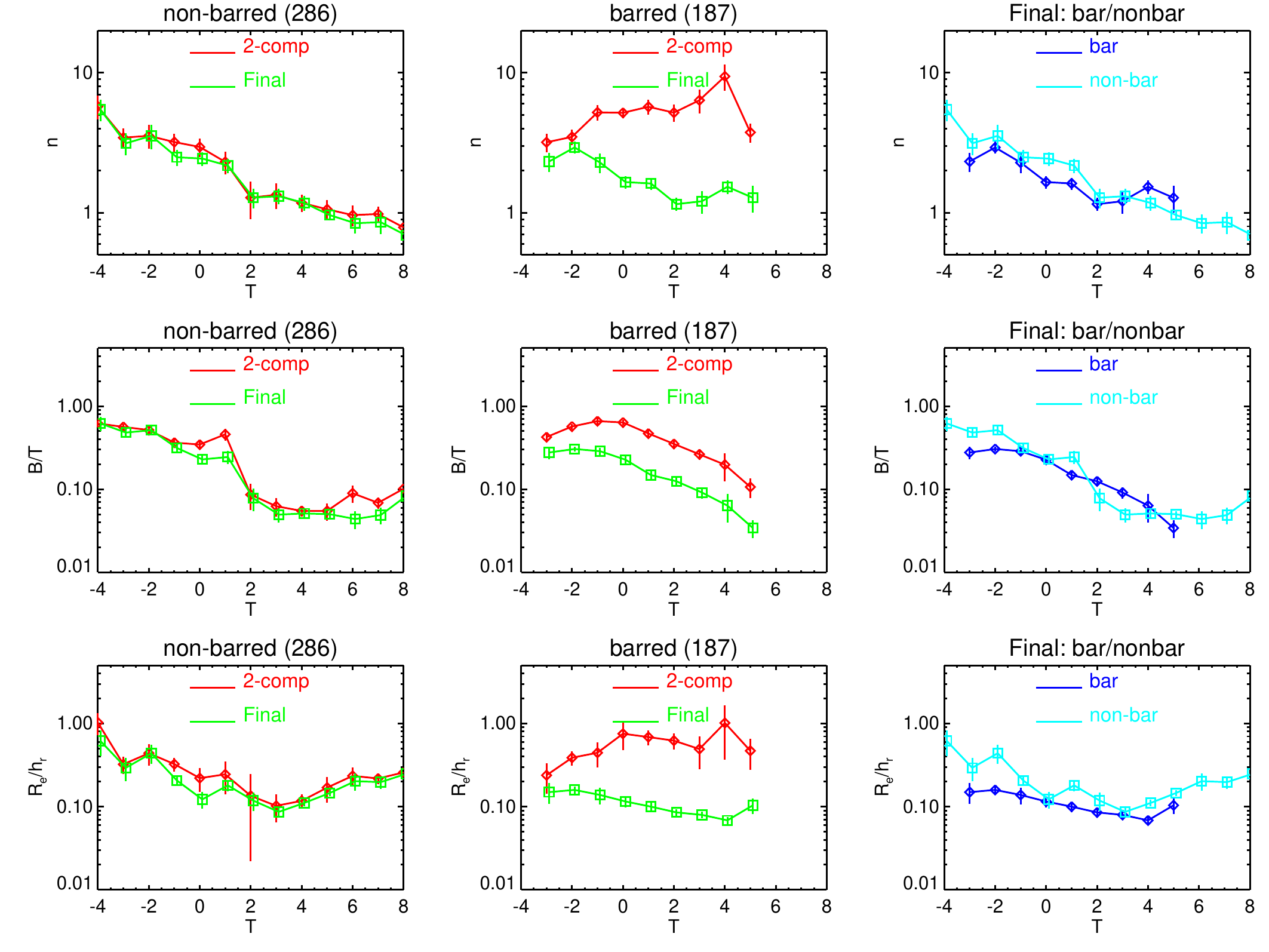}
\caption{ Comparison of bulge parameters between automatic 2-component
  and final multi-component decompositions.  The morphological type
  $T$ is from the Buta et al. (2014) mid-IR classification. Comparison is
  made for the galaxies for which the final model included both bulge
  and disk components. In the left frames decompositions for
  non-barred galaxies are compared, while the middle frames show those
  with a bar component in decompositions.  In the right frames the
  final decompositions for barred and non-barred galaxies are compared.
  The symbols stand for median values in bins with five or more
  galaxies, error bars are errors of the mean values in the bin. Note
  that the Buta et al. (2014) classification contains also half-integer
  values of $T$, resulting from averaging over two rounds of
  classification.  However, the number of galaxies with half-integer
  values is much less than those with integer $T$. Therefore, when
  binning the galaxies we have rounded the half-integer values
  randomly to the nearest smaller or larger integer value; same is
  done in Fig. \ref{fig32} below.
\label{fig26}
\label{fig_bd_final_bar_nonbar}}
\end{figure}

\clearpage

\begin{figure}

\hskip -1cm \includegraphics[angle=0,scale=.55]{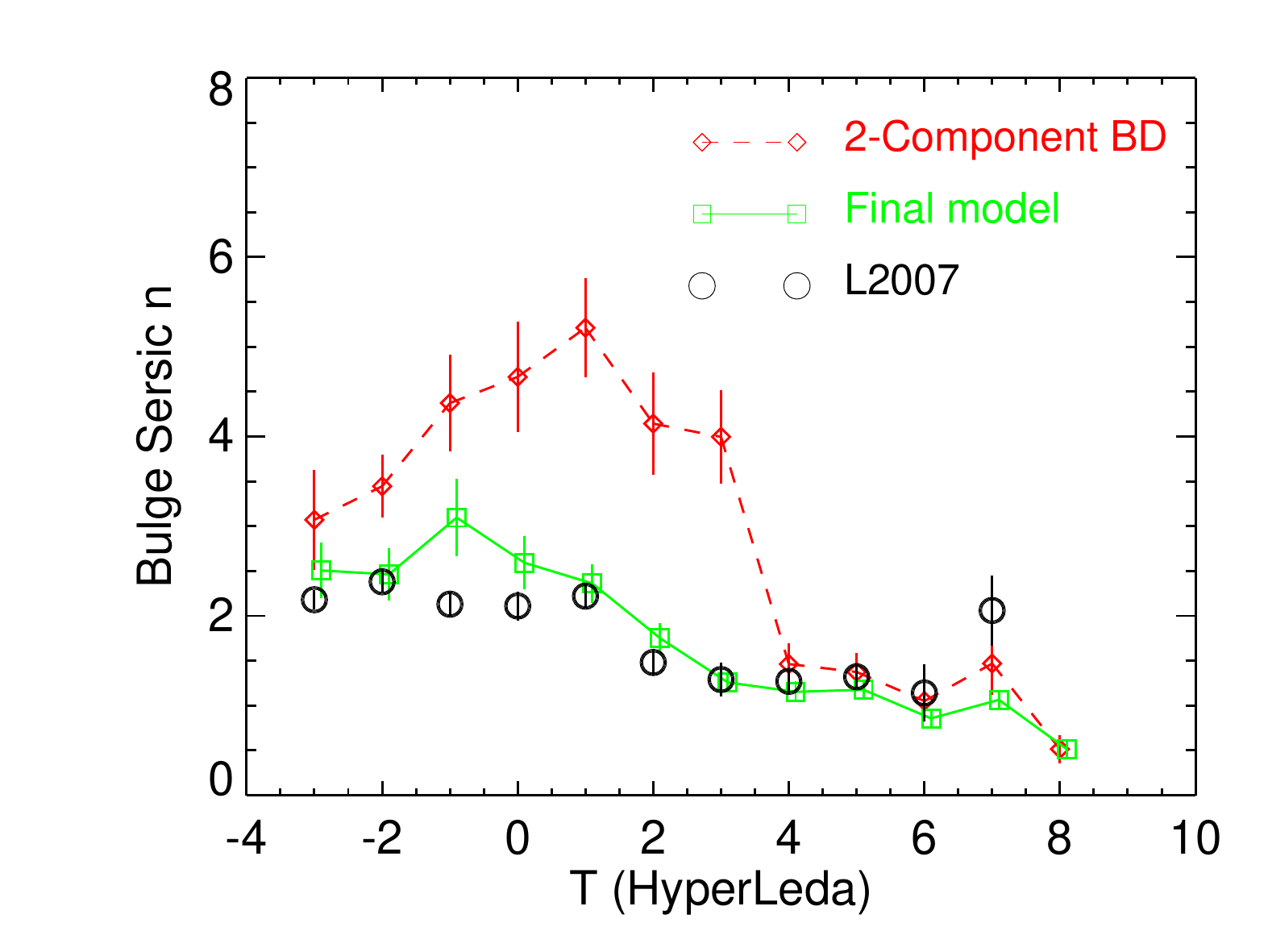}
\includegraphics[angle=0,scale=.55]{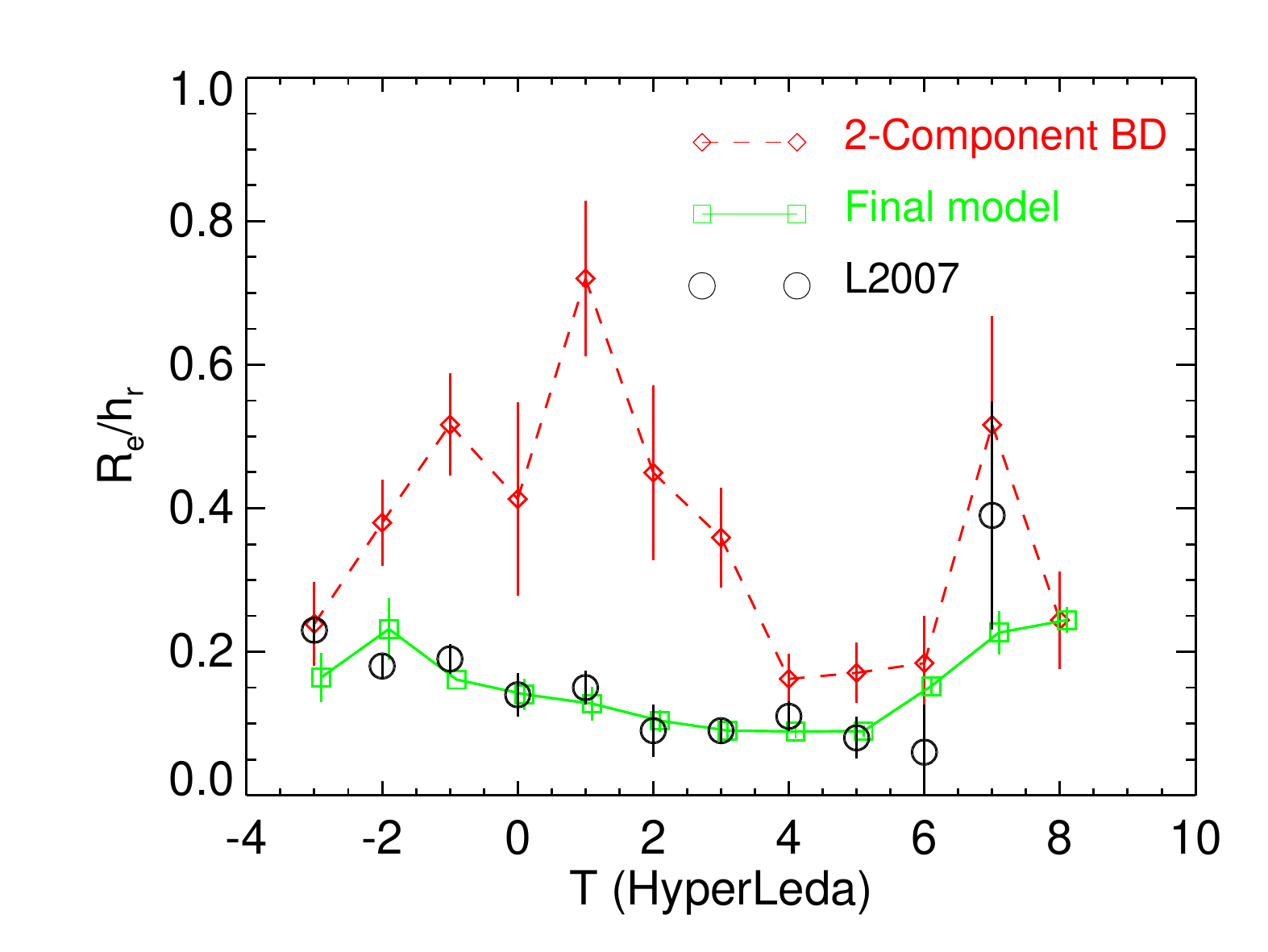}
\caption{Comparison of final S$^4$G decompositions parameters to those of
  Laurikainen et al. (2007; the values are tabulated in Table 2 in
  Laurikainen et al. 2010) multi-component decompositions for NIRS0S
  ($K_s$ band, 143 galaxies with $-3 \le T \le 1$ ) and OSUBSGS ({\em H} band, 129
  galaxies $2 \le T \le 7$) data.  The
S$^4$G results show 524 galaxies for which the final decomposition
model contained both a disk and a bulge component (excluding nearly
edge-one galaxies). For comparison, also the results of semi-automatic
2-component decompositions are shown.  The symbols stand for the
median values in each bin with 5 or more galaxies, while the error
bars denote the error of the mean. Note that here we use the optical
classifications from HyperLeda, to facilitate comparison with
previously published results.
\label{fig_final_nirsos}}
\end{figure}

\clearpage

\subsection{Disk breaks}
\label{sect:inner_outer}

One of the main goals of Pipeline 4 is to obtain measurements for the
galaxy size-magnitude scaling relations. 
In order to be consistent with earlier analysis (e.g. \cite{courteau2007})
the P4 final models as a default use single exponentials for the disk. 
However, deep optical and near-IR surveys \citep{erwin2005, pohlen2006, gutierrez2011, munoz_mateos2013} have shown
that only a fraction of galactic disks ($\sim 1/3$) are simple
exponentials (=Type I in \citealt{pohlen2006}
classification). Instead the typical brightness profiles consist of
two (sometimes three) exponential subsections with different radial
slopes. When the outer disk has a steeper slope, the galaxy is
classified as possessing a Type II break ('truncation'), and conversely if the
outer slope is more shallow, it is classified as a Type III break
('antitruncation').  \cite{kim2014} have recently made 2D
decompositions for 144 barred S$^4$G galaxies taking into account disk
breaks in their decompositions with the BUDDA code \citep{desouza2004, gadotti2008, gadotti2009}.  Their fitting function for the disk
consists of two exponential sections, with different scale-lengths
($h_{in}$ and $h_{out}$) inside and outside the break radius
$R_{break}$.  They also made decompositions where they fitted the disk
with a single exponential component. Their result indicate that the
inner scale lengths for two-component disks are typically about 40\%
longer than the scalelengths obtained in single disk fits; they thus
conclude that ``it is important to model breaks in Type II galaxies to
derive proper disk scale lengths.''

Nevertheless, it is not always obvious what is the "proper" disk scale
length estimate to use in various scaling relations, in case the galaxy
exhibits several exponential subsections.  For example, it is well
known \citep{pohlen2006} that Type II breaks are often
connected to outer rings associated with bar OLR resonances. Such
breaks are indeed dominant for early type barred disks ($ T < 3$;
\citealt{laine2014}).  Since the bar torques are able to push material
from the CR regions out toward OLR, this will promote a shallower
distribution inside the break radius. However, beyond the OLR the
effect of bar is insignificant, so that the underlying disk can remain
more or less intact. In such a case it might in fact be the outer,
rather than the inner scalelength that would better characterize the
original overall mass distribution. On the other hand, for later
Hubble types the Type II break is often connected with the end of
prominent spirals \citep{laine2014} and could be due to suppressed
star formation: for such a case the inner scale length might indeed be
more appropriate to characterize the disk as a whole.  \cite{laine2014}
also find that for such spiral-related breaks the ratio
$h_{inner}/h_{outer}$ is typically closer to unity than for OLR
related breaks.

Figure \ref{fig_inner_outer_fig1} compares Pipeline 4 decompositions
with several recent disk truncation studies, which use subsamples of
the same S$^4$G dataset. In this plot the disk scalelengths are
displayed against the stellar mass derived in P3
\citep{munoz_mateos2014}. Besides the above-mentioned \cite{kim2014}
2D decomposition study, we also compare with \cite{munoz_mateos2013}
and \cite{laine2014}, where fits to 1-dimensional profiles were
conducted.  First of all, the Figure (upper row) indicates a very good
agreement for the scale lengths of Type I profiles between all four
studies, conducted with independent methods. Secondly, it illustrates
the significant difference between the inner and outer slopes for Type
II (and III) profiles, amounting to roughly a factor of two (see
\citealt{munoz_mateos2013}, \citealt{kim2014}) .  The P4 single disk
scalelengths seem to fall quite close to being a geometric mean of
$h_{inner}$ and $h_{outer}$ derived in earlier studies.


To emphasize the possible 'unperturbed' nature of Type II outer disks,
we compare in Fig. \ref{fig_inner_outer_fig2} the P4 scale lengths
vs. stellar mass for Type I galaxies with the type II outer scale
lengths derived in the above mentioned disk break studies.  Indeed the
differences between the Type I single disk $h$ and the Type II
$h_{outer}$ are quite small, much smaller than the differences
compared to $h_{inner}$.  The fits to the data also give the
impression that $h_{inner}/h_{outer}$ ratio gets closer to unity for
less massive galaxies: this is in accordance with the above mentioned
dominance of spiral-related less abrupt truncations for later and thus
on average less massive spirals.

The Pipeline 4 single exponential fits have a convenient feature of
representing an effective average over inner and outer disks (when
both present). They thus provide a homogeneous set of robust scale
measurements, not sensitive to factors modifying the local slopes.
Nevertheless, a possible caveat is that the fitted effective single
$h$ might become dominated by different degrees by the inner/outer
parts, depending on the galaxy surface brightness. For example, the
estimated $h$ might be biased toward $h_{inner}$ when the disk central
surface brightness decreases toward less massive galaxies: this would
be the case if the image depth was not sufficient to cover the galaxy
regions beyond the break radius. Fig. \ref{fig_inner_outer_fig7}
addresses this potential problem by comparing the trends of the break
radii with respect to galaxy mass, to that of the galaxies' visual
outer extent ($R_{gal}$, see Sect \ref{sect_sky}; a similar trend
would result if $R_{25.5}$ were plotted instead of $R_{gal}$). The
figure indicates that a break, if present, should be detectable
through the whole range of S$^4$G galaxy masses.

In summary, we feel confident that the single disk fits provide a
useful overall estimate of the disk original scale length (and its
extrapolated surface brightness), though especially in case of barred
massive galaxies secular evolution might have led to significant
deviations from simple exponentials, important to include in detailed
models for individual galaxies. Moreover it is likely that the
slope differences associated with breaks are smaller
for later types, which form a vast majority of S$^4$G galaxies.

Nevertheless, as concluded by \cite{kim2014}, estimates of other
decomposition parameters, such as the B/T ratio for massive galaxies
would become more accurate if the inner slopes are accounted for (say,
leading to less disk light assigned to bulge).  The situation is
somewhat analogous to the benefit of including additional inner
components like bars \citep{laurikainen2005, gadotti2008}, lenses in
S0s \citep{laurikainen2010}, or barlenses \citep{laurikainen2014} into
decompositions.  However, for the goals of Pipeline decompositions,
the expected magnitude of changes (about 10\% relative change in B/T
according to Kim et al.) is quite small, compared to the uncertainties
related to choice of the decomposition model components (say,
including a bar versus ignoring it). The choice of the code might also
sometimes have a bigger effect. For example, \cite{kim2014} use NGC
936 as an example of Type II galaxy (see their Fig. 4). For this
galaxy they fit a break at $98\arcsec$ and derive $h_{inner}=
53\arcsec$ and $h_{outer}=28\arcsec$, all very close to the
measurements in both \cite{munoz_mateos2014} and \cite{laine2014}. On
the other hand, the Pipeline 4 single disk fit (see Fig. \ref
{ngc0936_eija}) gives $h=40\arcsec$. We verified that truncating the
disk in GALFIT decompositions at the break radius given in Kim et al.,
reproduces their inner slope quite well (we get $57\arcsec$). At the
same time, the $B/T$ we obtain increases slightly (from 0.19 to 0.22),
as anticipated by Kim et al.\footnote{We also checked the effect of
  letting the boxiness and shape parameters of the bar free but these
  turn out to be small}. Nevertheless, the $B/T$ we obtain after
accounting for the more shallow inner slope is still nearly $50\%$
smaller than the value obtained by Kim et al. ($B/T$=0.32), probably
because of some model/code dependent factors, such as how the image
pixels are weighted, or the PSF is treated.


\begin{figure}
\includegraphics[angle=0,scale=.65]{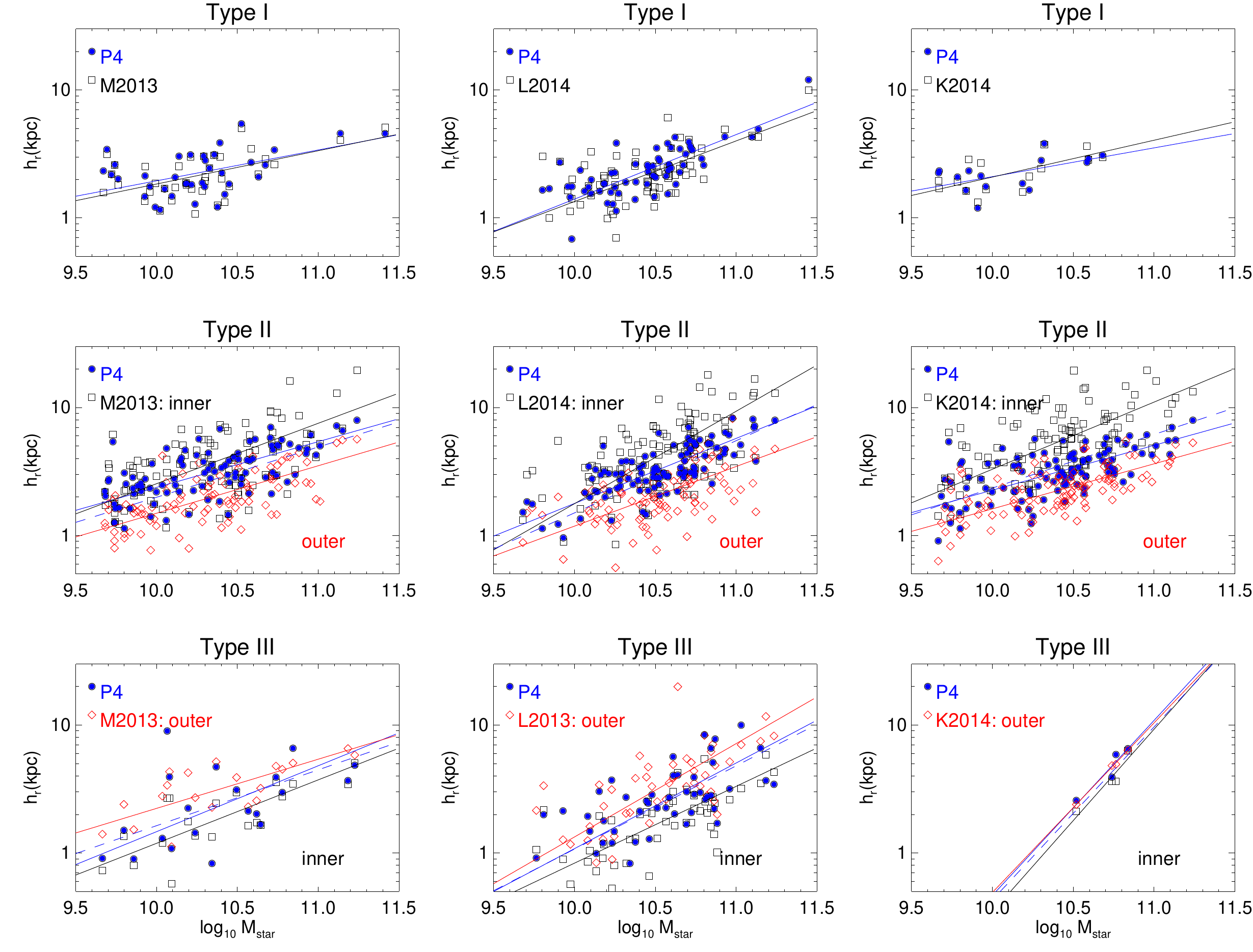}
\caption{Comparison of P4 single disk scalelengths with decomposition
  studies including disk truncations. In the left frames the $h_r$
  vs. stellar mass, obtained in P4 (blue symbols) are compared with
  the results in \citealt{munoz_mateos2013} (M2013) and
  \citealt{laine2014} (L2014), where the inner and outer scalelengths
  ($h_{inner}$ and $h_{outer}$, denoted with black and red symbols,
  respectively) were estimated for S$^4$G galaxies from fits to
  1-dimensional profiles.  On the right, similar comparison to \citealt{kim2014} (K2014), who used 2D BUDDA decompositions for 144 barred
  S$^4$G galaxies. In the uppermost frames Type I disks (no breaks)
  are compared: the lines show orthonormal fits to the measurements
  (orthogonal deviations minimized using the IDL PCOMP routine).  In
  the middle same for Type II profiles: the dashed blue line indicates
  a fit to geometric means of the inner and outer scalelengths. In the
  lowermost frame: galaxies with Type III breaks.}
\label{fig_inner_outer_fig1}
\end{figure} 
\clearpage

\begin{figure}
\includegraphics[angle=0,scale=.85]{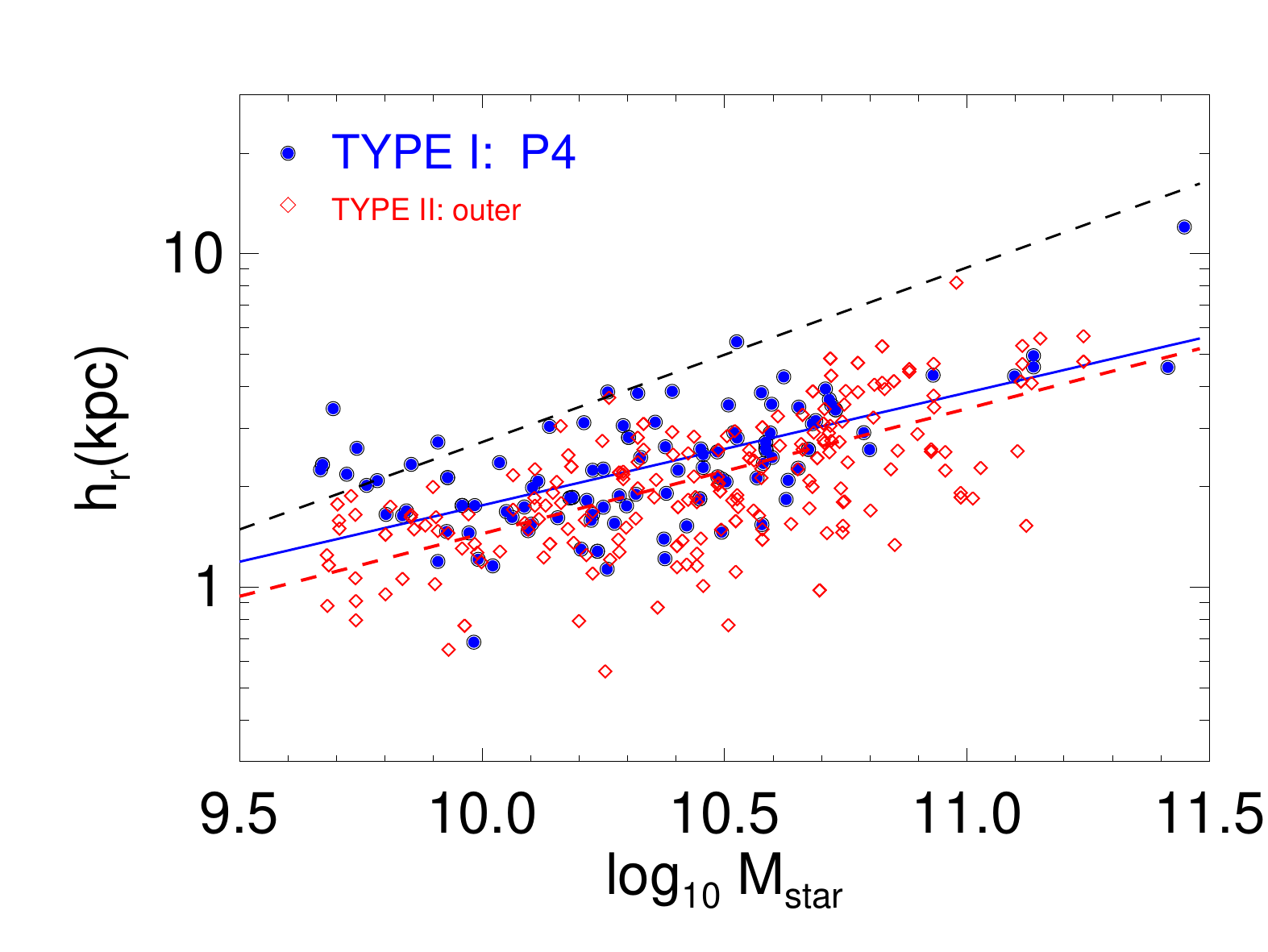}
\caption{ Comparison of P4 scalelengths for Type I galaxies (blue
  points), with the outer scalelengths of Type II profiles (red points;
  these combining \citealt{munoz_mateos2013}, Laine et al. (2014) and Kim et
  al. (2014) measurements). Also shown are the corresponding
  orthonormal fits (solid blue and dashed red lines, respectively). The black
  dashed line shows a fit to
  inner scalelengths derived in the above mentioned studies
  (individual measurement points not shown). Note that the trend of single
  scalelengths in Type I galaxies resemble much more the outer
  scalelengths in Type II's rather than the inner scalelengths.}
\label{fig_inner_outer_fig2}
\end{figure} 
\clearpage

\begin{figure}
\includegraphics[angle=0,scale=.95]{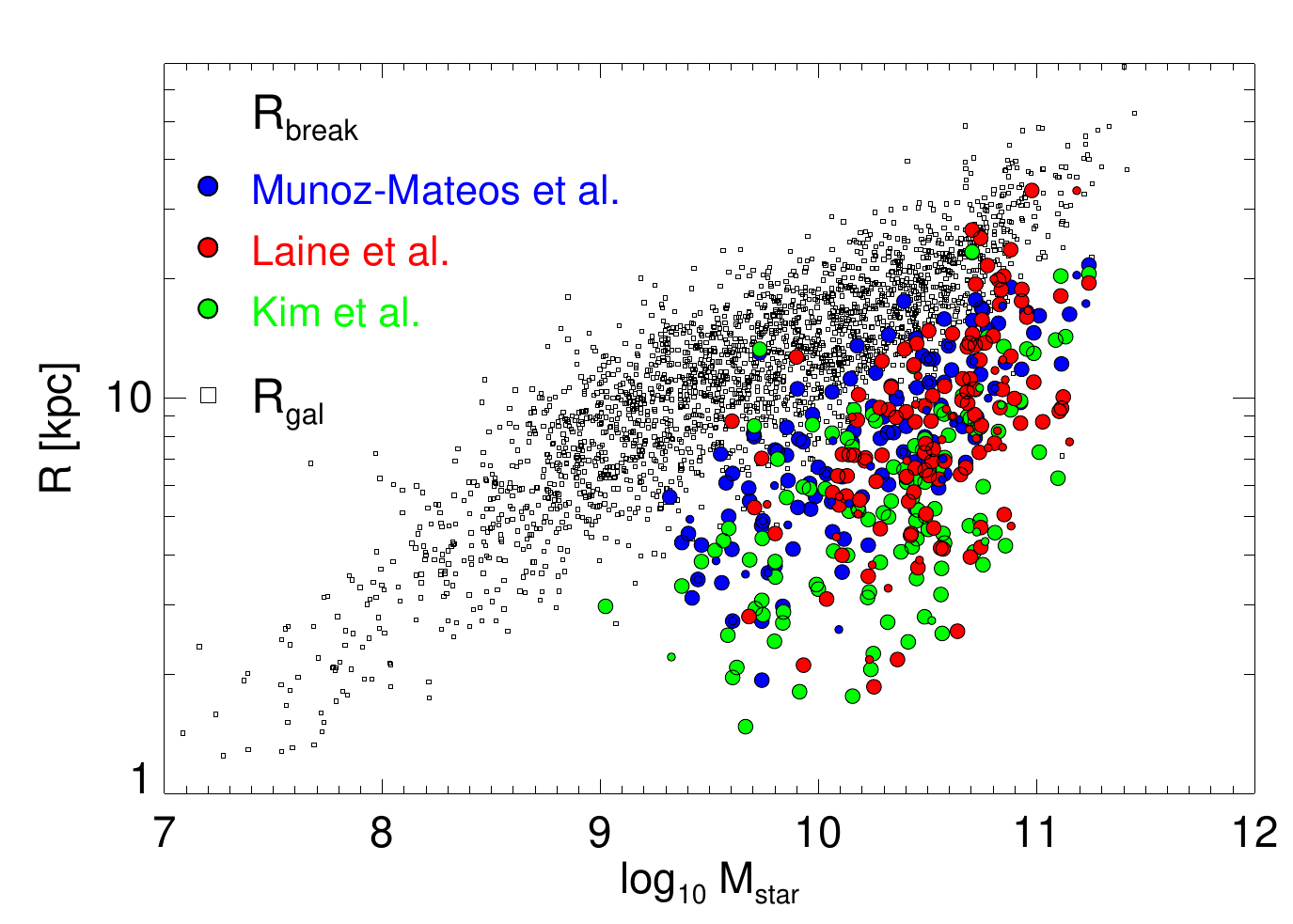}
\caption{Comparison of the break radii between the inner and outer
  disk segments in the studies of \cite{munoz_mateos2013}, \cite{laine2014}
and  \cite{kim2014}; large and small circles refer to
  Type II and Type III breaks, respectively. Note that the break
  radius drops rapidly with galaxy stellar mass. Also shown by small
  squares is $R_{gal}$, the P4 visual estimate of the galaxy extent in
  S$^4$G images.  The plot suggest that even in the case of low-mass,
  low surface brightness galaxies the depth of the S$^4$G images is
  sufficient to assure that the break, if present, is not likely to be
  buried in the sky background.}
\label{fig_inner_outer_fig7}
\end{figure} 
\clearpage

\section{Decomposition parameters}

In the current paper (paper 1) we provide all the 1-component and final
multi-component decomposition parameters in a tabular format (Tables 6
and 7), together with quality assignment flags.  A brief check of how
the major categories of the final models distribute among different
Hubble types is also shown. All actual analysis will be presented in
paper 2.

\subsection{Quality assessment}
\label{sect_quality}

The full S$^4$G sample contains 2352 galaxies, chosen according to
their internal extinction corrected B-magnitude ($M_{Bcorr}<15.5$),
apparent B-band 25-mag isophotal diameter ($D_{25}>60\arcsec$),
galactic latitude ($|b| > 30^\circ$), and HI recession velocity
($V_{radio} < 3000 km/s$), obtained from the HyperLeda database.
Due to its mag-limited character, it contains a
large number of low surface brightness late-type spirals and
irregulars.  Also, galaxies with peculiar morphology were not
specifically excluded.  In some cases the field-of-view (FOV) is not
large compared to the galaxy size (the new Spitzer observation mapped
regions covering at least 1.5 $D_{25}$, but this condition was not
fulfilled by all the archival galaxies included in the sample).  In
such cases the sky background is difficult to estimate reliably, and
for galaxies near the ecliptic, the background may have larger
gradients (see \citealt{munoz_mateos2014}).  Altogether, the sample contains a
number of galaxies for which decompositions are less reliable, or not
possible at all to carry out.

Because it is important to estimate the reliability of the derived
structural parameters, we have assigned to each galaxy a {\bf quality
  flag}, running from 1 (worst case) to 5 (most reliable). The
judgment was done partly by visual inspection of the original data and
partly by evaluating the decomposition models:


{\parskip 0.1cm
\vskip 0.5cm
\noindent $\bullet$ {\bf \em Quality=1}   \ (31 cases)

\noindent {\em Reasons:} Bad original data (very bright overlapping
stars, strongly varying background, image defects).

\noindent {\em Action:} Excluded from all analysis: galaxy
identifications are listed in parameter tables but no parameter
values are given. P4 web page illustrates the raw data + mask, but not any decomposition models. 

\vskip 0.25cm

\noindent $\bullet$ {\bf \em Quality=2}   \ (44 cases)

\noindent {\em Reasons:} Original data is more or less fine, but the FOV
is too small for reliable sky estimation.  Alternatively,
galaxies exhibit strongly distorted shapes which make even 1-component
fits unreliable (mergers, interacting, peculiar, strong warp, very
lopsided).

\noindent {\em Action:} 1-component S\'ersic fit was done and the parameters
are given in Table \ref{table_1comp} (and  in the web pages), together with a
comment indicating that they need to be taken with caution.
Multi-component decomposition parameters are not given.

\vskip 0.25cm

\noindent $\bullet$ {\bf \em Quality=3}   \ (61 cases)

\noindent {\em Reason:} Original data is fine but the galaxies have
complex structures that require detailed multi-component models
beyond the scope of the pipeline decompositions (which have a  maximum of
4 components).

\noindent {\em Action:} 1-component S\'ersic fit is considered reliable
(Table \ref{table_1comp}) . Multi-component decomposition was also made and
parameters are listed in Table \ref{table_final}, with a cautionary comment. Web
pages show both 1-component and multi-component decompositions.

\vskip 0.25cm

\noindent $\bullet$ {\bf \em Quality=4}   \ (406 cases)

\noindent {\em Reason:} 
Original data and decomposition are of good quality. 
However, the galaxy was either highly inclined (contained a 'z' component; 333 cases), 
or it had complicated structure, so that there might be a degeneracy between model components 
(such as between inner and outer disk components) (73 cases).

\noindent {\em Action:} All decomposition parameters given in Tables
\ref{table_1comp} and \ref{table_final}, and illustrated in the P4
web-pages. However, these are omitted from the analysis of disk
central brightness and scale length in paper 2.
\vskip 0.25cm
\noindent $\bullet$ {\bf \em Quality=5}   \ (1810 cases)

\noindent {\em Reason:} 
Both the original data and the decompositions are of good quality.

\noindent {\em Action:} 
All decomposition parameters are given in Tables \ref{table_1comp} and  \ref{table_final}, and illustrated
in the P4 web-pages. 

\vskip 0.25cm
Table \ref{table:quality} summarizes the number of galaxies in different quality categories.

\subsection{One-component fits}

The output parameters of 1-component S\'ersic fits are listed in Table
\ref{table_1comp}.  For 1-component fits the parameters are the
S\'ersic index $n$, effective radius $R_e$, integrated magnitude
$mag$, axial ratio $q$, and major axis position angle $PA$ (the
centers are fixed to those given in Table I; the isophotes are assumed
to be elliptical and to have fixed a shape and orientation with
radius). Additionally, there is a column indicating the reliability of
the fit.


Single-component S\'ersic-fits are routinely used in large data
surveys. This gives objective, easily reproducible results,
that  provide useful characterization of the galaxy global characteristics.
For example, \cite{cappellari2013} argued that
S\'ersic n$>$4 largely finds the most massive early-type galaxies
($ M  >  3\cdot 10^{11}M_\sun$), which are also the slow rotators in their
kinematic classification. Also, there are well-known correlations
between galaxy color and S\'ersic index.  Therefore, in paper 2 we
will present detailed analysis of the 1-component S\'ersic parameters for
the S$^4$G sample.  Here we report just the dependence of $n$ on the
morphological type.

Figure \ref{fig:sersic_histo} displays the histogram of S\'ersic
index-values in the 1-component models. Galaxies are divided into
three bins according to their mid-IR Hubble type (E with $T \le -4$,
S0 with $-3 \le T \le 0$, spiral with $1 \le T \le 9$, and irregulars
$T=10$).  Clearly, the distribution of galaxies peaks at $n \approx
1-2$, with a broad tail to larger values of $n$. There is also a
secondary peak close to $n\approx 4$ (corresponds to a de Vaucouleurs
profile), but this is not very prominent.  The overall distribution
reflects the nearly exponential profiles of many late type spirals,
which dominate the $S^4G$ sample. For irregulars, the distribution
peaks at $n\approx1$.  For S0's, the distribution is much broader.
The distributions remain essentially the same for less inclined
galaxies, for example if the sample is limited to those with apparent $b/a > 0.5$.

\subsection{Multi-component decompositions}
\label{sect_final}

For multi-component decompositions the tabulation is more complicated,
as model components/functions vary from one galaxy to another.  Also,
the same function can be used to describe different structure components in
different galaxies.  We have decided to present the multi-component
parameters in two different formats, one that is compact and easily
human-readable, another more suited to automatic reading.  In the first
format (Table \ref{table_final}) the first entry for each galaxy
indicates the used model and the number of components. The next lines,
for each component included in the model, indicate the physical
interpretation of the component ({\em B, D (or Z), bar, N}), and the
GALFIT function used ({\em sersic, \ expdisk, edgedisk, ferrer2,
  psf}), followed by the component parameters. The Table caption
specifies which parameters are listed for each function.
The other table (available via IRSA and P4 web page) lists for each galaxy
all possible components and their parameters: empty values indicate that this
component was not included in the decomposition of this galaxy.

{\reply The $S^4G$ sample contains 358 galaxies which} were considered to be close to an edge-on view and are
excluded from further analysis in paper 2.  Also, 26 are {\reply elliptical}
systems, modeled with a single S\'ersic function. This leaves 1855
moderately inclined disk systems.  As discussed in Section
\ref{sect_eija_examples} there are over 20 different combinations of
functions/components used in the final models, so that there is a need
to group the decompositions in to major categories.  A natural
approach is to base this grouping on whether the decomposition model
has a bulge component. Because the bulge can be modeled either with a
``sersic'' or ``psf'' component, depending on its apparent size, we
have two categories, {\em BD} and {\em ND} models, respectively.  When
there is no trace of a bulge, the system can be either a single disk
({\em D}), possess a bar-like component ({\em Dbar}), or contain both
inner and outer disk components ({\em DD}).

The numbers and relative fractions of galaxies in these categories as
a function of Hubble type are shown in Fig. \ref{fig_model_fractions}.
Here the mid-IR classification from Buta et al. (2014) is followed.
Apparently the relative fraction of {\em BD}-models
decreases gradually towards the late-type spirals.  However, taking
into account that most of the {\em ND} models describe small bulges
(support for this claim is given in paper 2), indicates a much smoother
distribution of galaxies with bulges, dropping rapidly only above $T
\sim 5$.
The fact that {\em ND} models cover a relatively large range of Hubble
types, including S0s, is quite interesting because it indicates
that S0s can possess very small bulges. This is in agreement with
\cite{laurikainen2010}, where the same conclusion was made based on
decompositions of NIRS0S data. The result is consistent with the idea
that at least some S0s might be former late-type spirals with
small bulges, devoid of gas, following quenching of star
formation. A very small $B/T$ ratio in an early-type spiral has also 
been
reported also by \cite{kormendy2010}. Another interesting
feature in Fig \ref{fig_model_fractions} is that many galaxies that lack
bulges, can still have bars. These {\em Dbar} galaxies peak at Hubble
types $T=7$.  Beyond $T=9$ they are replaced with single disks,
becoming almost the sole type of models for irregulars (T=10).  The
{\em DD}-models are most common (about 10\%) for $T=9$.

Finally, Fig. \ref{fig_montage_bd_nonbar}} gives examples of galaxies
in these major categories, indicating the model components for 4 low mass
and 4 large mass systems in each category. For {\em BD} and
{\em ND} categories the barred and non-barred models are also
distinguished.  Note that 'barred'/'non-barred' refers here to whether
or not a bar-component was included to the the final decomposition
model, not to any morphological classification; for example a
non-barred BD model has been adopted for NGC 5985, which has a SAB
family classification (see Fig. 33).  A detailed comparison to \cite{buta2014}
 classification will be presented in paper 2.

\begin{deluxetable}{ccccccccc}
\tablewidth{-20pt} 
\tabletypesize{\scriptsize} 
\tablecaption {Summary of decomposition quality flags}
\tablehead{ 
\colhead{Quality}  & \colhead{\#} & \colhead{\#($\ge $quality)}   &\colhead{1-component} & \colhead{Multi-component}     & \colhead{Disk $\mu_0$ and $h_r$}}
\startdata
1  &   31   &      &     -         &     -        &    - \cr
2  &   44   & 2321 &  uncertain    &     -        &    - \cr
3  &   61   & 2277 &     ok        &  uncertain   &    - \cr
4  &  406   & 2216 &     ok        &    ok        &    uncertain (or z) \cr
5  & 1810   & 1810 &     ok        &    ok        &    ok                 \cr
\enddata 
\label{table:quality}
\tablecomments{Quality flags (1-5) assess the reliability of
  decomposition parameters.
The second column indicates the number of
  galaxies in each class, while the third column indicates the
  number of galaxies with decomposition parameters of this or better
  quality.}
\end{deluxetable}


\begin{deluxetable}{ccccccc}
\tablewidth{-20pt}
\tabletypesize{\scriptsize}
\tablecaption {Parameters of 1-component S\'ersic fits}
\tablehead{\colhead{Identification}  & \colhead{Comment} & \colhead{mag}   &\colhead{q} & \colhead{PA}     & \colhead{$n$}  & \colhead{$R_e$}}
\startdata
ESO011-005 &   &   14.52 &   0.231 &   43.46 &   1.329 &   18.61 \cr
ESO012-010 &   &   13.31 &   0.493 &  156.97 &   1.592 &   56.38 \cr
ESO012-014 &   &   14.67 &   0.411 &   23.92 &   0.884 &   47.35 \cr
ESO013-016 &   &   12.75 &   0.508 &  168.64 &   1.314 &   38.85 \cr
ESO015-001 &   &   14.39 &   0.357 &  110.22 &   1.131 &   34.34 \cr
ESO026-001 &   &   12.52 &   0.689 &   58.26 &   3.126 &   62.93 \cr
ESO027-001 &   &   10.75 &   0.765 &   74.38 &   5.423 &   96.07 \cr
...\cr
UGC12856 &   &   14.04 &   0.316 &   18.48 &   1.339 &   41.22 \cr
UGC12857 &   &   12.99 &   0.195 &   33.50 &   1.316 &   19.09 \cr
UGC12893 &   &   13.58 &   0.861 &   94.74 &   1.179 &   33.55 \cr
\enddata
\label{table_1comp}
\tablecomments{{\em Comment} indicates cases where no fit was made, fit failed to converge, or its parameters are not reliable.
$mag$ is the the total magnitude, $q$ is the axial ratio and $PA$ the position angle of elliptical isophotes, $n$ is the S\'ersic index and
$R_e$ the effective radius (in arcsec).}
\end{deluxetable}
\clearpage


\begin{figure} [ht]
\includegraphics[angle=0,scale=.95]{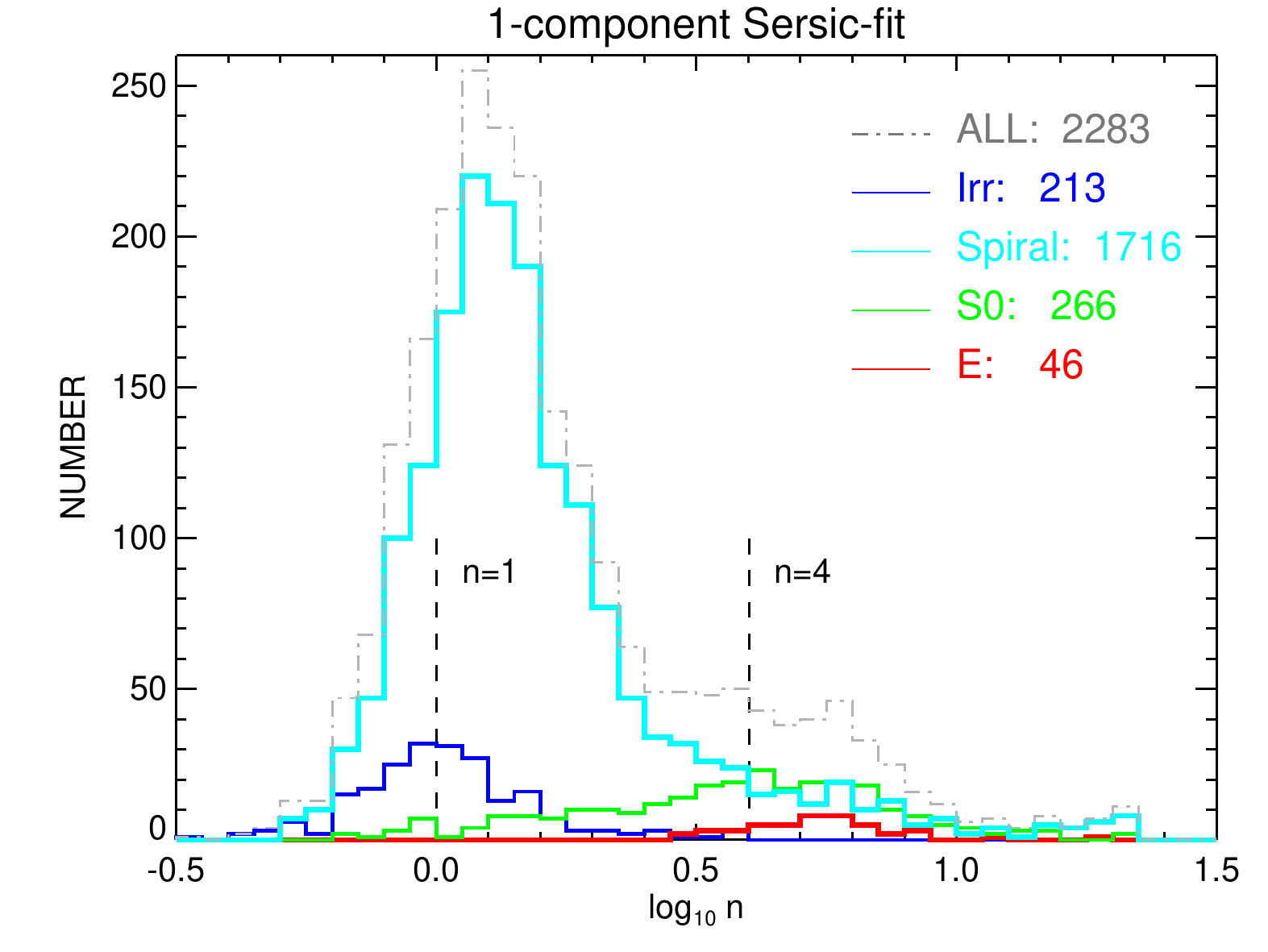}
\caption{Histogram of S\'ersic-index $n$ in 1-component S\'ersic-fits.
  Dashed line indicates the whole S$^4$G sample, after elimination of
  61 galaxies due to bad image quality, bright nearby star
  etc. Additionally, for 8 galaxies the decomposition failed to
  converge ($n=20$ in the plot).  The main morphological types are
  shown separately with different colors: Ellipticals ( $T < -3$), S0s
  ($ -3 \le T \le 0$), Spirals ($0 < T < 10$), and Irregulars ($T =
  10$). Additionally there are 8 galaxies without classification
  ($T=99$; the median value of their $n=2.3$), and 34 dwarf galaxies
  ($T=11$; median $n=1.3$) which are not shown.  The Hubble stage $T$
  is from the mid-IR morphological classification by Buta et
  al. (2014).  The vertical dashed lines indicate the values $n=1$ and
  $n=4$, corresponding to exponential and de Vaucouleurs profiles,
  respectively.  }
\label{fig:sersic_histo}
\end{figure} 
\clearpage



\begin{deluxetable}{llllccccccccc}
\tablewidth{-20pt}
\tabletypesize{\scriptsize}
\tablenum{7}
\tablecaption {Parameters of final multicomponent decompositions}
\tablehead{}
\startdata
\#1 & ESO011-005   & \_bz       & NCOMP=2   & quality=4 &  \cr
 & B      & sersic       &      0.571 &     14.822 &        0.354 &     40.447 &       48.937 &      3.001\cr
 & Z      & edgedisk     &      0.429 &     20.912 &       43.902 &      9.764 &        2.955\cr
\hline
\cr
\#2 & ESO012-010   & \_dbar     & NCOMP=2   & quality=5 &  \cr
 & D       & expdisk     &      0.944 &     13.445 &        0.458 &    -33.795 &       33.953 &     23.095\cr
 & BAR        & ferrer2  &      0.056 &     22.177 &        0.390 &     24.245 &       21.141\cr
\hline
\cr
\#3 & ESO012-014   & \_dbar     & NCOMP=2   & quality=4 &  \cr
 & D       & expdisk     &      0.835 &     14.795 &        0.420 &     31.032 &       33.152 &     24.393\cr
 & BAR        & ferrer2  &      0.165 &     24.704 &        0.249 &     11.102 &       83.009\cr
\hline
\cr

...\cr

\hline
\cr
\#2352 & UGC12893     & \_bd       & NCOMP=2   & quality=5 &  \cr
 & B      & sersic       &      0.021 &     17.826 &        0.786 &    102.674 &        6.559 &      0.515\cr
 & D       & expdisk     &      0.979 &     13.617 &        0.869 &     87.208 &       20.150 &     22.134\cr
\hline
\cr
\enddata
\tablecomments{The first row for each galaxy is the running number
  (1-2352). The second row gives the galaxy name, the type of final
  decomposition model (coded to all output file names together with underscore-prefix), the number of
  components in the model, and the quality flag.
  If no final decomposition was made ({\em quality}= 1 or 2) then for this galaxy
  we set type='-' and NCOMP=0. The next NCOMP entries give: (1) the physical
  interpretation of the component (B-bulge, D-disk, Z-edge-on disk, BAR-bar, N-unresolved central component), (2) the GALFIT function used for it,
  and (3) the component's relative fraction of the total model flux. The
  next entries depend on the GALFIT function.  For {\em sersic} they
  are: $mag, q, PA, n, R_e$, for {\em expdisk}: $mag, q, PA, h_r$,
  for {\em edgedisk}: $\mu_0, PA, h_r, h_z$, for {\em ferrer2}: $\mu_0,
  q, PA, R_{bar}$, and for {\em psf}: $mag$. Here $mag$ is the total
  3.6 $\mu$m AB magnitude, $\mu_0$ is the central surface brightness  in mag/arcsec$^2$
  (face-on brightness for {\em expdisk} and {\em edgedisk}, sky brightness for {\em ferrers2}), 
  $R_e,h_r,h_z$ are in arcsecs.
  All decompositions assume a fixed common center for all components and
  elliptical isophotal shape, constant over radius.  If there is an (outer)
  disk, its $q$ and $PA$ are kept fixed to those in Table I.}"
\label{table_final}
\end{deluxetable}

\clearpage

\begin{figure}
\includegraphics[angle=0,scale=.7]{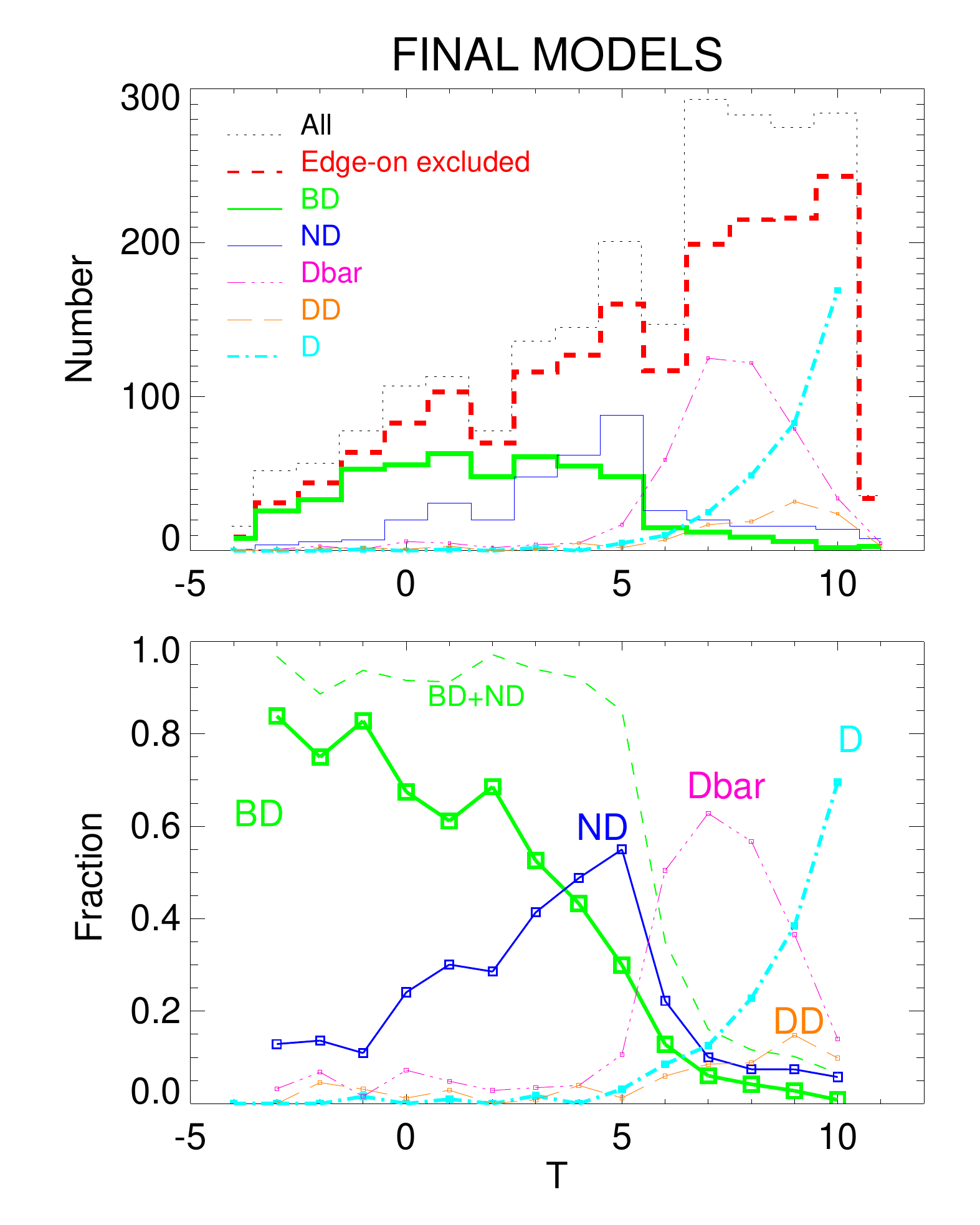}
\caption{
\baselineskip 0.5cm Distribution of final decomposition model
categories as a function of mid-IR Hubble stage from Buta et al (2014;
the half-integer values of T have been rounded before binning, see
caption of Fig. \ref{fig26}).
%
In the upper panel the distribution of the
original S$^4$G sample (dotted histogram), and of galaxies with final
models (red dashed line histogram; excluding the edge-on
galaxies). The green histogram is the distribution of models where {\em
  both} 'bulge' and 'disk' components where identified (BD). This is
also the subsample used in Section \ref{sect:2comp_multi} when
comparing automatic 2-component and final multi-component
decompositions. The dark blue line shows the distributions for
models with {\em both} 'nucleus' and 'disk' (excludes those with
'bulge') (ND). The three other lines are for models with neither
'bulge' nor 'nucleus': 'Dbar' stands for models where a Ferrers-bar
was included (together with one or more 'expdisk' components), 'DD'
stands for models with inner and outer disks, while 'D' stands for a
single 'expdisk' model.  In the lower panel the relative fractions of
different models are plotted, normalized to the total number of non
edge-on models (red dashed curve curve in the upper panel).  Here 'BD'
and 'ND' include both models with/without 'bar': for clarity, these
are not shown separately as the differences in the distributions are
small.
\label{fig_model_fractions}
\label{fig32}}
\end{figure}


\begin{figure}
\includegraphics[angle=0,scale=.85]{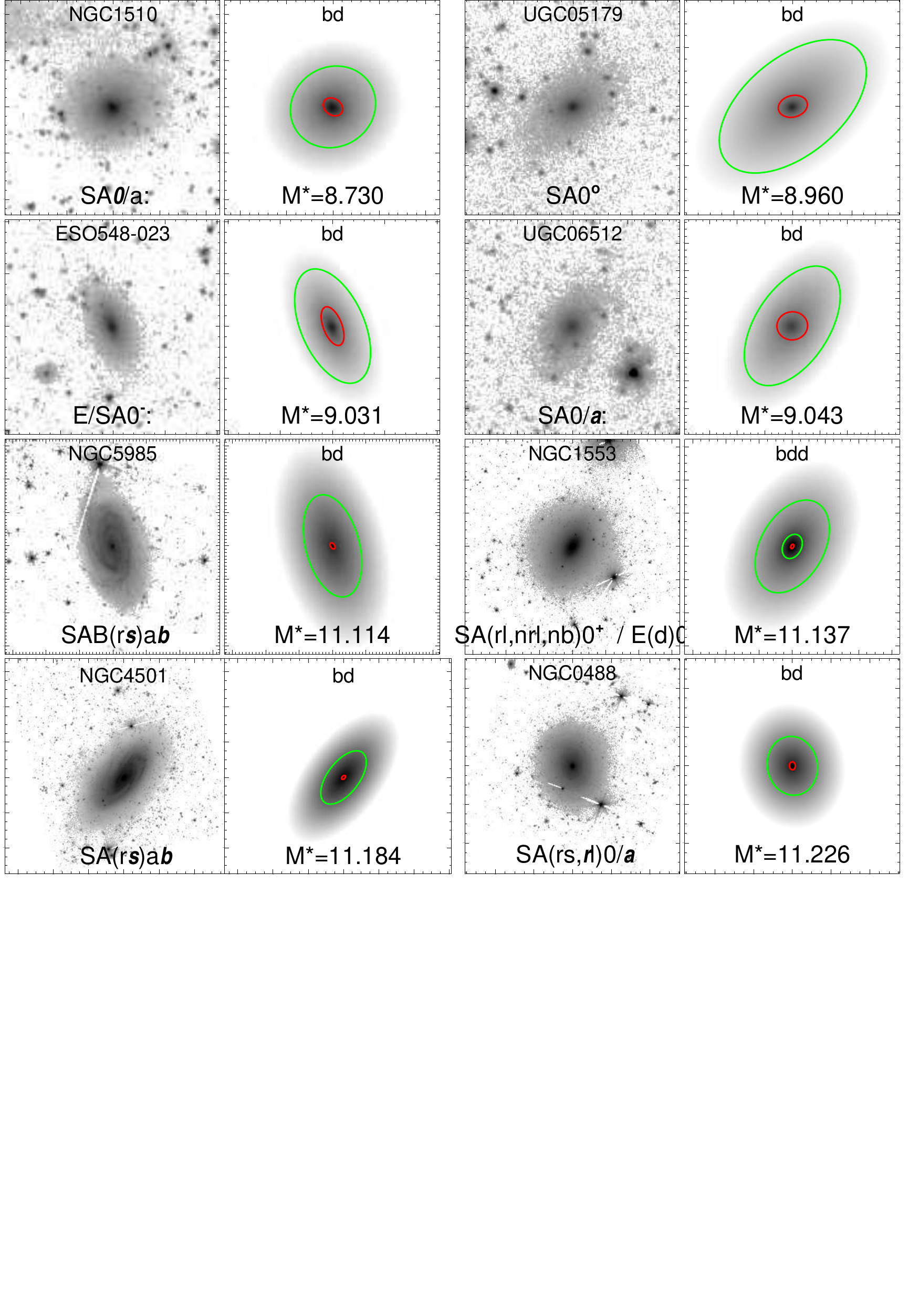}
\vskip -6cm \caption{
\baselineskip 0.55cm a) Examples of different main types of final
decomposition models. Four low mass and four large mass galaxies of
each type are displayed.  The left frames display the 3.6 $\mu$m
image, with fixed AB surface brightness range [18,27], while the right
indicates the model components: the semi-major axis corresponds to
$2R_{\rm eff}$ of the component. The labels in the left frame indicate
the galaxy name and the \cite{buta2014} mid-IR classification; in
the plots the Buta et al.  underline notation is indicated with
slanted characters.  Labels in the right frames give the physical
coding of the decomposition model components (same as used in the
names of the decomposition files), and the $\log_{10} (M_{star})$
(stellar masses are from \citealt{munoz_mateos2014}).  The colors of the
ellipses indicate the used functions: {\em expdisk} (green), {\em
  sersic} (red), {\em ferrer2} (blue). Similar plots for all galaxies
are given in the P4 web-page. In a) examples of {\bf BD} models (have
bulge \& disk but no bar component) are displayed.
\label{fig_montage_bd_nonbar}}
\end{figure} 

\begin{figure}
\figurenum{\ref{fig_montage_bd_nonbar}b}
\includegraphics[angle=0,scale=.85]{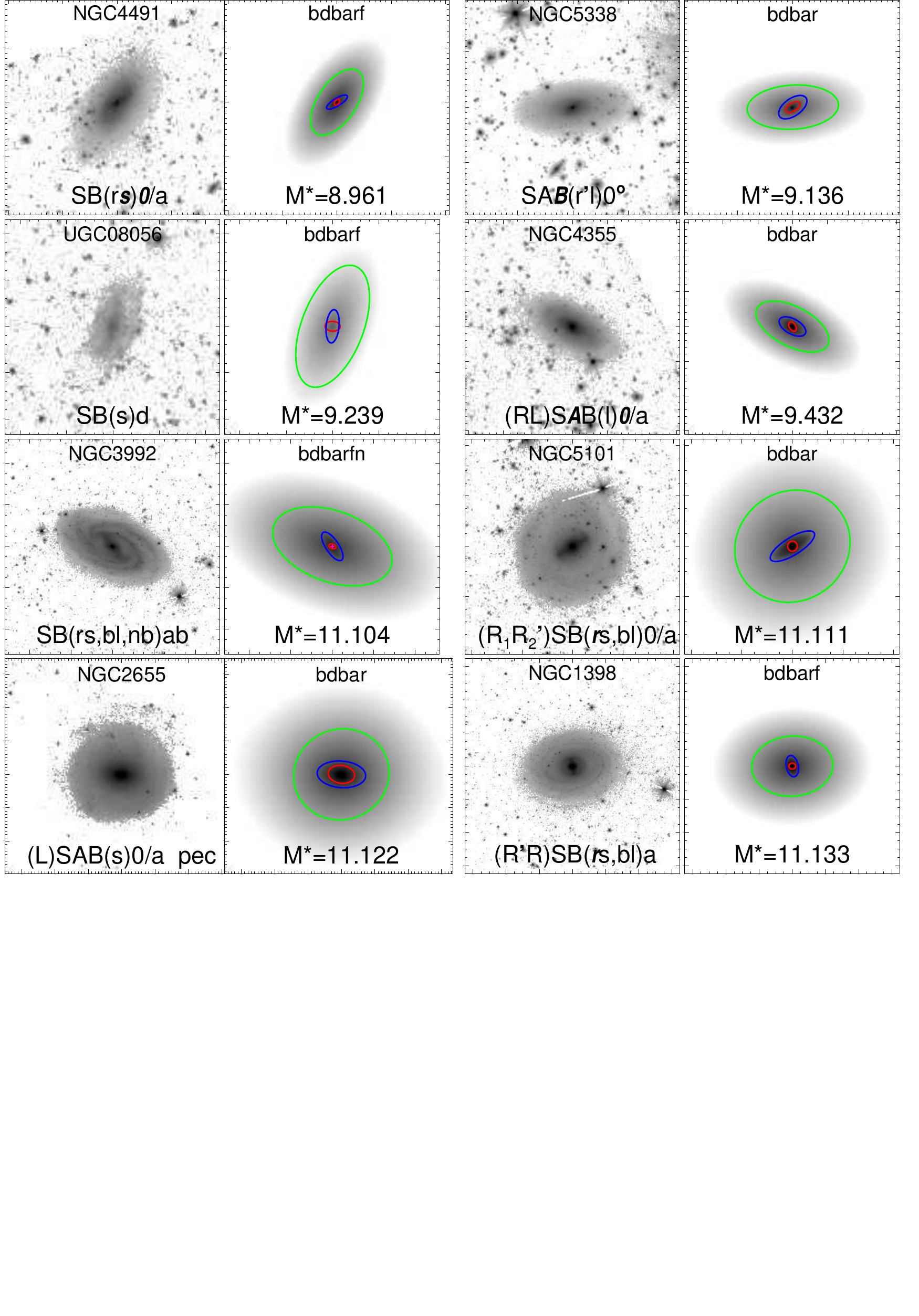}
\vskip -6cm \caption{Examples of {\bf BDbar} decomposition models (bulge \& disk with a bar component). 'barf' indicates that the length of the Ferrers bar was fixed in the decompositions.}
\label{fig_montage_bd_bar}
\end{figure}

\begin{figure}
\figurenum{\ref{fig_montage_bd_nonbar}c}
\includegraphics[angle=0,scale=.85]{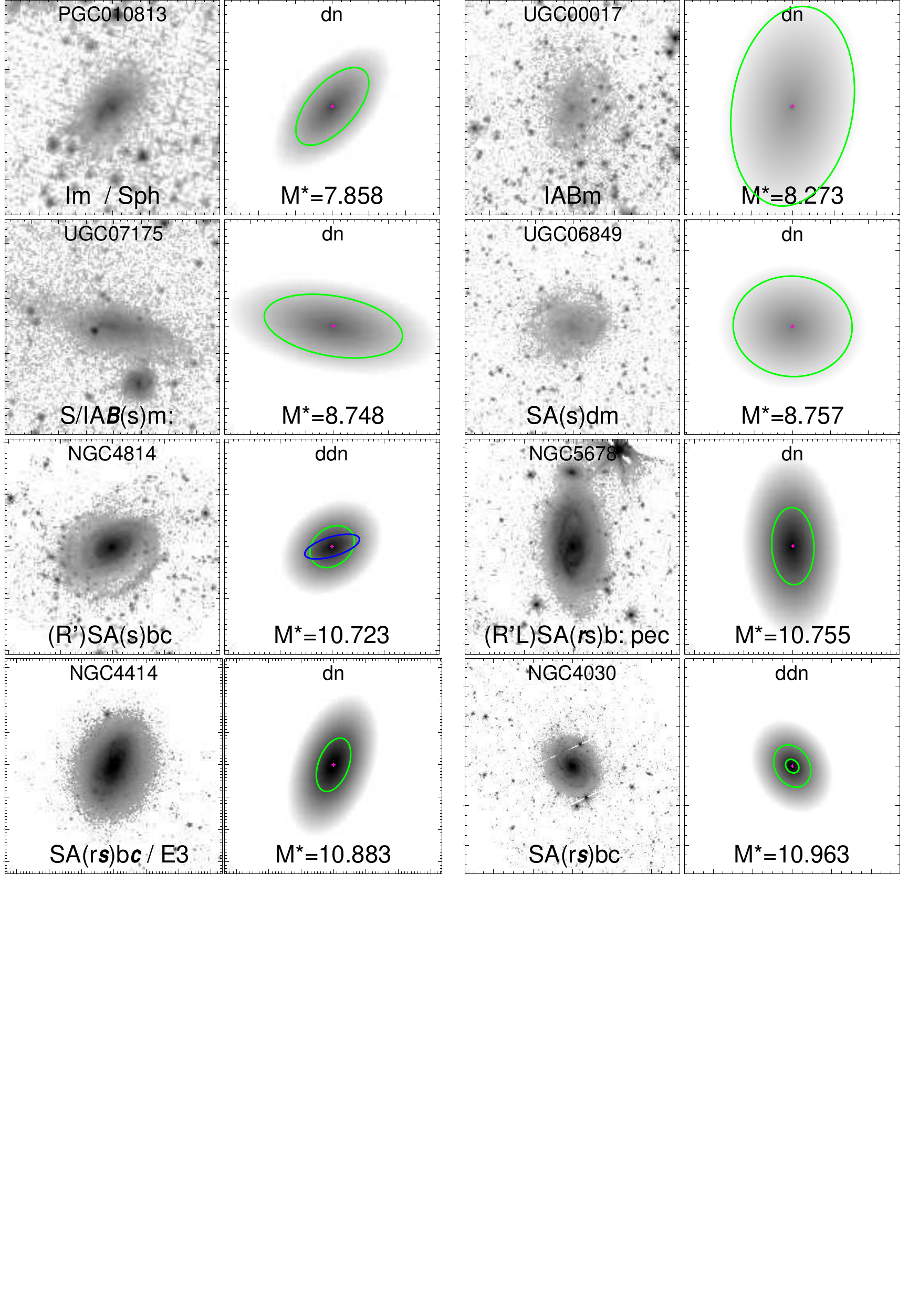}
\vskip -6cm \caption{Examples of {\bf ND} decomposition models (nucleus \& disk, no bar). The central component (unresolved in decomposition) is indicated with a red dot.}
\label{fig_montage_nd_nonbar}
\end{figure} 

\begin{figure}
\figurenum{\ref{fig_montage_bd_nonbar}d}
\includegraphics[angle=0,scale=.85]{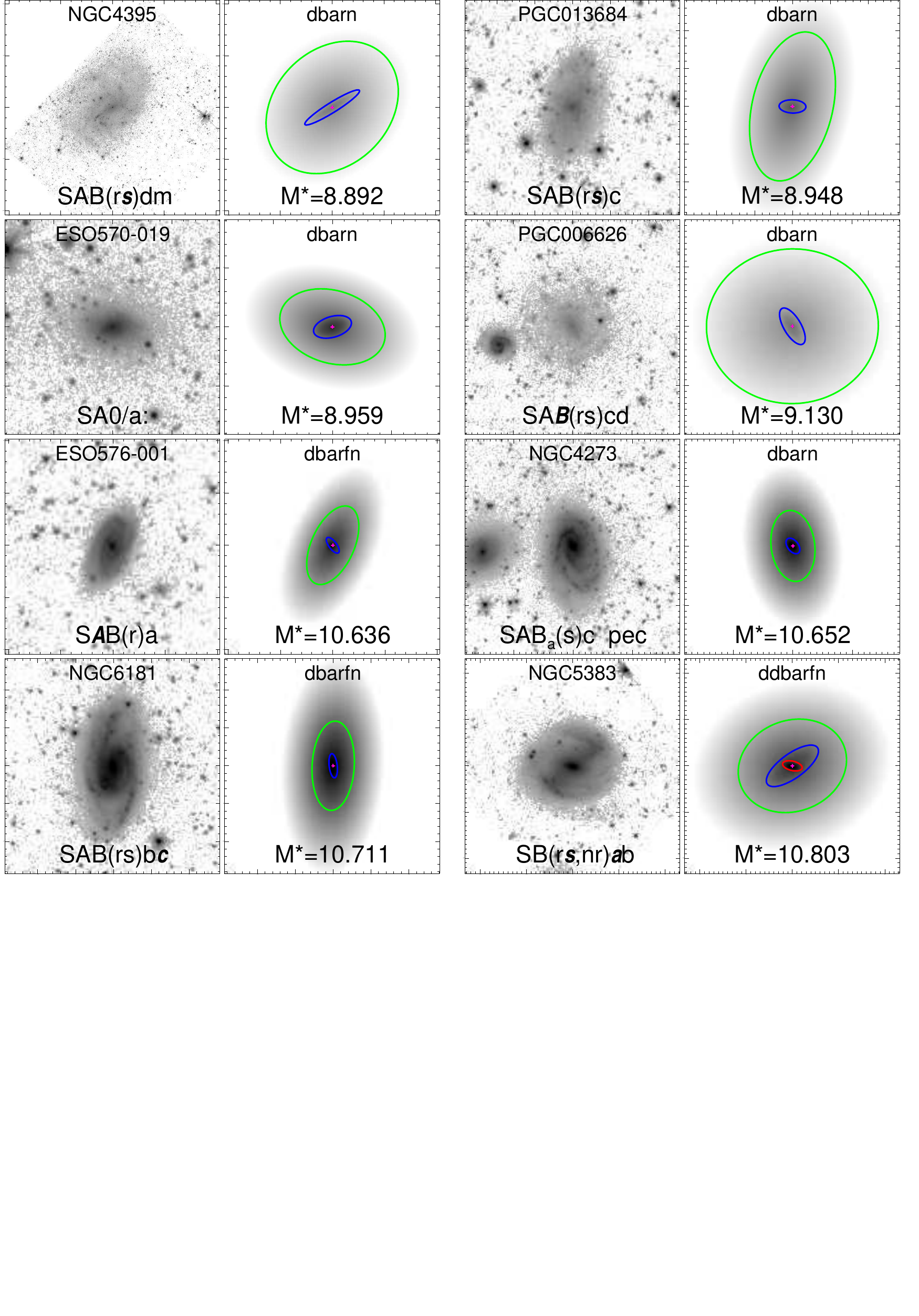}
\vskip -6cm \caption{Examples of {\bf  NDbar} decomposition models (nucleus \& disk, with a bar).}
\label{fig_montage_nd_bar}
\end{figure}

\begin{figure}
\figurenum{\ref{fig_montage_bd_nonbar}e}
\includegraphics[angle=0,scale=.85]{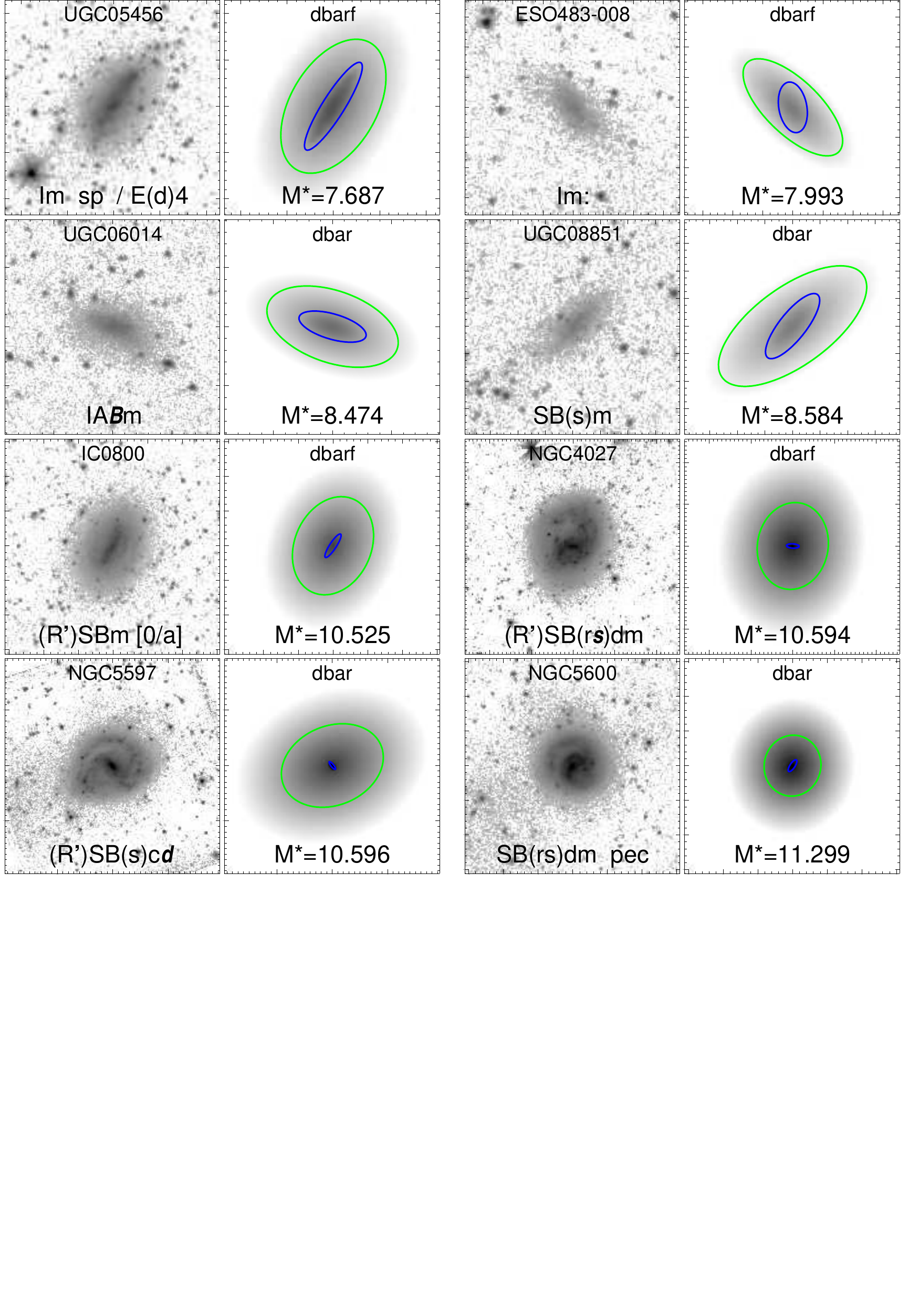}
\vskip -6cm \caption{Examples of {\bf Dbar} decomposition models (disk \& bar).}
\label{fig_montage_dbar}
\end{figure}

\begin{figure}
\figurenum{\ref{fig_montage_bd_nonbar}f}
\includegraphics[angle=0,scale=.85]{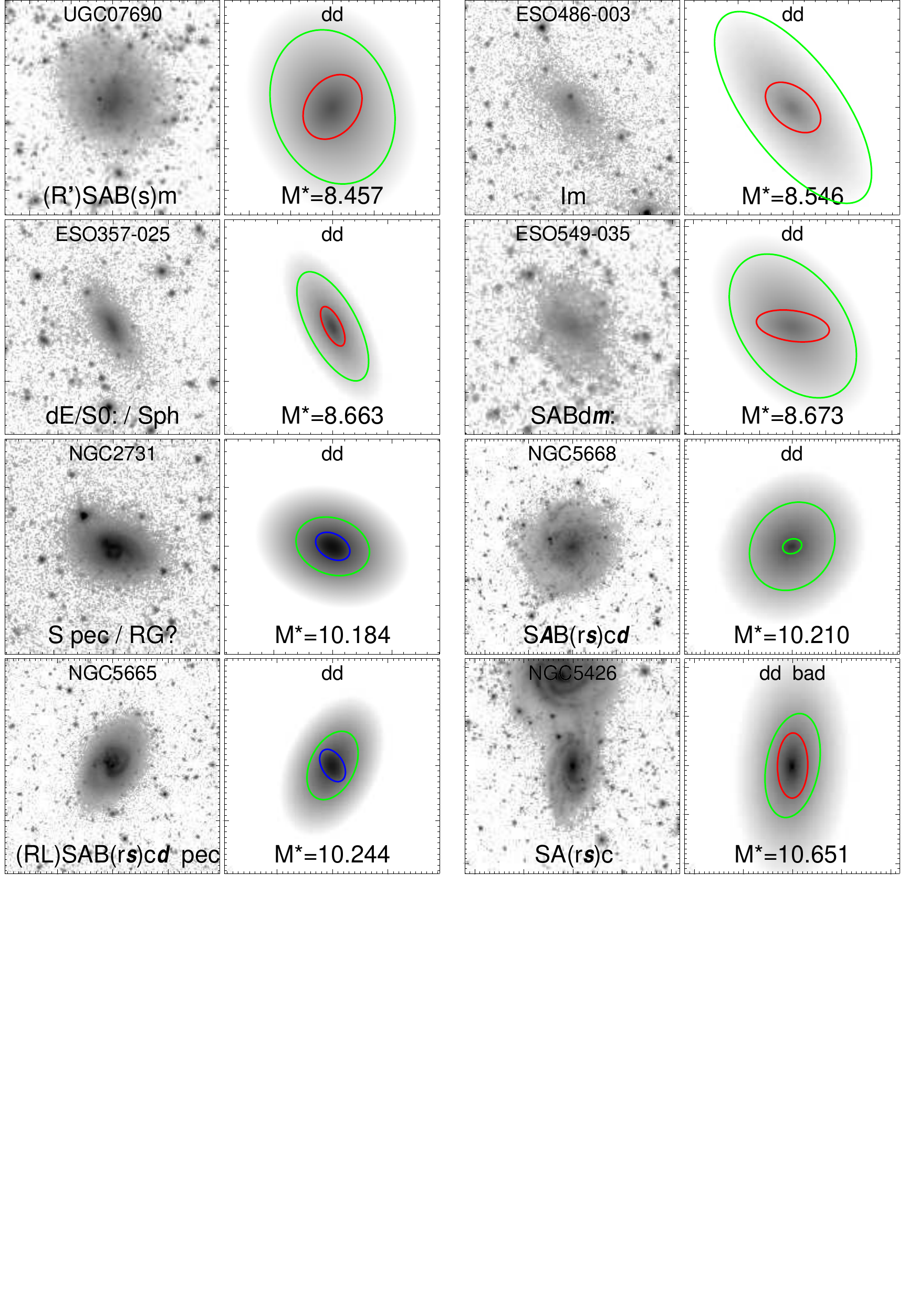}
\vskip -6cm \caption{Examples of {\bf DD} decomposition models (two disk components).}
\label{fig_montage_dd}
\end{figure} 

\begin{figure}
\figurenum{\ref{fig_montage_bd_nonbar}g}
\includegraphics[angle=0,scale=.85]{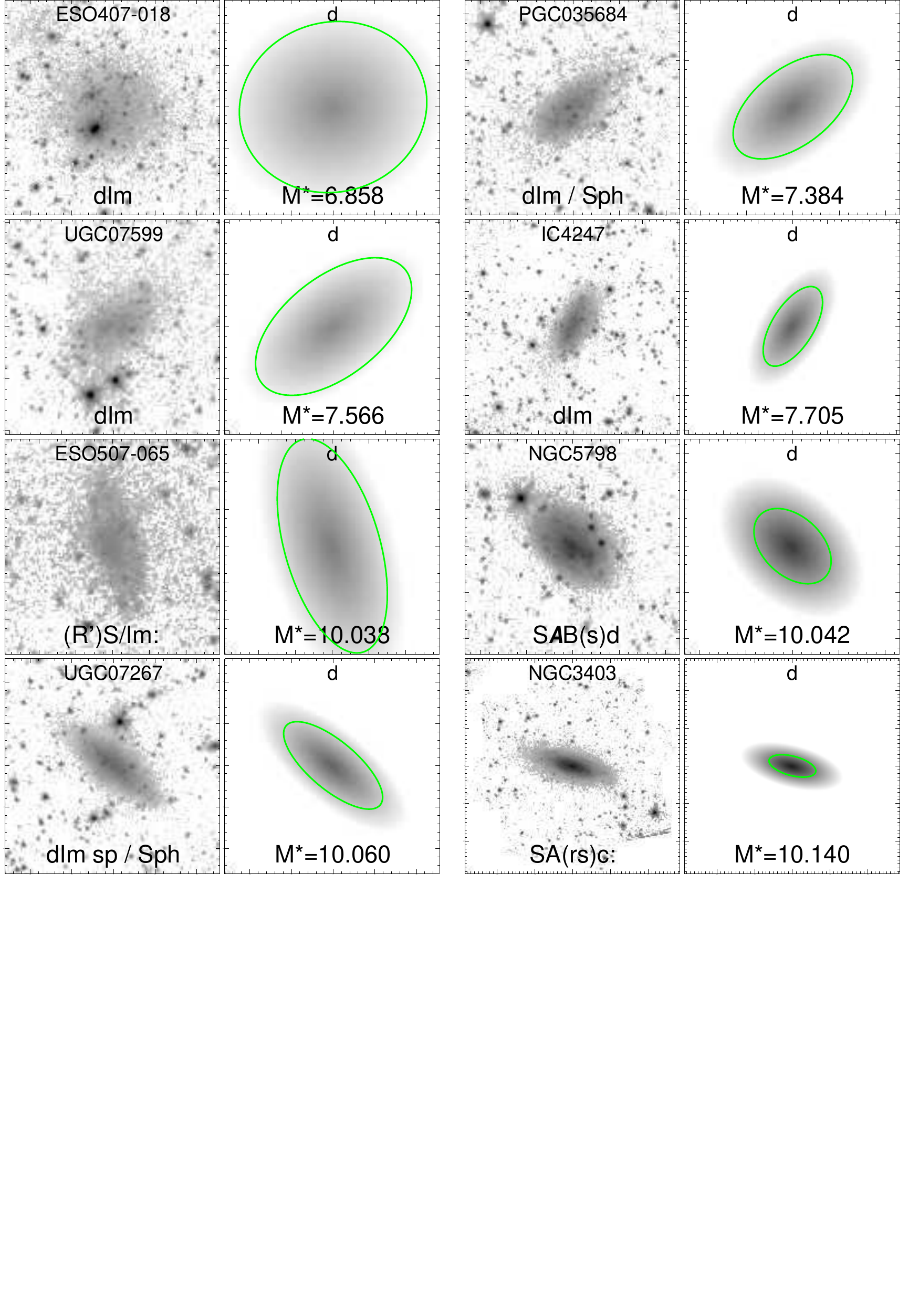}
\vskip -6cm \caption{Examples of {\bf D} decomposition models (single disk).}
\label{fig_montage_d}
\end{figure} 

\clearpage




\clearpage

\clearpage



\clearpage

\section{Summary and Conclusions}

Two-dimensional multi-component decompositions, using GALFIT3.0, have
been performed at 3.6 $\mu$m wavelength for the complete S$^4$G sample
(2352 galaxies). Reliable decompositions were possible for 2277
galaxies. Quality flags are given for each galaxy based on our
confidence on the model parameters.  The main goal of the
decompositions was to estimate the structural parameters of the bulges
and disks in a reliable manner, which dictated our decomposition
strategy.  Most importantly, a bar-component was included in the
decomposition model whenever present, to prevent its light from
biasing the derived bulge and disk parameters.  For the same reason,
the models sometimes included a central point source and additional
disk components.  However, no attempt was made to match the detailed
shape or length of the bar.

We present automated single S\'ersic, 2-component bulge-disk (S\'ersic +
exponential) decompositions, and human-supervised, individually
checked multi-component models. In the final multi-component models, a
maximum of four structural components are fit: bulge (S\'ersic), disk
(exponential), bar (modified Ferrers), and the nucleus
(PSF). Different combinations of component functions were used. For
example, in some barred galaxies it was convenient to fit the
underlying disk with two different functions.  As a first step, we
estimated the sky background levels, derived the orientation
parameters with ellipse fitting, and edited the masks to eliminate
foreground stars and image defects. In general we found an excellent
agreement with the independent P3 measurements \citep{munoz_mateos2014}.

The uncertainties related to the sky background, the adopted
PSF-function, and to the treatment of sigma-images were tested.  The
decomposition data are released in IRSA, and in the P4 web-page, where
the decomposition models, the ellipse fitting, and sky background
determinations are illustrated.  The IDL-based tool (GALFIDL) used in
visualization of GALFIT decompositions is also available on the web
pages.  Besides the decomposition output files, all input files needed
in re-doing the decompositions are given in IRSA.  All of this
provides the possibility to refine the pipeline models for the needs
of specific scientific goals.  

{\reply In particular, such refined models will be needed for
  early-type galaxies which often contain more structures than handled
  by current pipeline decompositions: for ellipticals such structures
  include nuclear point sources \citep{lauer1985, cote2006} and inner
  disk structures (\citealt{kormendy1996}, \citealt{buta2014}; modeled
  in \citealt{kormendy2010}, \citealt{weinzirl2014}); for S0s and
  early-type spirals the various lens structures \citep{kormendy1979,
    laurikainen2009} should be accounted for, including the barlens
  components \citep{laurikainen2011} recently identified as the more
  face-on counterparts of boxy/peanut bulges seen in nearly edge-on
  galaxies \citep{laurikainen2014, atha2014}. Such refined models
  become particularly important with the ongoing extension of S$^4$G data
  to include 465 additional gas-poor early-type galaxies
  \citep{sheth2013}, which were not part of the original sample which
  contained only galaxies with emission line velocity measurements.  }


The main results are the following:

(1) {\it Automatic single S\'ersic fits}{\reply . The} S\'ersic indexes 
peak at $n\sim$1.5, having only a minor peak at $n\sim$4, reflecting
the fact that a large majority of the sample galaxies are spirals with
extended disks.

(2) {\it Automatic 2-component bulge-disk decompositions}: Such
decompositions would suggest a large difference in the parameters of the bulges between
barred and non-barred galaxies. Since this is an artifact caused by the inadequate
decomposition model, we strongly caution against using
simple bulge-disk decompositions for barred galaxies.

(3) Final {\it multi-component decompositions}. In contrast to
2-component models, in our final models the differences in bulge
parameters between barred and non-barred galaxies disappear, leading
to the values of Sersic $n\sim 1-2 $ for bulges in both types of galaxies.
It means that if bars are not included in the fit, the flux of the bar is
erroneously mixed with the bulge flux. This conclusion is consistent
with several previous studies using a similar multi-component
decomposition approach.

(4) Small bulges containing at most a few percent of the galaxy flux at
3.6 $\mu$m appear in a large range of Hubble types, including
S0s. This is in agreement with Laurikainen et al. (2010) where
a similar result was obtained in near-IR.

(5) At intermediate Hubble types (T=5-7) the very small bulges
gradually disappear but the galaxies can still be either barred or non-barred.
At the very end of the Hubble sequence pure disks become dominant.

A detailed analysis of the properties of bulges, disks, and bars, as a function
of morphological type, and other overall properties (galaxy mass and global color)
will be presented in forthcoming paper (Paper 2 in preparation).

\section{Acknowledgments}

H. Salo, E. Laurikainen, and S. Comeron acknowledge the Academy of
Finland for support. J. Laine was supported by the V\"ais\"al\"a
Foundation Grant. E.A. and A.B acknowledge the CNES (Centre National
d'Etudies Spatiales - France) for financial support. The authors also
acknowledge support from the FP7 Marie Curie Actions of the European
Commission, via the Initial Training Network DAGAL under REA grant
agreement number 289313.

We thank J. Janz for discussions and for help with the initial
preparation of the web-pages.
We also thank the  Spitzer IRAC Instrument
Support Team for its help with the sigma-images.

\clearpage
\appendix

\section{Appendix: Decomposition pipeline products in IRSA}

\baselineskip 0.6cm

The results of pipeline 1-component and final multi-component
decompositions are available via the IRSA database.  For each galaxy,
decomposition output parameters ({\em outgal}-file), and input
fits-files are given. The user can refine/improve the given
multi-component models by including more components or by utilizing
additional GALFIT options for the component functions.\footnote{\baselineskip 0.5cm It is important to use the data and mask files from IRSA P4
  directories when refining the given decompositions, instead of using
  corresponding data products from P1 and P3 directories. Namely, the
  P4 {\em outgal}-files assume sky subtracted data images (with NaN
  image values removed) and EXPTIME keywords set to 1
  sec. Additionally, the (P4 vs (P1 $\&$ P3)) data and mask images may
  have small spatial shifts (a few pixels) and correspond to slightly
  different sky background levels, depending on when the various pipeline products were finalized.}




{\scriptsize
\baselineskip 0.4cm
\begin{verbatim}

------------------------------------------
1) Input data for GALFIT decompositions 
------------------------------------------

IDE is the galaxy designation (e.g. NGC1097)

fits-files:
  IDE.phot.1_nonan.fits     = 3.6 micron image used in decompositions,
                              Bad pixel values (NaN's) removed
                              header modified to make GALFIT work correctly
  IDE.1.finmask_nonan.fits  = corresponding mask-file
  IDE.phot.1_sigma.fits_ns  = -"- sigma-image
  PSF-1.composite.fits      = PSF-image 

------------------------------------------
2) Output from GALFIT decompositions
------------------------------------------

a) ascii-files:

  IDE_onecomp.outgal       =    Automatic best fit parameters for 1-component S\'ersic model
  IDE_twocomp.outgal       =    -"-  for 2-component sersic+expdisk (or sersic-edgedisk) fit
  IDE_MODEL.outgal         =    Final decomposition model with up to 4 different components

  IDE is the galaxy designation (e.g. NGC1097)
  MODEL-string identifies the components included in the final multi-component model:

        'b'    indicates bulge-component
        'd'    indicates disk -"-
        'bar'  indicates non-axisymmetric structure, mainly bars
               'barf' -> length of the bar was fixed in the decompositions  
        'n'    indicates nucleus (or nonresolved bulge)
        'z'    indicates edge-on disk

         e.g. "NGC1415_bdbar"  -> bulge+disk+bad final decomposition model

  These outgal-files contain the decomposition output parameters
  Together with the input fits-files, the user can immediately repeat/refine the decompositions
  starting from the outgal-file (e.g. outgal -o1 NGC1415_bdbar.outgal   -> re-creates the decomposition)


b) fits-file (for the FINAL model)

  IDE_MODEL.outgal_subcomps.fits      Final decomposition output images: 
                                      extension 1 =  OBS image
                                      extension 2,3,4... = model components 
                                      File header contains also final decomposition parameters.

c) jpg-files (for the FINAL MODEL)

   IDE_MODEL.outgal_profile.jpg        Decomposition model compared with observations:
                                       - shows surface brightness at each image pixel vs distance from galaxy center
                                         Observed image, model image, and model components displayed separately
                                       - Collects also decomposition input & output parameters
                                         Labels indicate relative contribution of model components  

   IDE_MODEL.outgal_residual.jpg       Model-observation comparisons:
                                         upper row: clipped 3.6 micron image, masked image
                                         lower row: model image, OBS-MODEL residual
   				         
   IDE_MODEL.outgal_1dprof.jpg         Decomposition model profiles compared with observations: 
                                       - shows surface brightness as a function of isophotal semi-major axis,
                                         comparing IRAF ellipse fits to the observed image and to the
                                         model image (using isophotes of the observed image)
					

   IDE_MODEL.outgal_components.jpg     Schematic plot of model components:
                                       - Upper row: Observed image and model image, with different model
                                         components marked: colors correspond to profile plots, and the semimajor-axis
                                         of the ellipse is 2 times the effective radius of the components
                                       - Lower row: same as upper row, except projected to the disk plane
                                         (assuming zero-thickness). Empty in case on edge-on final model.

====================================================================================================================
  The final decomposition models, as well as various intermediate steps involved, are illustrated in the web-page
  http://www.oulu.fi/astronomy/S4G_PIPELINE4/MAIN
====================================================================================================================

\end{verbatim}

}
\newpage

\section{Appendix: Decomposition Pipeline web-pages}

The input data used in decompositions, the various steps of the
decomposition pipeline, and the final decomposition
results are illustrated on the Pipeline 4 web site\footnote{
The website does not contain actual fits-files, which are
available via the IRSA server/P4.}

{\tt http://www.oulu.fi/astronomy/S4G\_PIPELINE4/MAIN}

\noindent The web site consists of three layers of pages:

1) Main page

2) Index pages

3) Decomposition pages

\noindent The {\bf main page} gives full instructions regarding  the contents
of the pages, followed by an alphabetical list of all 2352 galaxies in
the original sample 
Clicking on any of the
galaxy names opens a corresponding {\em index page}, with contains information
for 100 galaxies, near and including the chosen galaxy
(Fig. \ref{fig_screen1}).

\noindent The icons in the {\bf index page} indicate the data available for each galaxy

{\parskip 0.cm
1) P1 image mosaic

2) clipped P1 image, with mask

3) clipped P1 image

4) Deprojected image

5) Elliptical isophote profiles

6) Galaxy center/sky background plots

\vskip .1cm
What decompositions available: 

7)    1-component S\'ersic-model (icon shows residual plot)

8)    Final-model (icon shows residual plot)

9)    Final-model (icon shows model components)
}

An empty icon signifies no data/decomposition model, for instance if
the galaxy was discarded from further analysis because of a nearby
bright star, etc.  In particular, the absence of a final model indicates
that the galaxy was considered too problematic to fit reliably
(e.g. closely interacting, very peculiar or warped).

Clicking on the galaxy name links to its {\bf decomposition page}
(Fig. \ref{fig_screen2}), which summarizes the mask, ellipse fitting,
sky background, and galaxy center determinations, as described
in Section (\ref{sect_data}).

In addition, the pages display for all three types of models
(1-component S\'ersic model, 2-component bulge/disk model , final multi-component model) the

{\parskip 0cm
2D profile plot

Residual plot

1D profile plot

Model-components plot
}

Again, clicking on the plots displays the full size plots
(Figs. \ref{fig_screen3} \ref{fig_screen4}, \ref{fig_screen5}, \ref{fig_screen6}).


\begin{figure}[h]
  \centering
  \includegraphics[width=0.95\textwidth]{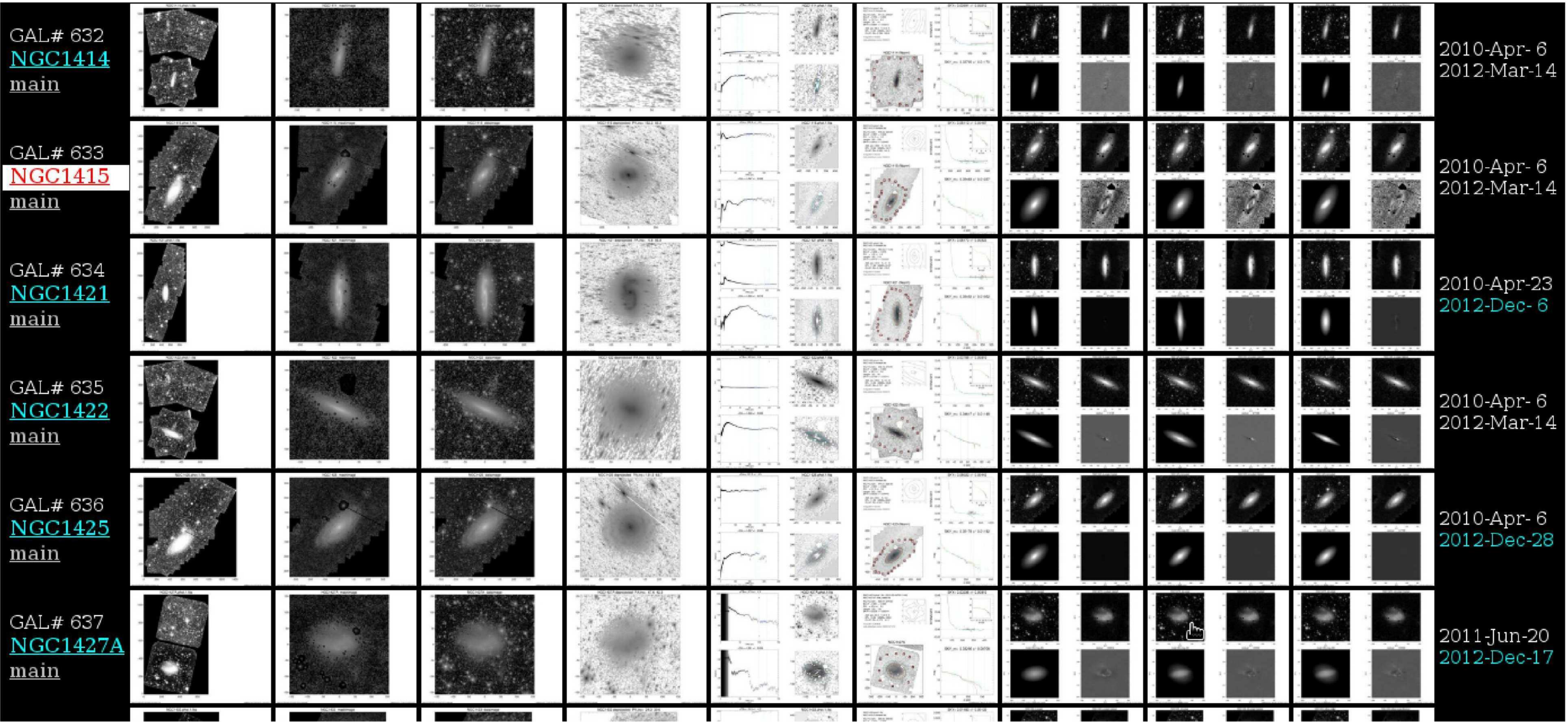}
  \caption{Screenshot of Pipeline 4 index page.
    \label{fig_screen1}}
\end{figure}

\begin{figure}[h]
\centering
  \includegraphics[width=0.95\textwidth]{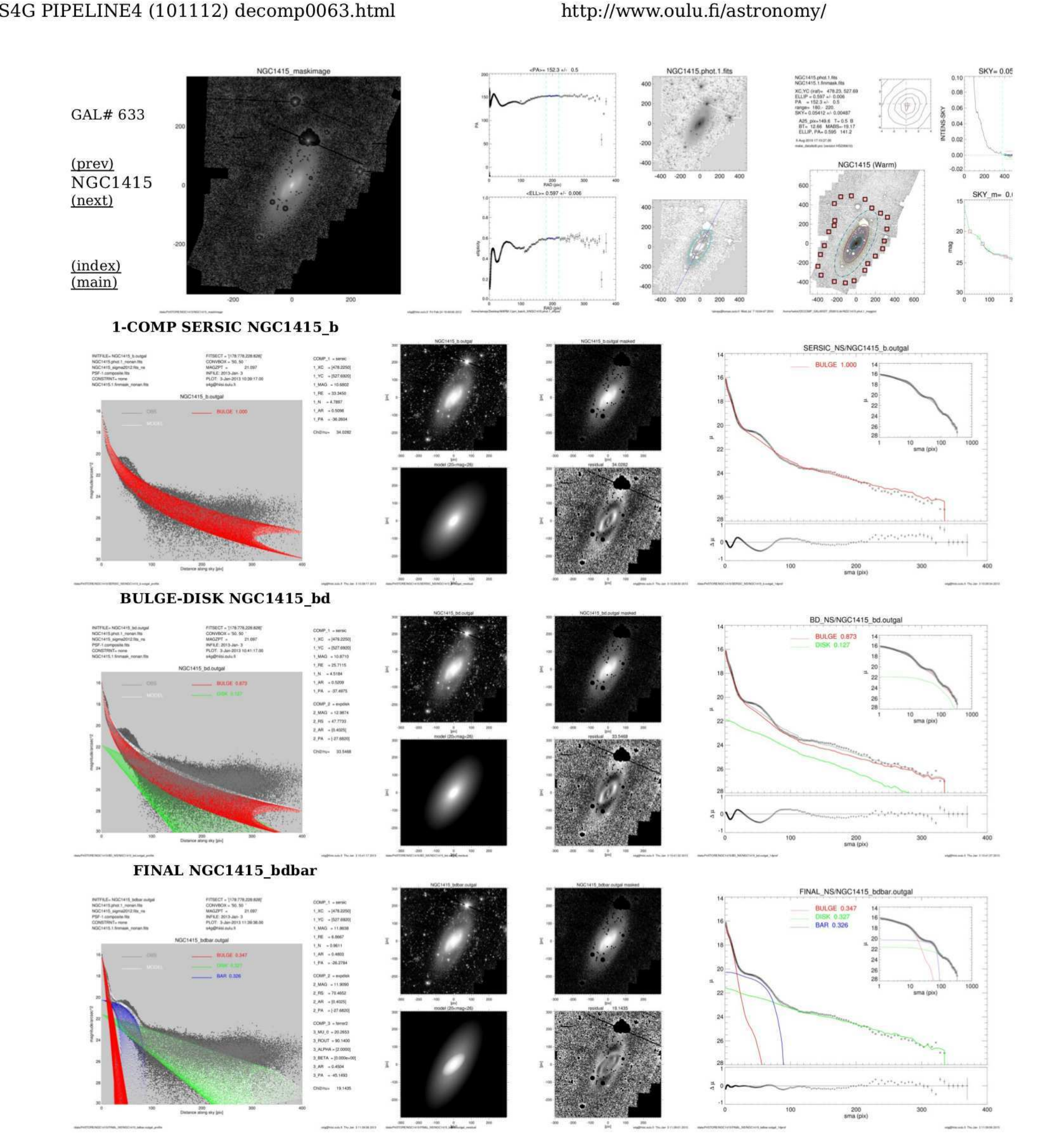}
  \caption{Screenshot of various decomposition models for this particular galaxy. Clicking on the
    image icons opens the enlarged image.
  \label{fig_screen2}}
\end{figure}

\begin{figure}[h]
\centering
  \includegraphics[width=0.95\textwidth]{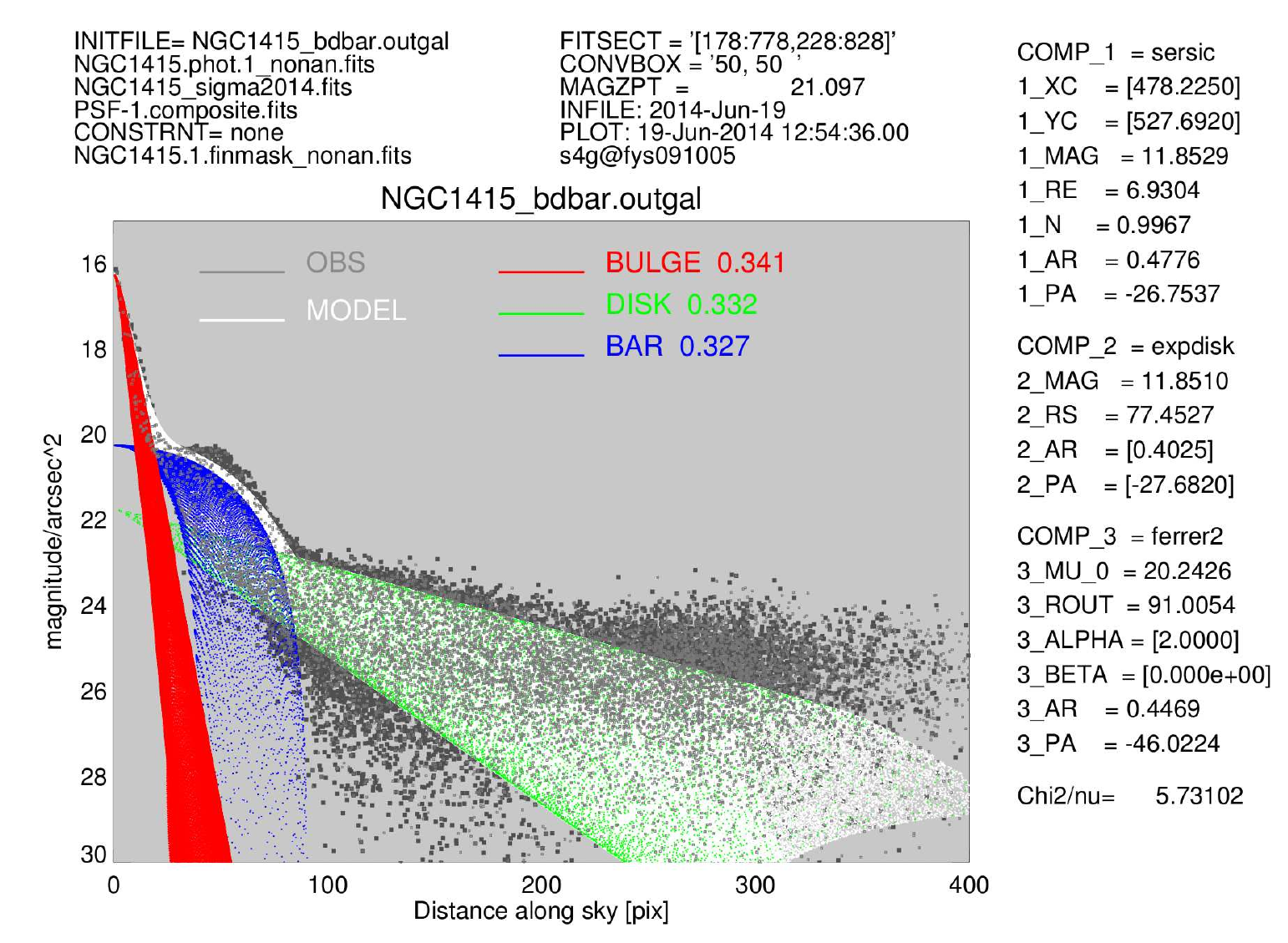}
  \caption{Final {\em pipeline} decomposition for NGC 1415. The
    decomposition includes bulge, disk, and bar components, indicated
    by different colors. The numbers after the labels indicate the
    relative fraction of light in each component.  This 2D-profile
    indicates the brightness of each pixel versus its distance from
    the galaxy center.  The frame also collects the names of the input
    data files and the final GALFIT decomposition parameter values.
  \label{fig_screen3}}
\end{figure}

\begin{figure}[h]
\centering
  \includegraphics[width=0.95\textwidth]{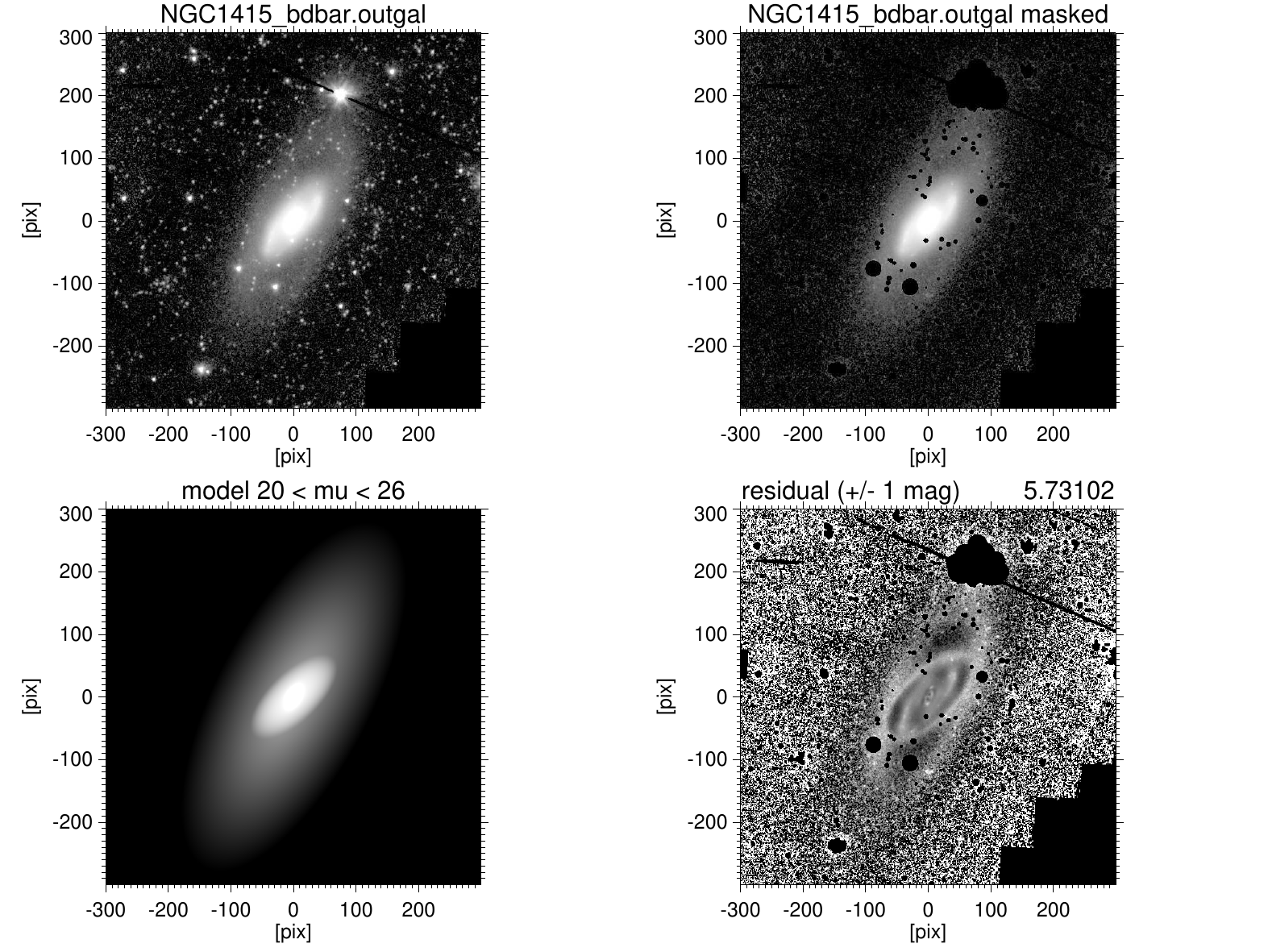}
  \caption{Final {\em pipeline} decomposition for NGC 1415,
    corresponding to Fig. \ref{fig_screen3}. The upper frames show the
    observed, clipped and sky-subtracted image (left) and the
    corresponding masked image (right). The lower frames display the
    model image (left), and the observed-model residual image
    (gray scale covers $\pm 1$ mag).
  \label{fig_screen4}}
\end{figure}

\begin{figure}[h]
\centering
\includegraphics[angle=90,width=0.95\textwidth]{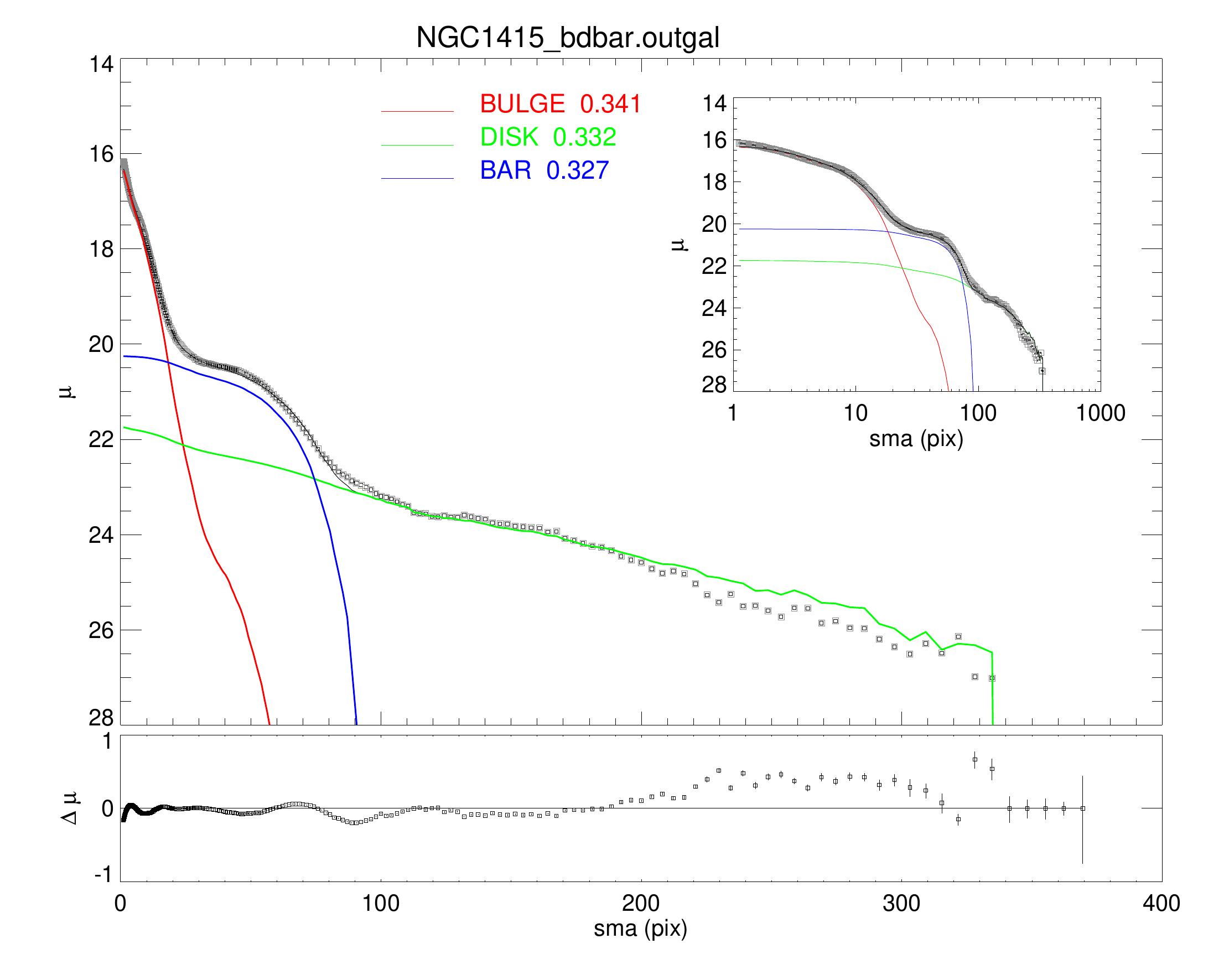}
  \caption{One dimensional profiles corresponding to
    Fig. \ref{fig_screen3}. The symbols indicate the azimuthally
    averaged surface brightness as a function of semi-major axis,
    obtained with IRAF {ellipse}-routine. The curves indicate the azimuthally averaged
profiles of the model components, using the same isophotes. 
  \label{fig_screen5}}
\end{figure}

\begin{figure}[h]
\centering
\includegraphics[angle=0,width=0.95\textwidth]{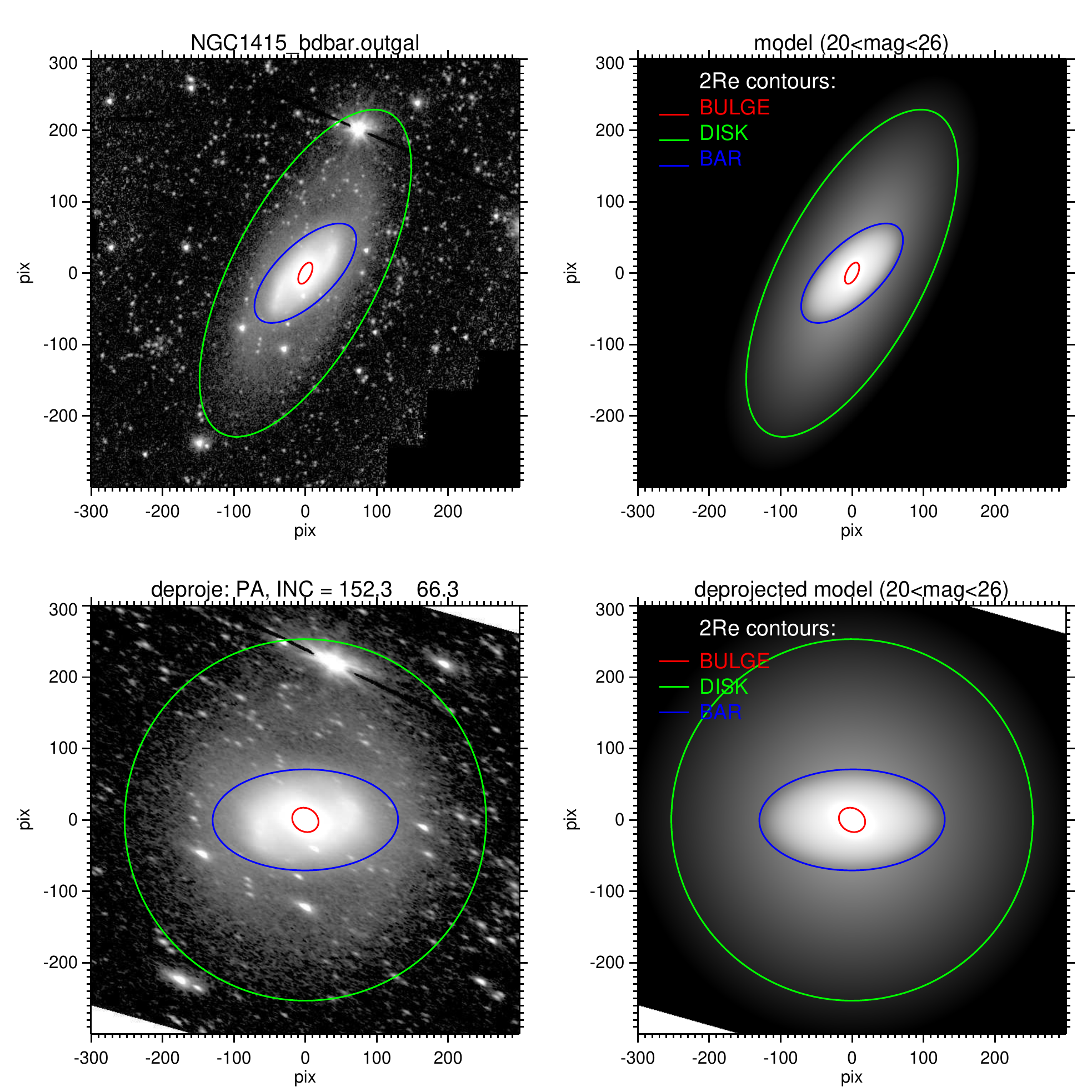}
  \caption{Illustration of the model components corresponding to
    Fig. \ref{fig_screen3}.  The upper left shows the observed image,
    together with superposed ellipses illustrating the various
    components of the final model: the size of the ellipse corresponds
    to 2 effective radii, the orientation corresponds to the
    components' axial ratio and PA. In the upper right, the same but
    showing the model image.  The lower frames are similar, except
    that the observed galaxy and the model have been deprojected,
    using the assumed disk orientation parameters. 
  \label{fig_screen6}}
\end{figure}

\clearpage

\newpage
}


\begin{thebibliography}{}
{\reply
\bibitem[Athanassoula (2005)]{atha2005} Athanassoula, E., 2005, \mnras, 358, 1477
}

{\reply
\bibitem[Athanassoula et al.(2014)]{atha2014} Athanassoula, E., 
Laurikainen, E., Salo, H., \& Bosma, A.\ 2014, arXiv:1405.6726 
}

  \bibitem[Allen et al. (2006)]{allen2006} Allen, P., Driver, S., Graham, A., Cameron, E., Liske, J., de Propris, R., 2006, \mnras, 371, 2
  \bibitem[Athanassoula et al. (1990)]{atha1990} Athanassoula, E., Morin, S., Wozniak, H., Puy, D., Pierce, M., Lombard, J., Bosma, A., 1990, \mnras, 245, 130
  \bibitem[Bacos \& Trujillo (2012)]{bacos2012} Bakos, J., Trujillo, I., 2012, MSAIS, 25, 21
  \bibitem[Barazza et al. (2008)]{barazza2008} Barazza, F., Jogee, S., Marinova, I., 2008, \apj, 675, 1194 
  \bibitem[Bertin \& Arnouts (1996)]{bertin1996} Bertin, E., Arnouts, S. 1996, \aaps, 117, 393
\bibitem[{{Buta} {et~al.}(2010){Buta}, {Sheth}, {Regan}, {Hinz}, {Gil de Paz},
  {Men{\'e}ndez-Delmestre}, {Munoz-Mateos}, {Seibert}, {Laurikainen}, {Salo},
  {Gadotti}, {Athanassoula}, {Bosma}, {Knapen}, {Ho}, {Madore}, {Elmegreen},
  {Masters}, {Comer{\'o}n}, {Aravena}, \& {Kim}}]{buta2010}
  {Buta}, R.~J., {Sheth}, K., {Regan}, M., {et~al.} 2010, \apjs, 190, 147
\bibitem[{Buta} {et~al.}(2014)]{buta2014} Buta R.~J. et al., 2014. Submitted to \apjs

{\reply
\bibitem[Combes \& Sanders(1981)]{combes1981} Combes, F., \& Sanders, R.~H.\ 1981, \aap, 96, 164 
}

  \bibitem[Cameron et al. (2009)]{cameron2009} Cameron, E., Driver, S., Graham, A., Liske, J., 2009, \apj, 699, 105
  \bibitem[Cappellari et al. (2013)]{cappellari2013}  Cappellari, M., McDermid, R., Alatalo, K., Blitz, L., Bois, M., Bournaud, F., Bureau, M., Crocker, A., Davies, R., Davis, T. and 14 coauthors, 2013,  \mnras, 432, 1862
%
\bibitem[Comer\'on et al. (2011)]{comeron2011} Comer{\'o}n, S., {Elmegreen}, B.~G., {Knapen}, J.~H., 
     {Sheth}, K., {Hinz}, J.~L., {Regan}, M.~W., {Gil de Paz}, A., 
     {Mu{\~n}oz-Mateos}, J.-C., {Men{\'e}ndez-Delmestre}, K., 
     {Seibert}, M., {Kim}, T., {Mizusawa}, T., {Laurikainen}, E., and 10 co-authors, 2011, 738, L17
%
 \bibitem[Comer\'on et al. (2012)]{comeron2012} Comer{\'o}n, S., {Elmegreen}, B.~G., {Salo}, H., 
     {Laurikainen}, E., {Athanassoula}, E., {Bosma}, A., 
     {Knapen}, J.~H., {Gadotti}, D.~A.,{Sheth}, K.  and 10 coauthors, 2012, \apj, 759, 98

\bibitem[Comer{\'o}n et al.(2014)]{comeron2014} Comer{\'o}n, S., 
Elmegreen, B.~G., Salo, H., et al.\ 2014, arXiv:1409.0466 


{\reply
\bibitem[C{\^o}t{\'e} et al.(2006)]{cote2006} C{\^o}t{\'e}, P., 
Piatek, S., Ferrarese, L., et al.\ 2006, \apjs, 165, 57 
}

\bibitem[\protect\citeauthoryear{Courteau et al.}{2007}]{courteau2007} Courteau S., Dutton A.~A., van den Bosch F.~C., MacArthur L.~A., Dekel A., McIntosh D.~H., Dale D.~A., 2007, ApJ, 671, 203 


\bibitem[{{de Souza} {et~al.}(2004){de Souza}, {Gadotti}, \& {dos
  Anjos}}]{desouza2004}
{de Souza}, R.~E., {Gadotti}, D.~A., \& {dos Anjos}, S. 2004, \apjs, 153, 411

\bibitem[{{de Vaucouleurs} {et~al.}(1991){de Vaucouleurs}, {de Vaucouleurs},
  {Corwin}, {Buta}, {Paturel}, \& {Fouqu{\'e}}}]{rc3}
{de Vaucouleurs}, G., {de Vaucouleurs}, A., {Corwin}, Jr., H.~G., {et~al.}
  1991, (RC3)


  \bibitem[Driver et al. (2006)]{driver2006} Driver, S., Allen, P., Graham, A., Cameron, E., Liske, J., Ellis, S., Cross, N., De Propris, R., Phillipps, S., Couch, W., 2006, \mnras, 368, 414

  \bibitem[Driver et al. (2013)]{driver2013} Driver, S., Robotham, A., Bland-Hawthorn, J., Brown, M., Hopkins, A., Liske, J., Phillipps, S., Wilkins, S.,  2013, \mnras, 430, 2622
  \bibitem[Draine \& Lee (1984)]{draine1984} Draine, B., Lee, H., 1984, \apj, 285, 89

  \bibitem[Erwin, Beckman \& Pohlen (2005)]{erwin2005} Erwin, P., Beckman, J., Pohlen, M., 2005,  \apj, 626, L81 

\bibitem[Eskridge et al.(2002)]{esk2002} Eskridge, P. B., Frogel, J. A., Pogge, R. W., Quillen, A. C., Berlind, A. A., Davies, R. L., DePoy, D. L., Gilbert, K. M., Houdashelt, M. L., Kuchinski, L. E., and 5 coauthors, 2002, \apjs, 143, 73



\bibitem[Fisher (2006)]{fisher2006} Fisher D., 2006, \apjl, 642, 17



{\reply
\bibitem[Fisher 
\& Drory(2010)]{fisher2010} Fisher, D.~B., \& Drory, N.\ 2010, \apj, 716, 942 
}
  \bibitem[Freeman (1970)]{freeman1970} Freeman K., 1970, \apj, 160, 811


\bibitem[{{Gadotti}(2001)}]{gadotti2001}{Gadotti}, D.~A.,  dos Anjos, S.,  2001, \aj, 122, 1298
\bibitem[{{Gadotti}(2008)}]{gadotti2008}{Gadotti}, D.~A. 2008, \mnras, 384, 420

  \bibitem[Gadotti (2009)]{gadotti2009} Gadotti D.~A., 2009, \mnras, 393, 1531
  \bibitem[Gadotti (2011)]{gadotti2011} Gadotti D.~A., 2011, \mnras, 415, 3308



  \bibitem[Graham \& Worley (2008)]{graham2008} Graham A., Worley C.C., \mnras, 388, 1708

\bibitem[\protect\citeauthoryear{Guti{\'e}rrez et al.}{2011}]{gutierrez2011} Guti{\'e}rrez L., Erwin P., Aladro R., Beckman J.~E., 2011, AJ, 142, 145 

  \bibitem[H\"au{\ss}ler (2013)]{hausler2013} H\"au{\ss}ler, B., Bamford, S., Vika, M., Rojas, A., Barden, M., Kelvin, L., Alpaslan, M., Robotham, A., Driver, S., Baldry, I., and 6 coauthors,  2013, \mnras, 430, 330

  \bibitem[Huang et al. (2013)]{huang2013} Huang, S., Ho, L., Peng, C., Li, Z. Barth, A., 2013, \apj, 766, 47


\bibitem[Janz et al.(2014)]{janz2014} Janz, J., Laurikainen, E., 
Lisker, T., et al.\ 2014, \apj, 786, 105 



  \bibitem[Jedrzejewski (1987)]{jerd1987} Jedrzejewski R. I., 1987, \mnras, 226, 747
  \bibitem[Lackner \& Gunn (2012)]{lackner2012} Lackner, C., Gunn, J., 2012, \mnras, 421, 2277
  \bibitem[Kim et al. (2014)]{kim2014} Kim, T., Gadotti, D.~A., Sheth, K., et al.,  2014, \apj, 782, 64


\bibitem[Knapen et 
al.(2014)]{knapen2014} Knapen, J.~H., Erroz-Ferrer, S., Roa, J., et al.\ 2014, \aap, 569, A91 

{\reply
\bibitem[Kormendy(1979)]{kormendy1979} Kormendy, J.\ 1979, \apj, 
227, 714 
}

{\reply
\bibitem[Kormendy(1982)]{kormendy1982} Kormendy, J.\ 1982, \apj, 
257, 75 
}

{\reply
\bibitem[Kormendy(1993)]{kormendy1993} Kormendy, J.\ 1993, Galactic 
Bulges (IAU Symposium 153), p. 209 
}


{\reply
\bibitem[Kormendy 
\& Bender(1996)]{kormendy1996} Kormendy, J., \& Bender, R.\ 1996, \apjl, 464, L119 
}


\bibitem[Kormendy et al.(2009)]{kormendy2009} Kormendy, J., Fisher, 
D.~B., Cornell, M.~E., \& Bender, R.\ 2009, \apjs, 182, 216 



  \bibitem[Kormendy \& Barentine (2010)]{kormendy2010} Kormendy, J., Barentine J.C., 2010, \apj, 715, 176


{\reply
\bibitem[Kormendy \& Bender(2012)]{kormendy2012} Kormendy, J., \& Bender, R.\ 2012, \apjs, 198, 2 
}

{\reply
\bibitem[Kormendy \& Kennicutt(2004)]{kormendy2004} Kormendy, J., \&
  Kennicutt, R.~C., Jr.\ 2004, \araa, 42, 603
}

\bibitem[Laine et al.(2014)]{laine2014} Laine, J., Laurikainen, 
E., Salo, H., et al.\ 2014, \mnras, 441, 1992 

\bibitem[Landsman (1993)]{landsman1993} Landsman 1993, in Astronomical Data Analysis Software and Systems II, A.S.P. Conference Series, Vol. 52, ed. R. J. Hanisch, R. J. V. Brissenden, and Jeannette Barnes,  p. 246.


{\reply
\bibitem[Lauer(1985)]{lauer1985} Lauer, T.~R.\ 1985, \apj, 292, 
104 
}

  \bibitem[Laurikainen et al.(2005)]{laurikainen2005} Laurikainen, E., Salo, H., Buta, R., 2005, \mnras, 362, 1319 

  \bibitem[Laurikainen et al.(2006)]{laurikainen2006} Laurikainen, E., Salo, H., Buta, R., Knapen, J., Speltincx, T., Block, D. L., 2006, \aj,
   132, 2634

  \bibitem[Laurikainen et al.(2007)]{laurikainen2007} Laurikainen, E., Salo, H., Buta, R., \& Knapen, J. H., 2007, \mnras, 381, 401

  \bibitem[Laurikainen et al.(2009)]{laurikainen2009} Laurikainen, E., Salo, H., Buta, R., \& Knapen, J. H., 2009, \apjl, 692, 34

  \bibitem[Laurikainen et al.(2010)]{laurikainen2010}  Laurikainen, E., Salo, H., Buta, R., Knapen, J. H., Comer\'on, S., 2010, \mnras, 405, 1089

\bibitem[Laurikainen et al.(2011)]{laurikainen2011} Laurikainen, E., 
Salo, H., Buta, R., \& Knapen, J.~H.\ 2011, \mnras, 418, 1452 

  \bibitem[Laurikainen et al.(2014)]{laurikainen2014}  Laurikainen, E., Salo, H., Athanassoula, E, Bosma, A, Herrera-Endoqui, M. 2014, \mnras, 444, L80


\bibitem[Mart{\'{\i}}n-Navarro et al.(2012)]{martin_navarro2012} 
Mart{\'{\i}}n-Navarro, I., Bakos, J., Trujillo, I., et al.\ 2012, \mnras, 
427, 1102 



\bibitem[{{Meidt} {et~al.}(2012{\natexlab{}}){Meidt}, {Schinnerer}, {Knapen},
  {Bosma}, {Athanassoula}, {Sheth}, {Buta}, {Zaritsky}, {Laurikainen},
  {Elmegreen}, {Elmegreen}, {Gadotti}, {Salo}, {Regan}, {Ho}, {Madore}, {Hinz},
  {Skibba}, {Gil de Paz}, {Mu{\~n}oz-Mateos}, {Men{\'e}ndez-Delmestre},
  {Seibert}, {Kim}, {Mizusawa}, {Laine}, \&
  {Comer{\'o}n}}]{meidt2012}
{Meidt}, S.~E., {Schinnerer}, E., {Knapen}, J.~H., {et~al.} 2012{\natexlab{}},
  \apj, 744, 17

\bibitem[Melvin et al.(2014)]{melvin2014} Melvin, T., Masters, K., 
Lintott, C., et al.\ 2014, \mnras, 438, 2882 




\bibitem[Mu{\~n}oz-Mateos et al.(2013)]{munoz_mateos2013} 
Mu{\~n}oz-Mateos, J.~C., Sheth, K., Gil de Paz, A., et al.\ 2013, \apj, 
771, 59 

  \bibitem[Mu\~{n}oz--Mateos et al. (2014)]{munoz_mateos2014} Mu\~{n}oz--Mateos J. C. et al.,
   2014, ApJS, submitted


  \bibitem[Nair et al. (2010)]{nair2010} Nair, P., Abraham, R.,  2010, ApJ, 714, 260

  \bibitem[Pahre et al. (2004)]{pahre2004} Pahre, M. A., Ashby, M. L. N., Fazio, G. G., Willner, S. P. 2004, \apjs, 154, 235
  \bibitem[Peletier et al. (2012)]{peletier2012} Peletier, R., Kutdemir, E., van der Wolk, G., Falc\'on-Barroso, J., Bacon, R., Bureau, M., Cappellari, M., Davies, R., de Zeeuw, P., Emsellem, E., and 8 coauthors, 2012, \mnras, 419, 2031
 \bibitem[Peng et al. (2002)]{peng2002}  Peng, C., Ho, L., Impey, C. Rix, H., 2002, \aj, 124, 266
 \bibitem[Peng et al. (2010)]{peng2010} Peng, C., Ho, L., Impey, C., Rix, H.,  2010 \aj, 139, 2097

\bibitem[\protect\citeauthoryear{Pohlen \& Trujillo}{2006}]{pohlen2006} Pohlen M., Trujillo I., 2006, A\&A, 454, 759 
  


\bibitem[Querejeta et al.(2014)]{2014arXiv1410.0009Q} Querejeta, M., Meidt, 
S.~E., Schinnerer, E., et al.\ 2014, arXiv:1410.0009 

  \bibitem[Regan et al. (2014)]{regan2014} Regan, M. et al.,
   2014, in prep. 


{\reply
\bibitem[Rest et al.(2001)]{rest2001} Rest, A., van den Bosch, 
F.~C., Jaffe, W., et al.\ 2001, \aj, 121, 2431 
}

 \bibitem[Sellwood (2008)]{sellwood2008} Sellwood, J.A., 2008, in 'Dynamical Evolution of Disk Galaxies', Astronomical Society of the Pacific Conference Series 396, eds. .~G.~Funes \& E.~M.~Corsini


\bibitem[Sheth (2008)]{sheth2008} Sheth, K. 2008 et al \apj 675, 1141

\bibitem[\protect\citeauthoryear{Sheth et al.}{2010}]{sheth2010} Sheth K., et al., 2010, PASP, 122, 1397 

{\reply
\bibitem[Sheth et al.(2013)]{sheth2013} Sheth, K., Armus, L., 
Athanassoula, E., et al.\ 2013, Spitzer Proposal, 10043 
}

\bibitem[Tasca 
\& White(2011)]{tasca2011} Tasca, L.~A.~M., \& White, S.~D.~M.\ 2011, \aap, 530, A106 


\bibitem[van der Kruit \& Searle (1981)]{vanderkruit1981} van der Kruit, P. C., Searle L., 1981 \aap, 95, 105


{\reply \bibitem[van der Wel(2008)]{vanderwel2008} van der Wel, A.\ 2008, 
\apjl, 675, L13 }


  \bibitem[Weinzirl et al. (2009)]{weinzirl2009} Weinzirl, T., Jogee, S., Khochfar, S., Burkert, A., Kormendy, J. 2009 \apj, ~696, 411.

{\reply
\bibitem[Weinzirl et al.(2014)]{weinzirl2014} Weinzirl, T., Jogee, 
S., Neistein, E., et al.\ 2014, \mnras, 441, 3083 
}


\bibitem[Zaritsky et al.(2013)]{zaritsky2013} Zaritsky, D., Salo, 
H., Laurikainen, E., et al.\ 2013, \apj, 772, 135 



\end{thebibliography}
\end{document}